%% file: manuscript.tex
\documentclass[%
11pt,
reprint,
onecolumn,
tightenlines,
superscriptaddress,
preprintnumbers,
nofootinbib,
amsmath,amssymb,amsthm,
physrev,
eqsecnum,tikz,
]{revtex4-2}

\input{preamble}

\newcommand{\vareps}{\varepsilon}
\newcommand{\bfvareps}{\mathbold {\varepsilon}}

\definecolor{hidecolor}{cmyk}{0.10,0.1,0,0.1}

\newcommand{\Wdlin}{W_{\mathrm{d},\mathrm{lin}}} 

\usepackage{comment}
\usepackage{xcolor}
\usepackage{siunitx}

\begin{document}


\preprint{To appear in International Journal for Numerical and Analytical Methods in Geomechanics (\url{https://doi.org/10.1002/nag.3933}).}

\title{Phase-Field Modeling of Fracture under Compression and Confinement in Anisotropic Geomaterials}

\author{Maryam Hakimzadeh}
    \email{mhakimza@andrew.cmu.edu}
    \affiliation{Department of Civil and Environmental Engineering, Carnegie Mellon University}

\author{Carlos Mora-Corral \orcidlink{0000-0002-2412-5508}}
    \affiliation{Departamento de Matem\'aticas, Universidad Aut\'onoma de Madrid, 28049 Madrid, Spain}
    \affiliation{Instituto de Ciencias Matem\'aticas, CSIC-UAM-UC3M-UCM, 28049 Madrid, Spain}
    
\author{Noel Walkington}
    \affiliation{Center for Nonlinear Analysis, Department of Mathematical Sciences, Carnegie Mellon University}

\author{Giuseppe Buscarnera}
    \affiliation{Department of Civil and Environmental Engineering, Northwestern University}

\author{Kaushik Dayal \orcidlink{0000-0002-0516-3066}}
    \affiliation{Department of Civil and Environmental Engineering, Carnegie Mellon University}
    \affiliation{Center for Nonlinear Analysis, Department of Mathematical Sciences, Carnegie Mellon University}
    \affiliation{Department of Mechanical Engineering, Carnegie Mellon University}

\date{\today}


\begin{abstract}

    Strongly anisotropic geomaterials, such as layered shales, have been observed to undergo fracture under compressive loading.
    This paper applies a phase-field fracture model to study this fracture process.
    While phase-field fracture models have several advantages --- primarily that the fracture path is not pre-determined but arises naturally from the evolution of a smooth non-singular damage field --- they provide unphysical predictions when the stress state is complex and includes compression that can cause crack faces to contact.

    Building on a recently-developed phase-field model that accounts for compressive traction across the crack face, this paper extends the model to the setting of anisotropic fracture.
    The key features of the model include:
    (1) a homogenized anisotropic elastic response and strongly-anisotropic model for the work to fracture;
    (2) an effective damage response that accounts consistently for compressive traction across the crack face, that is derived from the anisotropic elastic response;
    (3) a regularized crack normal field that overcomes the shortcomings of the isotropic setting, and enables the correct crack response, both across and transverse to the crack face.
    
    To test the model, we first compare the predictions to phase-field fracture evolution calculations in a fully-resolved layered specimen with spatial inhomogeneity, and show that it captures the overall patterns of crack growth.
    We then apply the model to previously-reported experimental observations of fracture evolution in laboratory specimens of shales under compression with confinement, and find that it predicts well the observed crack patterns in a broad range of loading conditions.
    We further apply the model to predict the growth of wing cracks under compression and confinement.
    Prior approaches to simulate wing cracks have treated the initial cracks as an external boundary, which makes them difficult to apply to general settings.
    Here, the effective crack response model enables us to treat the initial crack simply as a non-singular damaged zone within the computational domain, thereby allowing for easy and general computations.

\end{abstract}

\maketitle


\section{Introduction}

Many geomaterials in the environment are anisotropic. 
In particular, transverse isotropy is a characteristic feature of various types of rock, including sedimentary and metamorphic rocks. 
Many foliated metamorphic rocks, namely schists, slates, gneisses, and phyllites, generally have a distinct directional dependence on their strength \cite{thomas1993book}. 
Furthermore, stratified / layered sedimentary structures, such as sandstone and shale, have layers that are closely spaced, resulting in a mechanical response that is transverse isotropic, that is, the response is isotropic in the bedding plane of the layers and distinct in the out-of-plane direction \cite{amadei2012rock, wittke1990rock, tien2001failure, cui2022role,chang2022crack}.
A key aspect that is the focus of this paper is modeling the failure of such rocks in settings that include compression and confinement, which is a typical stress state in natural and engineered settings.

\paragraph*{Related prior work.}

Many laboratory studies have been carried out to determine the impact of anisotropy on the strength of various transversely isotropic rocks \cite{niandou1997laboratory, bennett2015instrumented, semnani2017quantifying, tien2006experimental}. 
To evaluate the strength of the rock sample, triaxial compression tests with different confining pressures were used in these studies. 
To model this behavior, some studies have developed failure models to study the strength of rocks and its dependence on the orientation of the bedding plane. 
For example, Jaeger \cite{jaeger1960shear} extrapolated the Coulomb-Navier criterion to introduce a single plane of weakness theory, recognizing two predominant failure modes: sliding along the weak plane and failure traversing the rock matrix. 
Tien and Kuo \cite{tien2001failure} extended the failure criterion proposed by Jaeger and substituted the Mohr-Coulomb approach with the Hoek-Brown model, and then tested their model with experimental data \cite{tien2006experimental}. 

In addition to developing failure criteria, the inclusion of anisotropy within the constitutive law is important to predict the effect of the orientation of the bedding plane on the elastoplastic response and fracture of the material. 
To include the anisotropy of the elastic response in the constitutive law, it is straightforward to use classical homogenization methods, e.g. \cite{backus1962long}.
To model the anisotropic inelastic response, many studies have extended the existing isotropic criteria to the anisotropic setting, such as generalizing the von Mises criterion \cite{hill1948theory} and the Mohr-Coulomb criterion \cite{pariseau1968plasticity} to account for anisotropy; a summary is provided in \cite{semnani2016thermoplasticity}.
In addition, recent work, for example, \cite{zhao2018strength, zhao2022double, mader2022gradient}, has investigated the effect of anisotropy on the strength of transversely isotropic rocks. 
In \cite{zhao2018strength}, they have used finite element methods to simulate the anisotropic elastoplastic constitutive model proposed by \cite{semnani2016thermoplasticity}, thereby providing insight into the deformation and strength of transversely isotropic material under axial and lateral confinements, including capturing the zigzag failure mode observed in layered rocks.  

While these prior works have focused on the role of anisotropy in the elastic and inelastic responses, the role of anisotropy in the fracture behavior has been much less studied.
However, the anisotropy of fracture is an important aspect of the overall behavior, and plays a key role in a vast range of phenomena ranging from the fracture of apple flesh \cite{khan1993anisotropy}, to single crystals \cite{qiao2003cleavage}, to geomaterials \cite{nasseri2008fracture}. 
Recent work that aims to model the anisotropy in the fracture behavior include strategies that make the fracture energy anisotropic, e.g., \cite{hakim2005crack, hakim2009laws}. However, these studies are restricted to weakly twofold anisotropic settings and cannot capture the behaviour of material with strongly anisotropic surface energies \cite{takei2013forbidden}. 

\paragraph*{Phase-field fracture models.}

Phase-field fracture models introduce a new field, in addition to the deformation, that indicates if a material element is intact or fractured, and then regularize this phase-field using higher-order derivatives.
The higher-order derivatives endow the phase-field with sufficient smoothness to enable the use of standard numerical methods to predict crack growth without specifying crack paths beforehand.
That is, rather than sharp singular cracks as in the classical approach to fracture, cracks are now regularized to be non-singular variations of the phase-field, making them amenable to standard numerical approaches.
The elastic response is coupled to the phase-field by setting that the regions that are fractured cannot carry any load, that is, the elastic energy density goes to zero in fractured regions.
However, the combination of crack regularization to a finite volume and setting the fractured elastic energy to zero causes unphysical behavior, such as crack growth under compression.

This issue, which also allows unphysical contact and interpenetration of crack faces, motivated efforts such as \cite{miehe2010thermodynamically, amor2009regularized}, to improve the phase-field model for compressive loading.
In  \cite{miehe2010thermodynamically}, they proposed a model that decomposes the elastic energy into tensile and compressive parts, and allows the crack to grow only due to the tensile part of the energy.
In \cite{amor2009regularized}, they proposed decomposing the elastic energy into compressive hydrostatic, tensile hydrostatic, and deviatoric parts, and allowed the crack to sustain only the compressive hydrostatic loads.
This has been extended to the finite deformation setting in \cite{tian2022mixed,xing2023adaptive,najmeddine2024efficient,tang2019phase}; however, these works also do not consider the orientation of the crack.
Other modifications of the energy decompositions include \cite{vicentini2023energy} that proposed the star-convex model as a modification of the volumetric-deviatoric decomposition; 
\cite{wang2021phase} that extended the spectral decomposition to account for mixed mode fracture;
and \cite{van2020strain} to account for anisotropy.
These approaches are reviewed and compared in \cite{zhang2022assessment,gupta2024damage,rahaman2022open,clayton2021nonlinear} and elsewhere.
However, these approaches have a key shortcoming: they do not account for the orientation of the crack.
Compressive stresses along the crack face and across the crack face are both treated in exactly the same way; i.e., tension/compression across the crack faces is indistinguishable from tension/compression along the crack faces in these models.

An important departure from the idea of energy decompositions is the work by \cite{steinke2019phase}, that introduced the idea that the crack orientation plays an essential role in determining the response.
Based on this idea, \cite{steinke2019phase,fei2020phase-cmame,fei2020phase-ijnme,agrawal2016multiscale} proposed modifications of the stress tensor to account for the difference in crack-normal and crack-parallel tractions.
While our approach in \cite{hakimzadeh2022phase} builds on the ideas of \cite{steinke2019phase}, an important distinction is that we modify the energy and hence our model is hyperelastic.
In contrast, as shown in \cite{hakimzadeh2022phase}, the stress-based approach is not generally hyperelastic.

To account for the role of material anisotropy of the fracture process, taking inspiration from phase-field models of crystal growth, \cite{li2015phase} proposed a strongly anisotropic surface energy model by using higher derivatives of the fracture phase parameter.
This results in a fourth-order system of partial differential equations (PDEs) for the displacement and crack (phase-field) variables, which they solve using a finite element method with additional smoothness.
Building on this, \cite{teichtmeister2017phase, bijaya2023consistent, zhang2021phase} developed a comprehensive framework to account for general anisotropies in the phase-field modeling framework.

\paragraph*{Contributions of this paper.}

In this paper, we build on our previous work \cite{hakimzadeh2022phase}, wherein the issue of compressive stresses across the crack face was addressed.
That work proposed a model that gave the correct intact response of the material when crack faces came in contact, while giving the appropriate soft response for crack opening or sliding of the crack faces past each other.
An important element of that approach was a vector phase-field parameter, whose orientation provided the crack normal and whose magnitude provided the level of fracture.
However, as we discuss further below, the vector phase-field parameter is only appropriate for the isotropic setting that was studied in \cite{hakimzadeh2022phase}.

Our prior work \cite{hakimzadeh2022phase} was focused on homogeneous isotropic solids. 
In this work, we study inhomogeneous solids and their homogenized response\footnote
{
    We use homogenized in this paper to mean an equivalent homogeneous material that provides approximately the same response in terms of load-displacement and overall crack path.
} 
which is effectively anisotropic. 
To be applicable to general anisotropic settings, we therefore use in this work a scalar phase-field variable and introduce a variational method to extract a regularized crack normal field from the primary scalar phase-field variable.
Further, we integrate the proposed phase-field model with the general model for anisotropic fracture energy developed in \cite{li2015phase} and \cite{teichtmeister2017phase}.
The model that we develop is then able to model the unguided growth of fractures in anisotropic homogenized specimens under complex loading that includes compression and lateral confinement.
We first compare the homogenized model predictions to phase-field fracture evolution calculations in a fully-resolved layered specimen with spatial inhomogeneity, i.e., a specimen wherein the individual layers are accounted for by spatial inhomogeneity in the elastic and fracture properties (Fig. \ref{fig:Fully resolved and mesh}).
We show that the overall patterns of crack growth are similar.
We then apply the model to previously-reported experimental observations of fracture evolution in laboratory specimens of shales under compression with confinement \cite{tien2006experimental}, and find that it predicts well the observed crack patterns in a broad range of loading conditions.
We finally apply the model to predict the growth of wing cracks under compression and confinement.
Prior approaches to simulate wing cracks have treated the initial cracks as an external boundary \cite{li2021phase, spetz2021modified, bryant2018mixed}, which makes them difficult to apply to general settings.
Here, the effective crack response model enables us to treat the initial crack simply as a non-singular damaged zone within the computational domain, thereby allowing for easy and general computations.

\paragraph*{Structure of the paper.}

In Section \ref{sec:formulation}, we describe the formulation, numerical implementation, and validation against a fully-resolved specimen.
In Section \ref{sec:comparison}, we compare the predictions of the homogenized fracture model with those of fully-resolved calculations under compressive loading with Mode I and Mode II fracture modes.
In Section \ref{sec:wing cracks}, we apply the model to predict the formation of wing cracks under compression and confinement.
In Section \ref{sec:Application to Layered Rock}, we apply the model to configurations corresponding to the experiments of \cite{tien2006experimental}.

\section{Formulation} \label{sec:formulation}

Section \ref{sec:basic phase-field} introduces the basic phase-field fracture model that is widely used.
Section \ref{sec:phase field contact} enhances the model to incorporate the contact mechanics of crack faces following \cite{hakimzadeh2022phase}.
Section \ref{Elastic Anisotropy} discusses the homogenized elastic response for a layered rock.
Section \ref{frac aniso} accounts for anisotropy in the work to fracture (i.e., the energy required to create a crack surface).
Section \ref{sec:helmholtz} describes a variational approach to define the regularized crack normal.
Section \ref{sec:nonhealing} describes our approach to prevent the crack from healing.
Section \ref{sec:model-summary} summarizes the model that we use.

\subsection{Basic Phase-Field Model} \label{sec:basic phase-field}

A variational (energy-based) model for quasistatic brittle fracture was proposed by \cite{francfort1998revisiting}. 
This model is characterized by the minimization of the potential energy $E$ given by:
\begin{equation}\label{eq:introEnergy}
     E[\bfy, \Gamma] = \int_{\Omega \setminus \Gamma} W(\nabla \bfy) \dm\Omega + G_c \mathcal{H}^{n-1} (\Gamma) + \text{ work due to external loads }
\end{equation}
where $\bfy$ and $\Gamma$ represent the deformation and the crack surface; $\Omega$ represents the body in the reference configuration; $W$ is the elastic energy density; and $G_c$ and $\mathcal{H}^{n-1}$  represent the work to fracture and the Hausdorff area of $\Gamma$, respectively.
The model asserts that the deformation and the crack surface are obtained by minimizing $E$ subject to irreversibility, i.e., the crack cannot heal.

This model is very challenging to use to solve realistic boundary-value problems.
A regularized phase-field version of this model was given by Ambrosio and Tortorelli \cite{AmTo90,AmTo92}, inspired by \cite{MoMo77,Modica87}:
\begin{equation}\label{eq:introAT1}
    E[\bfy, \phi] 
    = 
    \int_{\Omega} \left(\phi^2 + \eta_\epsilon\right) W (\nabla \bfy) \dm\Omega + G_c \int_{\Omega} \left( \frac{(1-\phi)^2}{4\epsilon} + \epsilon |\nabla\phi|^2 \right) \dm\Omega  + \text{ work due to external loads }
\end{equation}
where $\phi(\bfx)$ is a scalar phase field that quantifies the level of damage at every referential point $\bfx$, and is sufficiently smooth to allow standard FEM approximation.
Specifically, $\phi$, bounded by $0 \leq \phi \leq 1$, indicates that the material is intact at $\phi(\bfx) \simeq 1$ and completely fractured at $\phi(\bfx) \simeq 0$.
The small positive quantity $\eta_{\epsilon}$ is used to ensure that the overall problem remains positive definite even if $\phi$ is $0$ in finite regions; for brevity, we will not write it out explicitly further. In this paper, a value of $\eta_{\epsilon} = \SI{1e-3}{}$ is used for numerical calculations.

In physical terms, if the elastic energy is large, the solid can relieve this energy by reducing the value of $\phi$ in regions with large elastic energy density in the first integral in \eqref{eq:introAT1}, i.e., it fractures to reduce the elastic energy.
However, the fracture process requires work, and that is accounted for by the second integral in \eqref{eq:introAT1}.

\subsection{Accounting for the Contact of Crack Faces} \label{sec:phase field contact}

\begin{figure}[htb!]
    \centering
    \includegraphics[width=0.7\textwidth]{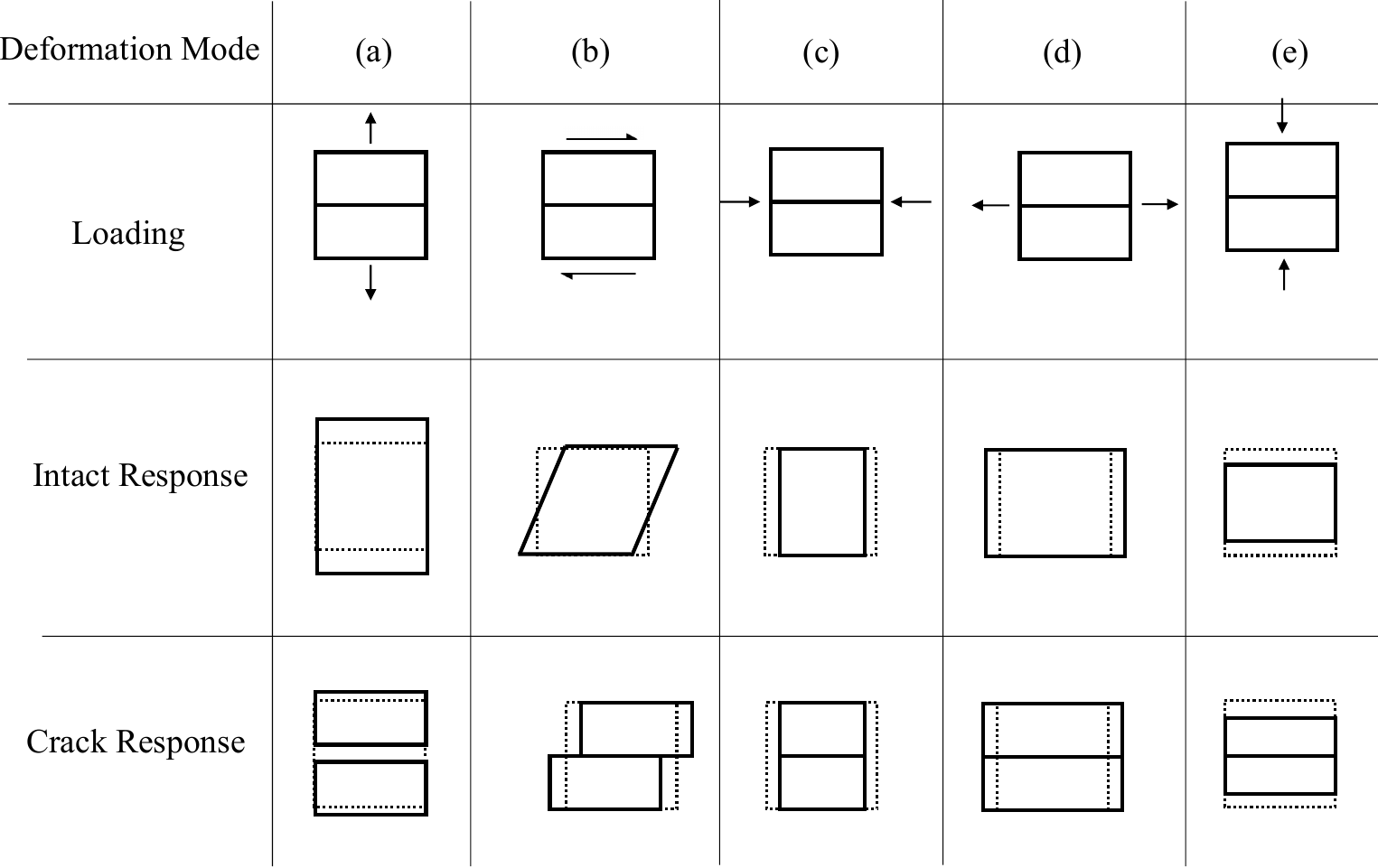}
    \caption{\label{fig:deform modes} The top row shows different loadings, and the middle and lower rows show the idealized deformation for intact and cracked specimens, respectively. Based on this idealization, we assign zero energy to modes (a) and (b). The dotted lines in the second and third row show the undeformed configuration. This decomposition follows \cite{steinke2019phase}.}
\end{figure}

A key shortcoming of the model in Section \ref{sec:basic phase-field} is that it does not distinguish between compressive, shear, and tensile loadings across the crack.
Therefore, all of these types of loadings can lead to fracture which is unphysical.
Various works have proposed decompositions of the energy in different ways to try to prevent this, e.g. \cite{de2022nucleation, clayton2023phase, clayton2021nonlinear, vicentini2023energy, tang2019phase, van2020strain, wang2021phase, zhang2022assessment}.
However, as noted by \cite{steinke2019phase}, and discussed further with some examples in \cite{hakimzadeh2022phase}, it is essential to account for the crack orientation.
For instance, tension across the crack face cannot be resisted by a crack, whereas tension along the crack face will be resisted as well as if the material was intact (Fig. \ref{fig:deform modes}).

Following \cite{hakimzadeh2022phase}, we use a model of the form:
\begin{equation}
\label{eqn:model}
    E[\bfy,  \phi]  
    = 
    \int_{\Omega} \left( \phi^2 W (\nabla \bfy) + \left( 1-\phi^2 \right)  W_d (\nabla \bfy,\bfd) \right)  \dm\Omega + G_c \int_{\Omega} \left( \frac{(1-\phi)^2}{2\epsilon} + \frac{\epsilon}{2}  |\nabla\phi|^2 \right) \dm\Omega  + \text{ work due to external loads. }
\end{equation}
The key feature of this model is the introduction of the effective crack energy density $W_d$ that is active in the fractured volumes.
It imparts the properties of an idealized sharp crack to the regularized (finite volume) phase-field crack. 
Using the QR (upper triangular) decomposition of the deformation gradient tensor in the crack basis, it enables us to transparently separate out the deformation modes shown in Figure \ref{fig:deform modes}.
Subsequently, we relax (i.e., minimize the energy) over the opening and shearing modes, and assign the intact energy to crack closing modes.
$W_d$ takes as argument the deformation, as well as a vector field $\bfd$ that corresponds to the unit normal to the crack face.
However, $\bfd$ is not an independent field; rather it is a functional of $\phi$ as described in Section \ref{sec:helmholtz}.
The use of $\bfd$ is essential to decompose the deformation into the modes shown in Figure \ref{fig:deform modes}.

It is important to highlight here that $W_d$ is necessarily derived from a given $W$ to ensure that it reproduces the intact elastic response when the crack faces close, as described in detail with examples in \cite{hakimzadeh2022phase}. 
In this work, for the specific case of a linear elastic transversely isotropic material with a given linear elastic stiffness tensor with components $C_{ijkl}$, we have from \cite{hakimzadeh2022phase} the following expression for the effective crack energy density:
\begin{equation}\label{eq:aniso Wd}
\begin{split}
 \Wdlin & \left( \bfvareps , \bfd \right) \\
 & = \begin{cases} \textstyle \frac{1}{2}\left( C_{1111} - \frac{C_{1112}^2 C_{2222}-2 C_{1112} C_{1122} C_{1222}+C_{1122}^2 C_{1212}}{C_{1212} C_{2222} - C_{1222}^2} \right) (\vareps_{11}^{\bfd})^2 \qquad \qquad \text{if } \vareps^{\bfd}_{22} > \frac{C_{1112} C_{1222} - C_{1122} C_{1212}}{C_{1212} C_{2222} - C_{1222}^2} \vareps^{\bfd}_{11} , \\
 \textstyle \frac{1}{2}\left( C_{1111}-\frac{C_{1112}^2}{C_{1212}} \right) (\vareps_{11}^{\bfd})^2 + \left( C_{1122} -\frac{C_{1112} C_{1222} }{C_{1212}}\right) \vareps^{\bfd}_{11} \vareps^{\bfd}_{22} + \frac{1}{2} \left( C_{2222} -\frac{C_{1222}^2}{C_{1212}} \right) (\vareps_{22}^{\bfd})^2 \qquad \text{otherwise} ,
 \end{cases}
\end{split}
\end{equation}
where we have used $\bfvareps$ to denote the linear strain tensor; and $\vareps_{11}^{\bfd}$ and $\vareps_{22}^{\bfd}$ are the normal strains in the crack basis, given by:
\begin{equation}\label{eq:Hn}
     \vareps_{11}^{\bfd} = \vareps_{22} d_1^2 - 2 \vareps_{12} d_1 d_2 + \vareps_{11} d_2^2 , \qquad
     \vareps_{22}^{\bfd} = \vareps_{11} d_1^2 + 2 \vareps_{12} d_1 d_2 + \vareps_{22} d_2^2 .
\end{equation}

\subsection{Transverse Isotropic Homogenized Elastic Response} \label{Elastic Anisotropy}

Layered rocks can be modeled as transversely isotropic, i.e., given a layer normal or structural director denoted by $\bfa$, they are isotropic within the plane with normal $\bfa$.
The transverse isotropic homogenized elastic response is a classical result in micromechanics, and we summarize the result here following \cite{backus1962long}. For a linear elastic material, 
the stiffness tensor can be represented as a matrix $\left[\mathbf{C}^e\right]$ in Voigt notation:
\begin{equation}  
    \left[\mathbf{C}^e\right]=\left[\begin{array}{cccccc}
        c_{11} & c_{12} & c_{12} & 0 & 0 & 0 \\
        c_{12} & c_{33} & c_{13} & 0 & 0 & 0 \\
        c_{12} & c_{13} & c_{33} & 0 & 0 & 0 \\
        0 & 0 & 0 & c_{66} & 0 & 0 \\
        0 & 0 & 0 & 0 & c_{44} & 0 \\
        0 & 0 & 0 & 0 & 0 & c_{44}
    \end{array}\right], \qquad
    \begin{cases}
        c_{11}=\left\langle H^{-1}\right\rangle^{-1} \\
        c_{12}=\left\langle H^{-1}\right\rangle^{-1}\left\langle\frac{H-2 \mu}{H}\right\rangle \\
        c_{13}=\left\langle H^{-1}\right\rangle^{-1}\left\langle\frac{H-2 \mu}{H}\right\rangle^2 + 2 \left\langle \frac{ (H-2 \mu)\mu}{H}\right\rangle \\
        c_{44}=\left\langle\mu^{-1}\right\rangle^{-1} \\ 
        c_{66}=\langle\mu\rangle \\
        c_{33} = c_{13} + 2c_{66}
    \end{cases}
\end{equation}
The structural director has been oriented along $\bfe_1$ in this representation; $\mu$ is the shear modulus, and $H = K + \frac{4}{3} \mu$, where $K$ is elastic bulk modulus; and $\langle\cdot\rangle$ denotes the Voigt average.

\subsection{Anisotropy of the Work to Fracture} \label{frac aniso}

In a general anisotropic material, the work to fracture depends on the fracture direction. 
Specifically in stratified rock, crack propagation is much easier along the layers rather than across the layers.
In a homogenized model with structural director $\bfa$, this is reflected by the work to fracture for crack propagation along $\bfa$ being much larger than the work to fracture for propagation in the plane normal to $\bfa$.

In the context of phase-field fracture, a simple way to include anisotropy of the fracture is to replace $|\nabla\phi|^2$ in \eqref{eqn:model} by $\nabla\phi\cdot\bfA\nabla\phi$, where $\bfA$ is some anisotropic 2-nd order tensor.
However, it was shown by \cite{li2015phase} that this does not provide a sufficient level of anisotropy to model numerous relevant settings, include stratified systems.
To increase the level of anisotropy, they suggested adding a further term $\nabla\nabla\phi : \mathbb{A} : \nabla\nabla\phi$, where $\mathbb{A}$ is a 4-th order tensor, ``:'' denotes a 2-nd order contraction between tensors, and $\nabla\nabla\phi$ is the Hessian of $\phi$.
Following the systematic formulation of this idea from \cite{teichtmeister2017phase}, we define the anisotropic work to fracture $\gamma$:
\begin{equation}
\label{eqn:fracture-aniso}
    \gamma(\phi, \nabla \phi, \nabla\nabla\phi) 
    =
    G_c \left( \frac{\phi ^2}{2\epsilon} + \frac{\epsilon}{4} \nabla \phi \cdot \bfA \nabla \phi + \frac{\epsilon ^3}{32} \nabla\nabla\phi : \mathbb{A} : \nabla\nabla\phi \right), \qquad
    \begin{cases}
        \bfA = \bfI + \alpha \bfa_{\perp}  \otimes \bfa_{\perp}    \\
        \mathbb{A} = \mathbb{I} + \alpha_1 \left(\bfa_{\perp}  \otimes \bfa_{\perp} \right) \otimes \left(\bfa_{\perp}  \otimes \bfa_{\perp} \right)   
    \end{cases}
\end{equation}
where $\mathbb{I} = \frac{1}{2} \left(\delta_{ik} \delta_{jl} + \delta_{il} \delta_{jk}\right) \bfe_i \otimes \bfe_j \otimes  \bfe_k \otimes  \bfe_l$ is the 4-th order identity tensor, and $\alpha$ and $\alpha_1$ are the anisotropy coefficients that determine the level of anisotropy.
Following the usual convention: $\bfe_i$ is the $i$-th unit vector in a Cartesian basis; $\otimes$ is the outer/tensor product; $\delta_{ik}$ is the $i,k$-component of the identity matrix; and $\bfa_{\perp}$ is a unit vector that is normal to $\bfa$.

\begin{figure}[htb!] 
    \subfloat[]
    {\includegraphics[width=0.19\textwidth]{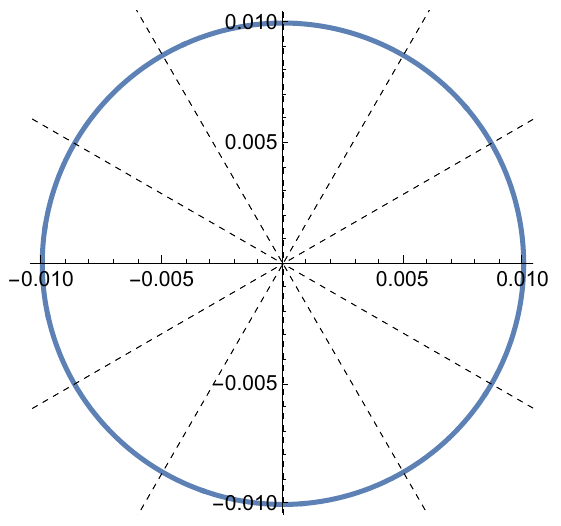}}
    \hfill
    \subfloat[]
    {\includegraphics[width=0.38\textwidth]{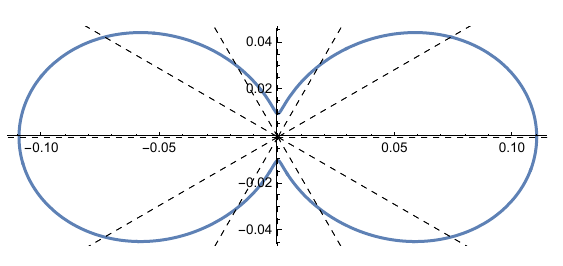}}
    \hfill
    \subfloat[]
    {\includegraphics[width=0.38\textwidth]{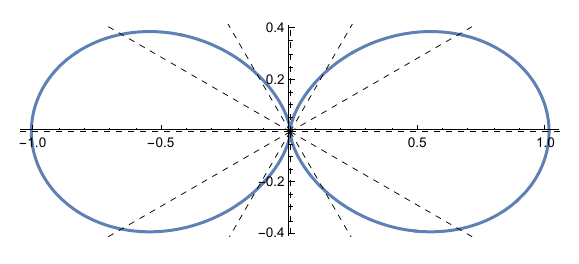}}
    \caption{
        Polar plot of $\left(\frac{\epsilon}{4} \nabla \phi \cdot \bfA \nabla \phi + \frac{\epsilon ^3}{32} \nabla\nabla\phi : \mathbb{A} : \nabla\nabla\phi\right)$ as a function of $\theta$, where we set $\nabla\phi \sim \begin{bmatrix} \cos\theta \\ \sin \theta \end{bmatrix}$ and $\nabla\nabla\phi \sim \begin{bmatrix} \cos^2\theta & \cos\theta\sin\theta \\ \cos\theta\sin\theta & \sin^2 \theta \end{bmatrix}$, following \cite{li2015phase}.
        We use $\epsilon = 0.04$, 
        and (a) $\alpha = \alpha_1 = 0$, (b) $\alpha = \alpha_1 = 10$, and (c) $\alpha = \alpha_1 = 100$. We use $\alpha = \alpha_1 = 100$ throughout our calculations.
        We note that $\alpha$ and $\alpha_1$ must lie in the open range $(-1,\infty)$ for $\bfA$ and $\mathbb{A}$ to be positive-definite \cite{teichtmeister2017phase}.
        The structural director is set to be $\bfa = \bfe_2$ (i.e., $\bfa_{\perp} = \bfe_1$). 
    }
    \label{fig:POLAR ENERGY}
\end{figure}

Figure \ref{fig:POLAR ENERGY} shows the anisotropy in the work to fracture, with $\bfa=\bfe_2$, and we see a clear reduction in this quantity for the crack propagating with normal $\pm\bfe_2$.

In the fracture model \eqref{eqn:model}, we replace the entire second integral by $\int_\Omega \gamma(\phi, \nabla \phi, \nabla\nabla\phi) \dm\Omega$, with $\gamma$ obtained from \eqref{eqn:fracture-aniso}.
We highlight that this model requires more smoothness than is available with standard FEM.
\cite{li2015phase} solve this by using finite elements with more smoothness.
We instead use a mixed FEM approach as described in Section \ref{sec:Numerical Implementation and Calculations}.

\subsection{Variational Regularized Definition of the Crack Normal} \label{sec:helmholtz}

The phase-field model that accounts for crack face closure in \eqref{eqn:model} has the crack face normal vector field $\bfd(\bfx)$ as an essential element; this is required to be able to distinguish the modes in Figure \ref{fig:deform modes}.
In principle, $\phi(\bfx)$ has sufficient information to obtain $\bfd$ simply by computing $\pm\nabla\phi$.
In practice, this leads to severe difficulties related to the numerical evaluation: for instance, the numerical evaluation is unstable near the crack tip, and it is not well-defined in the interior of the crack away from the crack faces \cite{hakimzadeh2022phase,agrawal2016multiscale}.

\begin{figure}[h!] 
    \subfloat[]{\includegraphics[width=0.5\textwidth]{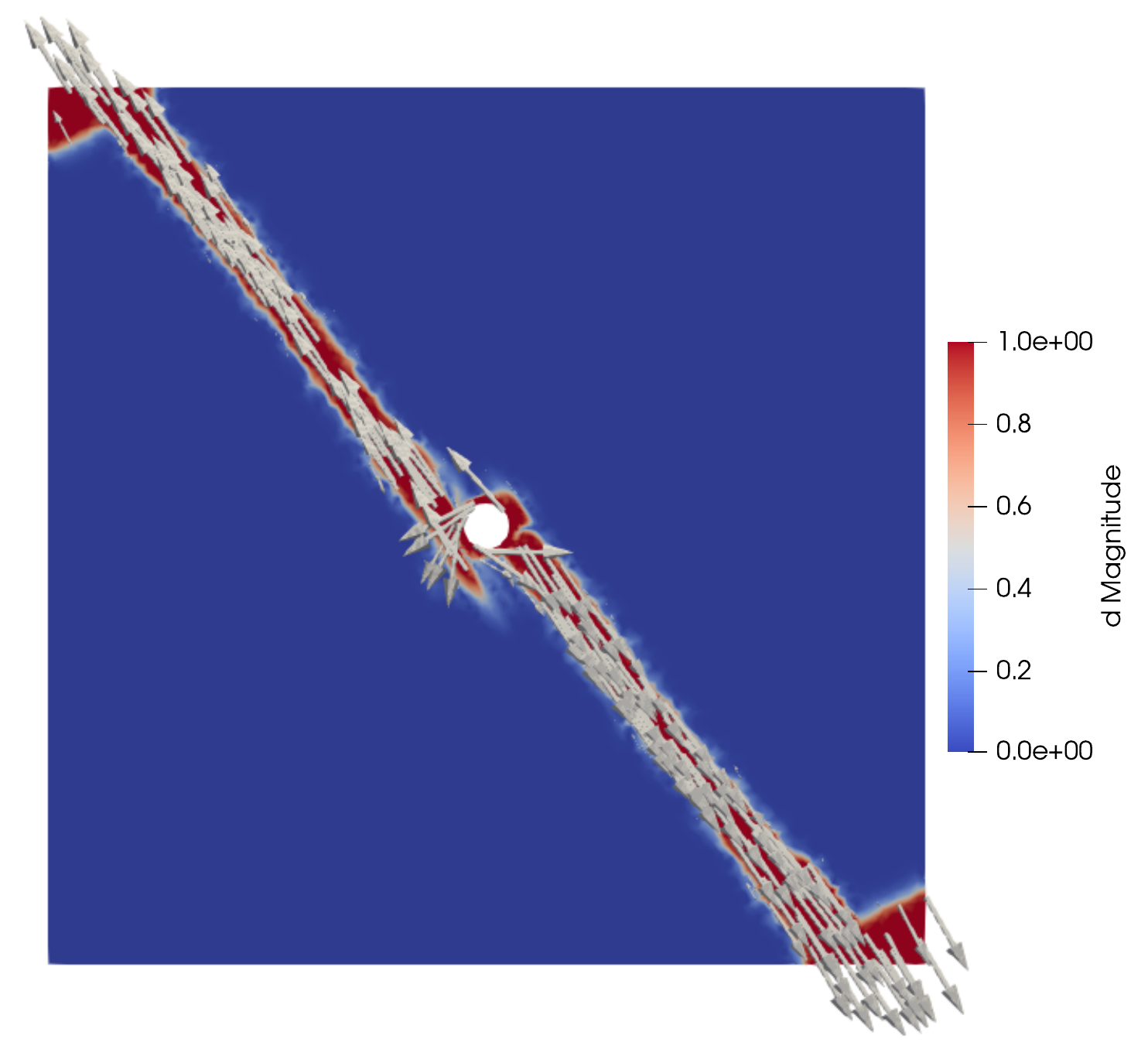}}
    \hfill
    \subfloat[]{\includegraphics[width=0.5\textwidth]{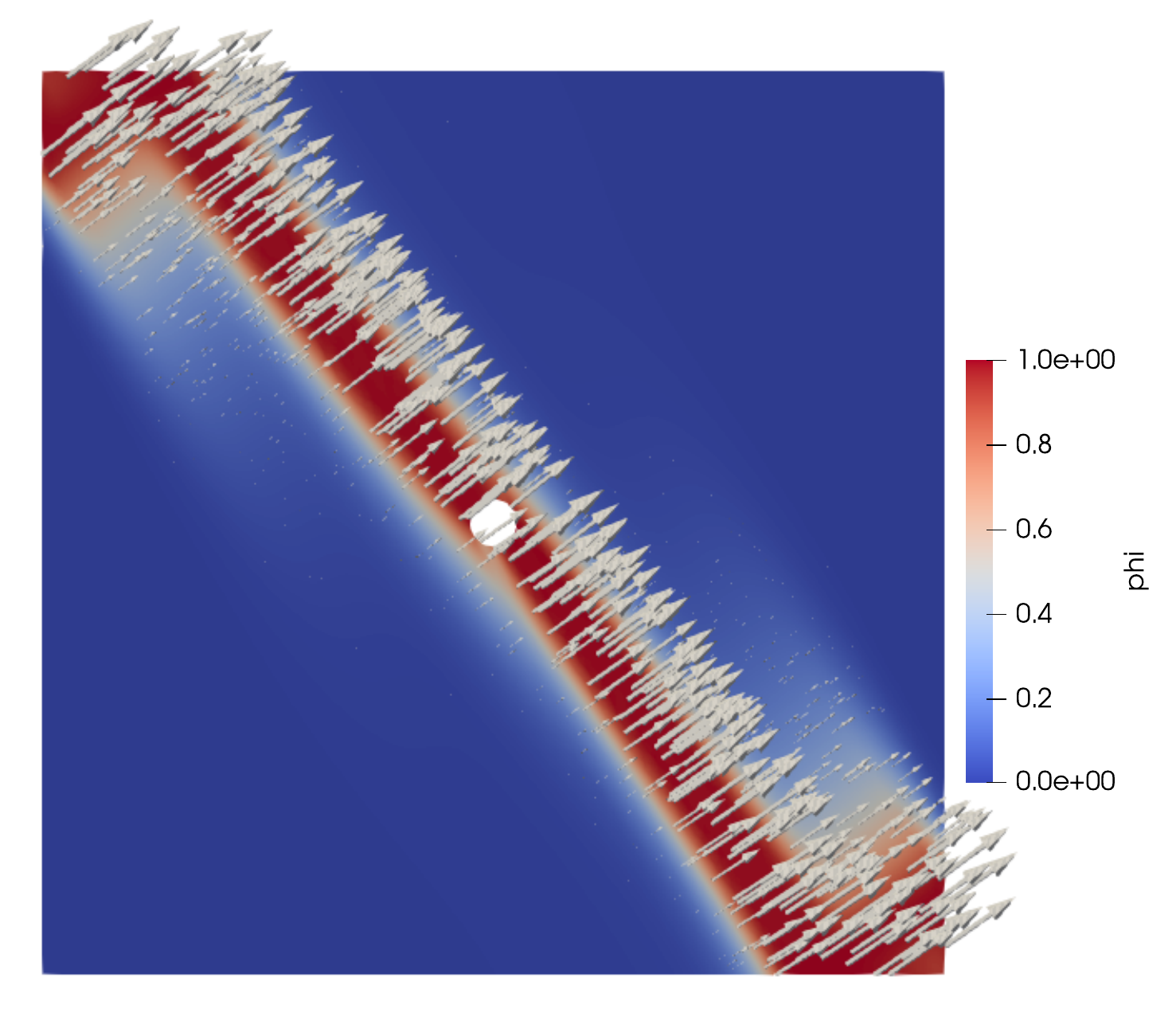}}
    \caption{
        Left: the vector field $\bfd$ is not normal to the crack faces when we incorporate an anisotropic penalization of $\bfd$ in a model that treats $\bfd$ as a primary variable, generalizing the isotropic model of \cite{hakimzadeh2022phase}.
        Right: the model proposed in this paper where the scalar field $\phi$ is a primary variable and $\bfd$ is computed from the kinematic constraints \eqref{eq:bulk1}.
        It is clear that (a) does not work and (b) does.
    }
 \label{fig:CrackNotPerpendicular}
\end{figure}

In our previous work \cite{hakimzadeh2022phase} where this model was formulated, we used $\bfd(\bfx)$ as an independent vector field that replaced the scalar field $\phi(\bfx)$; in short, the magnitude of $\bfd$ related to the level of fracturing and the orientation provided the crack face normal.
That is, the free energy in \cite{hakimzadeh2022phase} had as arguments $(\bfy,\bfd)$ and hence we avoided completely the challenge of computing $\nabla\phi$.
While this worked well in \cite{hakimzadeh2022phase} when we studied fracture in isotropic materials, it is not useful beyond this specific simple setting.
Considering the general anisotropic setting, we would account for anisotropy through terms of the form $\bfd\cdot\bfA\bfd$ and $\nabla\bfd:\mathbb{A}:\nabla\bfd$, and these terms would drive $\bfd$ to orient in preferred directions.
However, this clashes with the primary kinematic role of $\bfd$ as the crack face normal; Figure \ref{fig:CrackNotPerpendicular} shows a crack configuration calculated with such a model, where the elastic terms drive a certain direction of crack propagation and the anisotropic terms drive an orientation of $\bfd$ that does not align with the crack face normal.

In summary, while the vector field $\bfd$ as an independent descriptor of crack propagation was a useful convenience for the isotropic setting, it is not useful for the general setting.
In physical terms, the scalar field $\phi$ has all the information required to describe the configuration of the system and there is simply no need for an (independent) vector field $\bfd$.
It is essential to use the scalar field $\phi$ as the independent descriptor, and obtain $\bfd$ as a derived kinematic quantity.
We formulate the kinematic constraint relation between $\phi(\bfx)$ and $\bfd(\bfx)$ as the Euler-Lagrange equation of a variational principle below.

For a given field $\phi(\bfx)$, we define $\bfd$ as the minimizer of the functional:
\begin{equation}\label{eq:bulk1}
    \bfd = \argmin_\bfp \int_{\Omega} \left(
        H_{l_1}(|\nabla \phi|, g_m) \left|\bfp \otimes \bfp - \phi ^2 \widehat{\nabla\phi} \otimes \widehat{\nabla\phi}\right|^2 + \beta_1 H_{l_2}(\phi, \phi_m) (|\bfp| - \phi)^2 + \beta_2 |\nabla \bfp|^2  
    \right) \dm\Omega, \quad \text{ given } \phi(\bfx)
\end{equation}
where $H_l(x, x_m)$ is a smooth approximation of the step function centered at $x_m$, for which we use $H_l(x, x_m) = 0.5 + 0.5 \tanh\left(\frac{x - x_m}{l}\right)$, and $\widehat{\nabla\phi}$ is the unit vector in the direction $\nabla\phi$.

We understand this functional term-by-term below.
As a preliminary, we recall from studies of liquid crystals, e.g. \cite{ericksen1989liquid}, that unit vector fields often lead to singularities.
Therefore, our aim is to obtain $\bfd$ as the normal in the vicinity of the crack, but allow it to take any other value in the undamaged solid --- where the direction of $\bfd$ is unimportant --- to prevent the formation of singularities.
Similarly, since we do not care about the orientation of $\bfd$ away from the crack, we apply natural boundary conditions to \eqref{eq:bulk1}.
Further, we highlight that we want $\bfd$ to not flip as we go from one face of a crack to the other.

Consider the first term in \eqref{eq:bulk1}.
The square term aims to align $\bfd$ with $\pm\nabla\phi$ on the crack faces, and aims to make the magnitude of $\bfd$ equal to $\phi$ in accord with the second term.
Further, the leading step function aims to enforce this only at high values of $|\nabla\phi|$ to select the crack faces, and not within the crack volume or away from the crack.

Consider the second term in \eqref{eq:bulk1}.
The square term aims to set the magnitude of $\bfd$ proportional to $\phi$, but the leading step function aims to enforce this only within the crack volume.
This ensures that the normal vector does not vanish and ``flip around'' within the crack volume.

Finally, the third term in \eqref{eq:bulk1} is a standard regularization to provide some smoothness and prevent singularities in regions such as crack tip or elsewhere.
While this is used to impose regularity, it is kept small.

In \eqref{eq:bulk1}, we use $l_1 = 0.01$, $g_m = 10$, $l_2 = 0.001$, $\phi_m = 0.98$, $\beta_1=0.8$, and $\beta_2=0.001$.
While the specific values are obtained from numerical experiments to get good performance, the heuristics are as follows.
First, the regularization term is small and meant only to preserve smoothness, and hence $\beta_2 \ll \beta_1$.
Second, the first and second contributions are both of relatively equal importance, and hence $\beta_1 \simeq 1$.
Third, we want $l_1 \ll 1$, $l_2 \ll 1$ to have rapid transitions in the step function.
Fourth, we use $\phi_m$ close to $1$ to ensure that only the fractured regions contribute to the second term.
Finally, we use $g_m=10$ to ensure that only the crack faces contribute to the first term.

\subsection{Equilibrium, Evolution, and Irreversibility of Fracture} \label{sec:nonhealing}

In the standard approach to phase-field modeling, given the free energy $E[\bfy,\phi]$, we find the equilibrium state as the configuration that corresponds to the energy-minimizing state, by solving the Euler-Lagrange equations $\variation{E}{\bfy} = {\bf 0}$ and $\variation{E}{\phi} = 0$.
Further, evolution --- for instance, driven by time-varying external loads --- is modeled simply by using the equilibrium equations at all times, when we neglect effects such as inertia and time-dependent material response \cite{agrawal2017dependence,chua2022phase,naghibzadeh2021surface,naghibzadeh2022accretion}.
Under fairly general conditions, this can be shown to satisfy the constraints of thermodynamics \cite{karimi2022energetic,penrose1990thermodynamically}.
This approach, however, does not permit history-dependence.

In fracture, on the other hand, the growth of cracks is irreversible.
A standard approach to enforce this is as follows.
Given $\phi_0$ as the solution at a given time, the solution $\phi$ for the next time is obtained by minimizing over all admissible fields that satisfy the pointwise constraint that $\phi \ge \phi_m$ at all points at which $\phi_0 \ge \phi_m$. 

While this is generally applied as an inequality constraint, we use the following robust and effective variational penalization.
We augment $E$ with a term of the form:
\begin{equation}
    E \to E + m \int_\Omega H_{l_2}(\phi_0, \phi_m) (\phi - \phi_0)^2 \dm\Omega
\end{equation}
where $m$ is a large penalty parameter, set as $\SI{4e5}{\giga\pascal}$ in our calculations.
The regularized step function $H_l(\phi_0, \phi_m)$ is activated only at points where $\phi_0 \gtrsim \phi_m$, and hence the overall penalty forces $\phi$ at those points to not evolve further (i.e., neither to decrease nor increase). As stated in the description of \eqref{eq:bulk1}, $l_2 = 0.001$, and $\phi_m = 0.98$.

\subsection{Summary of the Complete Model}
\label{sec:model-summary}

The free energy that we minimize is given by the expression:
\begin{equation}
\label{eqn:final-model}
\begin{split}
    E[\bfy,\phi]  
    = 
    &
    \int_{\Omega} \left( \phi^2 W (\nabla \bfy) + \left( 1-\phi^2 \right)  W_d (\nabla \bfy,\bfd[\phi]) \right)  \dm\Omega 
    + 
    \int_{\Omega} G_c \left( \frac{\phi ^2}{2\epsilon} + \frac{\epsilon}{4} \nabla \phi \cdot \bfA \nabla \phi + \frac{\epsilon ^3}{32} \nabla\nabla\phi : \mathbb{A} : \nabla\nabla\phi \right) \dm\Omega 
    \\
    &
    + 
    m \int_\Omega H_{l_2}(\phi_0, \phi_m) (\phi - \phi_0)^2 \dm\Omega
    +
    \text{ work due to external loads. }
\end{split}    
\end{equation}
where we have accounted for the energy of the crack volume $W_d$ in the first integral (Section \ref{sec:phase field contact}); the strong anisotropy in the second integral (Section \ref{frac aniso}); and the irreversibility of fracture in the third integral (Section \ref{sec:nonhealing}).
Given an initial crack configuration $\phi_0(\bfx)$, we minimize $E$ over admissible $\bfy$ and $\phi$ to find the updated crack configuration.

The crack normal vector field $\bfd(\bfx)$ is kinematically constrained to $\phi$, and given by the expression (Section \ref{sec:helmholtz}):
\begin{equation}
\label{eqn:final-constraint}
    \bfd[\phi] = \argmin_\bfp 
        \underbrace{
            \int_{\Omega} \left(
            H_{l_1}(|\nabla \phi|, g_m) \left|\bfp \otimes \bfp - \phi ^2 \widehat{\nabla\phi} \otimes \widehat{\nabla\phi}\right|^2 + \beta_1 H_{l_2}(\phi, \phi_m) (|\bfp| - \phi)^2 + \beta_2 |\nabla \bfp|^2  
            \right) \dm\Omega
        }_{=:D[\bfp,\phi]}
\end{equation}

\subsection{Numerical Implementation} \label{sec:Numerical Implementation and Calculations}

Our model is implemented using the finite element method (FEM) in FEniCS, an open-source tool for FEM \cite{logg2012automated, barchiesi2021computation}, based on the implementation from \cite{natarajan2019phase}.
The primary unknowns are $\bfy (\bfx)$, $\phi(\bfx)$, and $\bfd(\bfx)$.
Further, to account for the higher-order derivatives due to anisotropy in \eqref{eqn:final-model}, we used a mixed method that introduces $\nabla \phi(\bfx)$ as an additional variable.
All fields are discretized using linear shape functions.

To solve for the evolution under a time-dependent external loading, we need to minimize \eqref{eqn:final-model} with respect to $\bfy$ and $\phi$ given $\phi_0$, while satisfying the constraint \eqref{eqn:final-constraint} that relates $\phi$ to $\bfd$.
We do this by individually minimizing over $\phi$ and $\bfy$ first, then solving the constraint equation for $\bfd$, and then repeating this procedure until we reach convergence. 
That is, 
\begin{align}
    \bfy^{i+1} & = \argmin_{\bfy} E[\bfy, \phi^i], \text{ given } \phi_0 \\
    \phi^{i+1} & = \argmin_{\phi} E[\bfy^{i+1}, \phi] , \text{ given } \phi_0 \\
    \bfd^{i+1} &= \argmin_\bfp D[\bfp,\phi^{i+1}]  
\end{align}
where the subscript $0$ indicates initial values from the previous load increment, and superscripts indicate iterations.

In the calculations described in Sections \ref{sec:comparison} and \ref{sec:Application to Layered Rock}, the elastic constants corresponded to those used in \cite{tien2006experimental}.
Further, we use $G_c = \SI{8e4}{\newton\per\meter}$ for the homogenized material. 
This value is consistent with those reported in \cite{zhang2022energy} which indicate $G_c$ values range from $\SI{3.2e3}{}$ to $\SI{9.4e3}{\newton\per\meter}$ for sandstone under biaxial compression \cite{liu2020dynamic}, and from $\SI{1.3e4}{}$ to $\SI{5.1e4}{\newton\per\meter}$ for granite under triaxial compression\cite{wong1982shear}.

\begin{figure}[h!]
    \centering
    \subfloat[]{\includegraphics[width=0.35\textwidth]{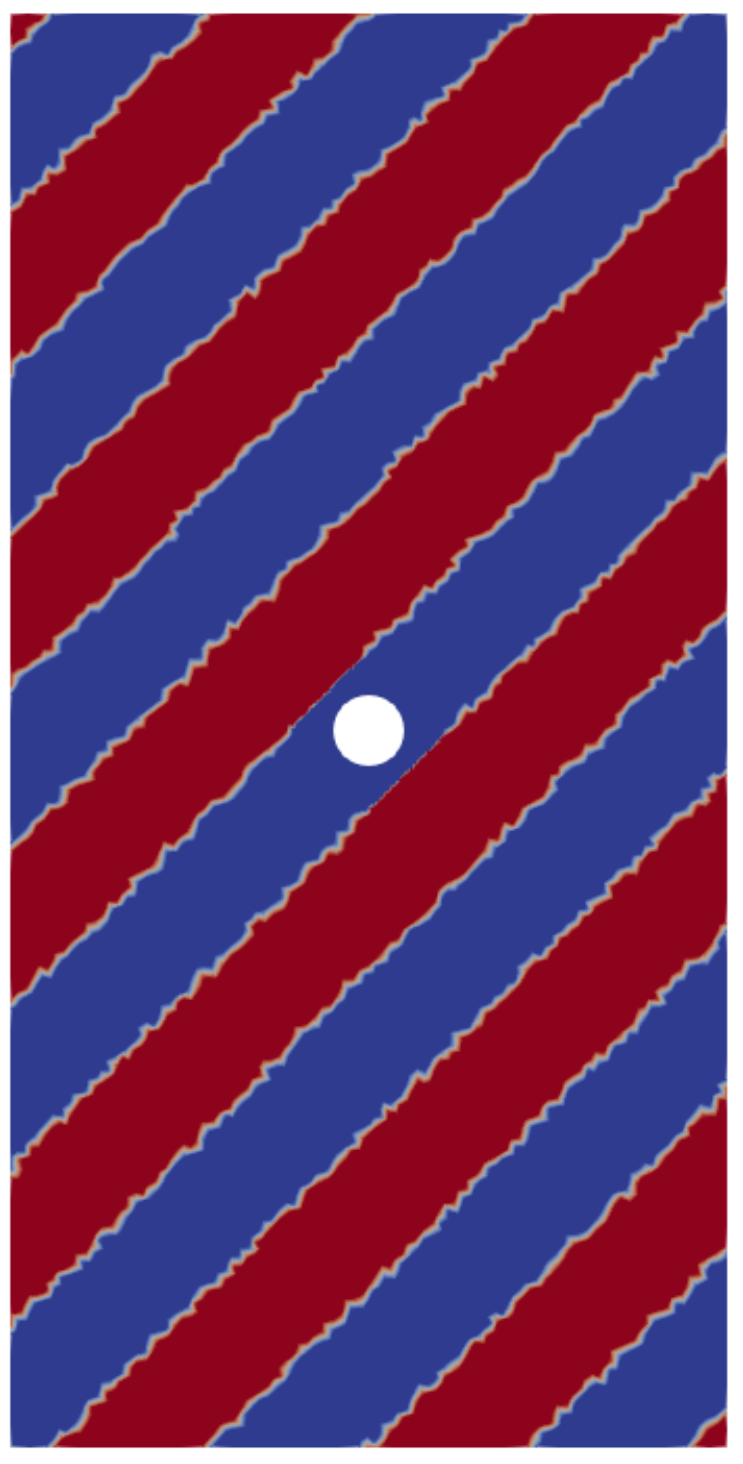}}
    \hspace{0.2\textwidth}
    \subfloat[]{\includegraphics[width=0.35\textwidth]{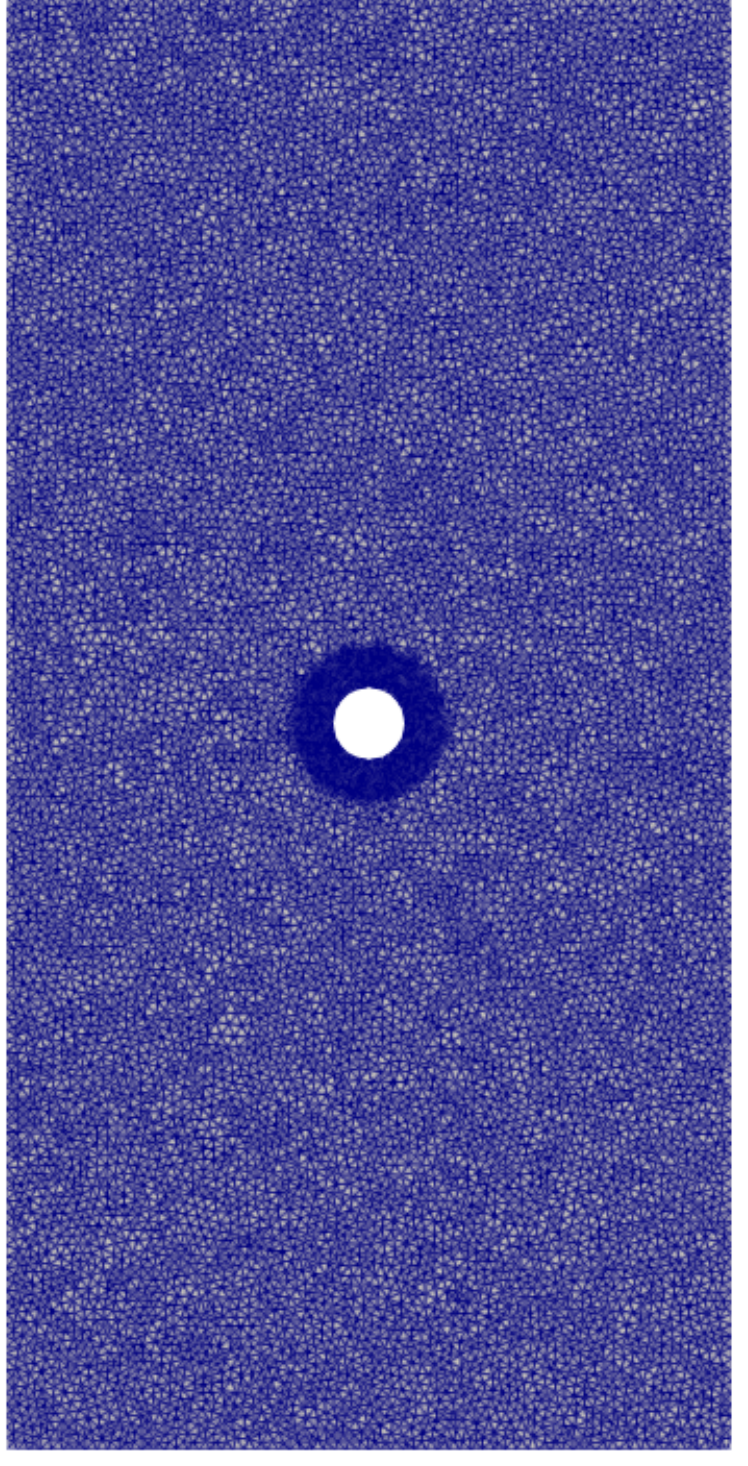}} 
    \vspace{0.1cm}
    \subfloat[]{\includegraphics[width=0.99\textwidth]{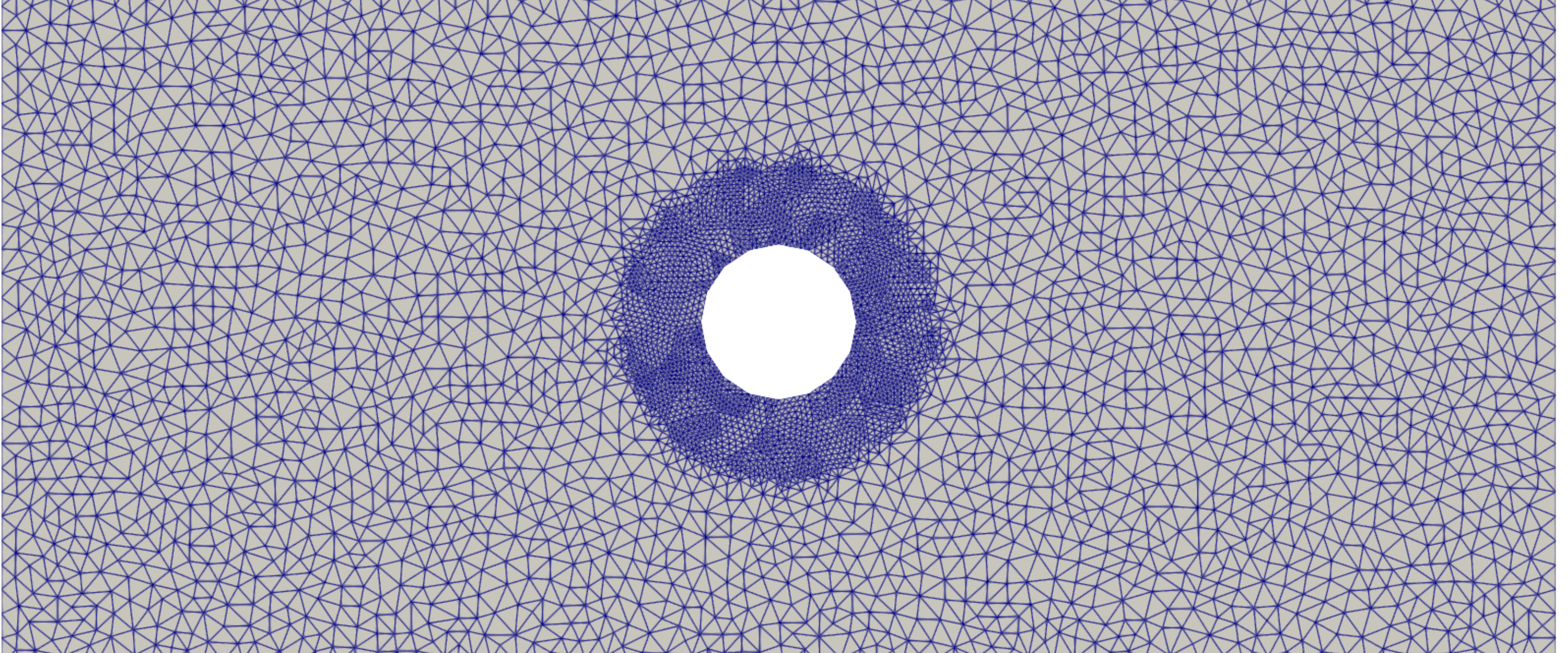}}
    \caption{
        (a) The geometry of the fully-resolved specimen showing the inhomogeneous layers. Red indicates Material A (the ``strong'' material) and blue indicates Material B (``weak'' material). The layers have flat interfaces between them, but they appear rugged due to the meshing.
        (b) A typical finite element mesh used in the calculations. 
        (c) A close-up of the mesh around the center hole.
    }
 \label{fig:Fully resolved and mesh}
\end{figure}

We employed a phase-field regularization parameter $\epsilon = \SI{4e-2}{\meter}$, which is relatively large, and all computations are performed on domains of size $\SI{1}{\meter}\times\SI{2}{\meter}$. 
We also performed calculations with $\epsilon = \SI{1.5e-2}{\meter}$ and saw no appreciable differences.
Figure \ref{fig:Fully resolved and mesh} shows the geometry that is used for the fully-resolved calculations and a typical finite element mesh.
We use a fine mesh in the vicinity of the crack, with element dimension on the order of $\SI{4e-3}{\meter}$, and a less fine mesh further away with element dimension on the order of $\SI{1.6e-2}{\meter}$.
This mesh is sufficient to resolve the phase-field for the chosen value of $\epsilon$.

\section{Comparing Homogenized and Fully-Resolved Models}
\label{sec:comparison}

We first compare our homogenized model against phase-field fracture simulations in a fully-resolved specimen wherein the individual layers are resolved.
In the fully-resolved setting, we treat the layered specimen as composed of isotropic constituent layers with different properties, and hence overall the specimen is heterogeneous.
The fully-resolved model is both much more expensive and much more accurate than the homogenized model; hence, our goal is to test the less expensive homogenized model.

The layers of equal thickness are taken to consist of Material A (``strong'') with $E =\SI{21.7}{\giga\pascal}$, $\nu = 0.25$, and $G_c = \SI{4e5}{\newton\per\meter}$, and Material B (``weak'') with $E = \SI{10.85}{\giga\pascal}$, $\nu = 0.25$, and $G_c = \SI{4e4}{\newton\per\meter}$. 
These numbers roughly correspond to \cite{tien2006experimental}.

The elasticity tensor for the homogenized anisotropic material is calculated using the expressions in Section \ref{Elastic Anisotropy}.
While there exist clear and systematic procedures for the homogenization of linear elasticity, there exist no such procedures for the nonlinear setting generally, especially for the challenging case of fracture. 
We therefore obtain the values of the anisotropy parameters by using trial-and-error and comparing the homogenized and fully-resolved calculations for a {\em small and limited} set of calculations, and then use these values for the further calculations in this section and in Section \ref{sec:Application to Layered Rock}.
Specifically, we compared crack growth for layers oriented at $45^{\circ}$ in the for crack opening mode under uniaxial compressive loading.
We find that $\alpha=100$ and $\alpha_1=100$ best reproduces the fully-resolved crack patterns, and these numbers are in line with those used elsewhere, e.g. \cite{clayton2023phase}.
We note that these chosen values for $\alpha$ and $\alpha_1$ are very close to the allowed limit of anisotropy that still retains the positive definiteness of the surface energy (Fig. \ref{fig:POLAR ENERGY}).

\begin{figure}[htb!]
\centering
    \subfloat[Crack growth for the layers at $45^{\circ}$ in the fully-resolved domain.]
	{\includegraphics[width=0.25\textwidth]{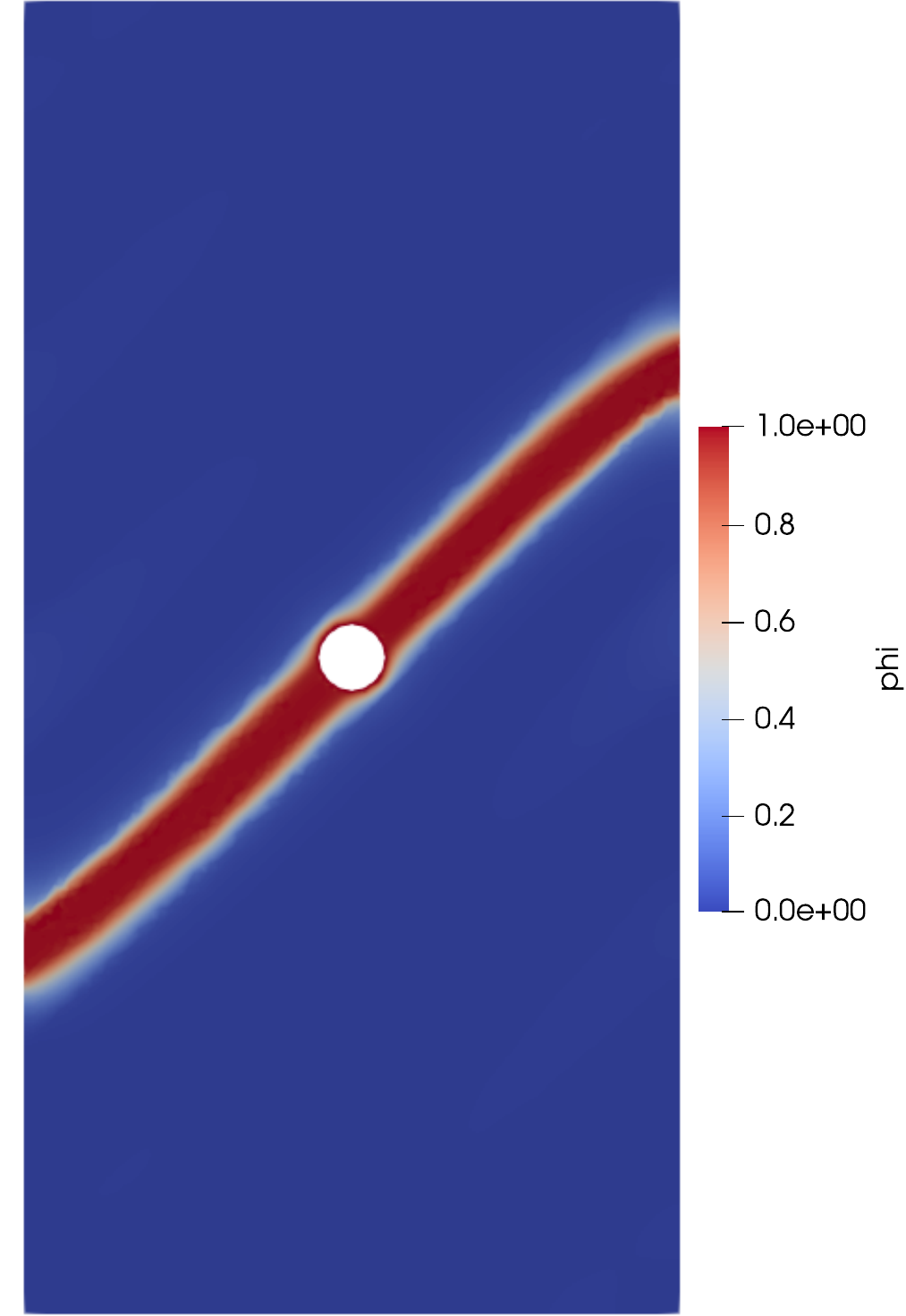}}
    \hfill
    \subfloat[Crack growth for the layers at $60^{\circ}$ in the fully-resolved domain.]
	{\includegraphics[width=0.25\textwidth]{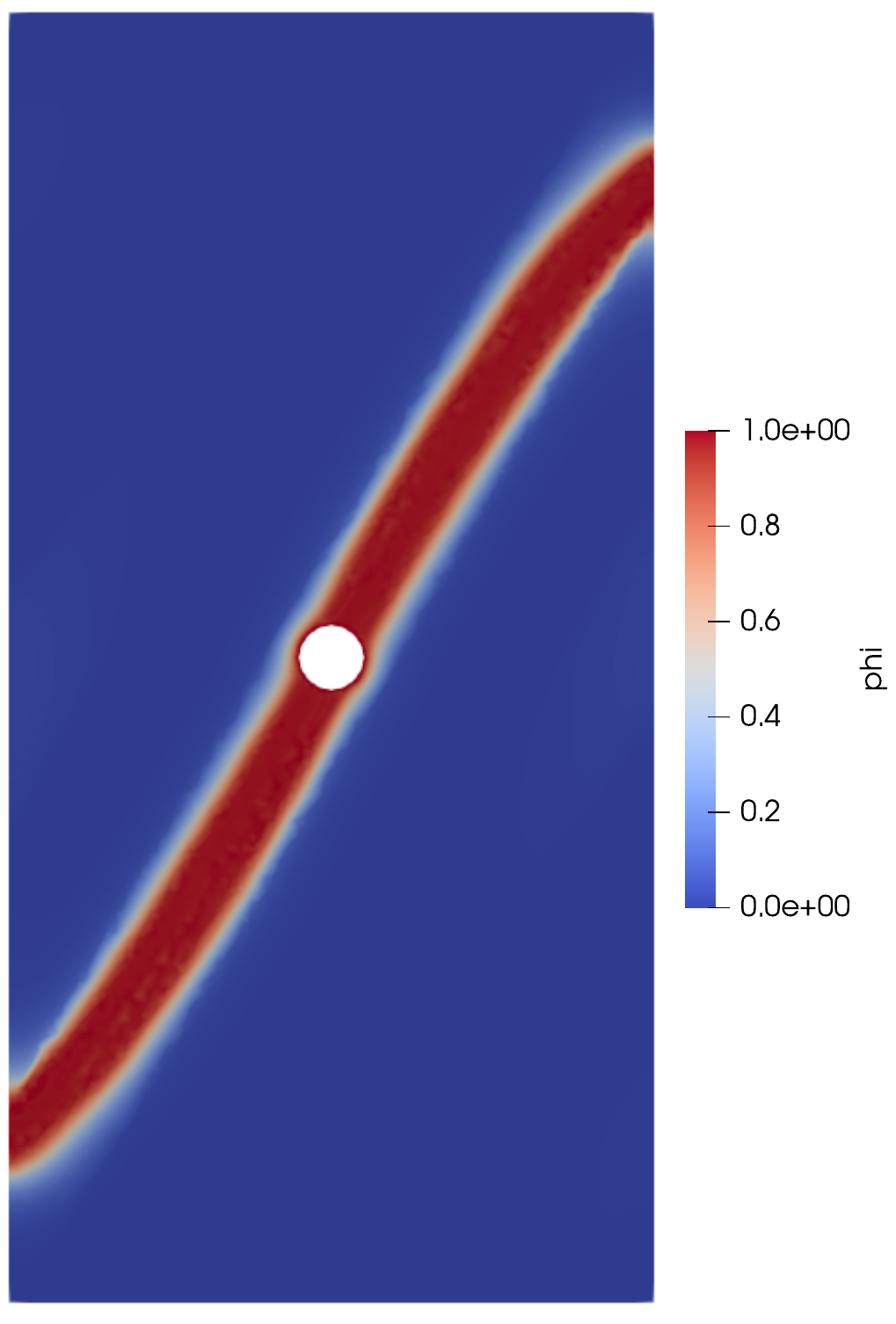}}
    \hfill
    \subfloat[Crack growth for the layers at $75^{\circ}$ in the fully-resolved domain.]
	{\includegraphics[width=0.25\textwidth]{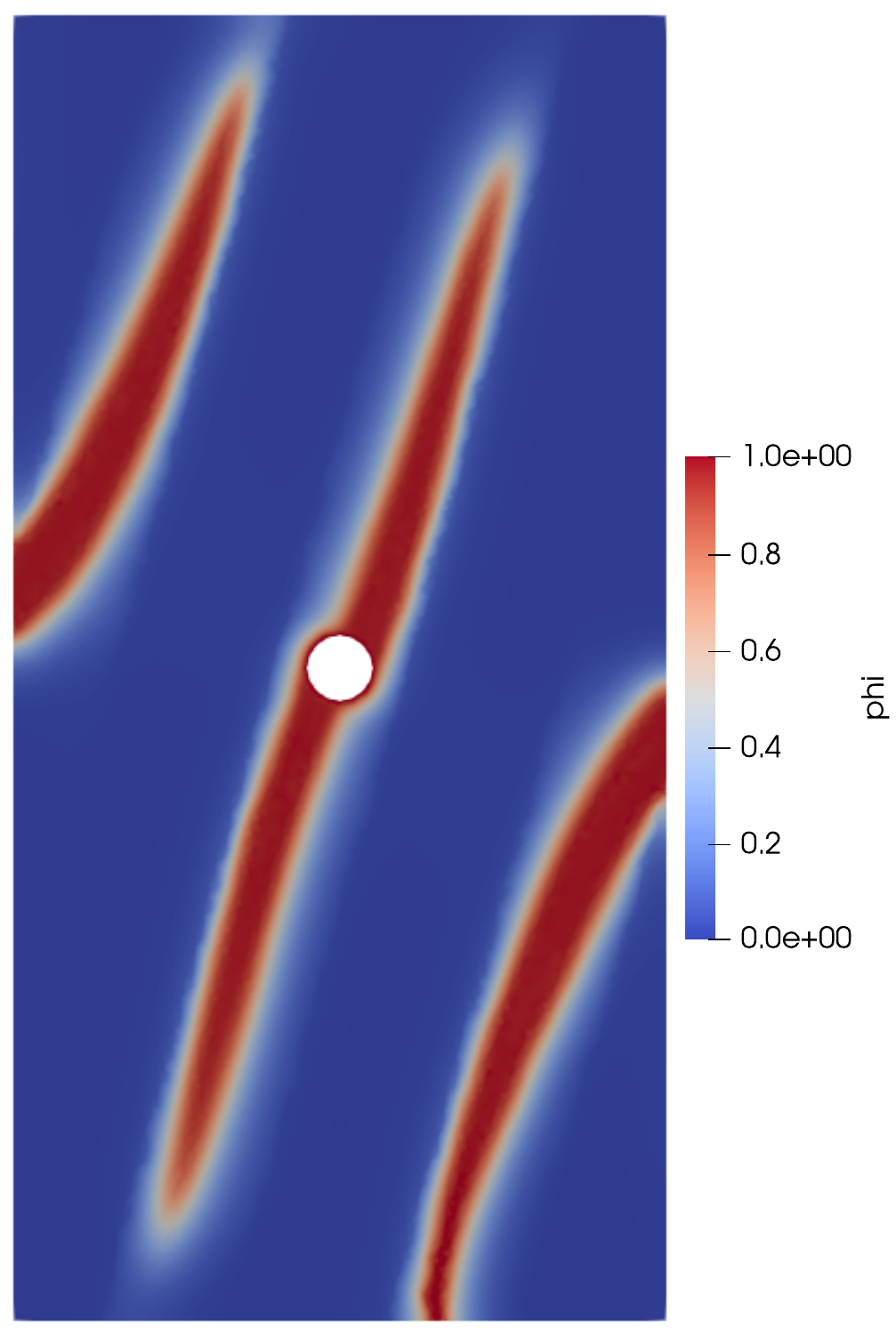}}
    \\
    \subfloat[Crack growth for the layers at $45^{\circ}$ in the homogenized domain.]
	{\includegraphics[width=0.25\textwidth]{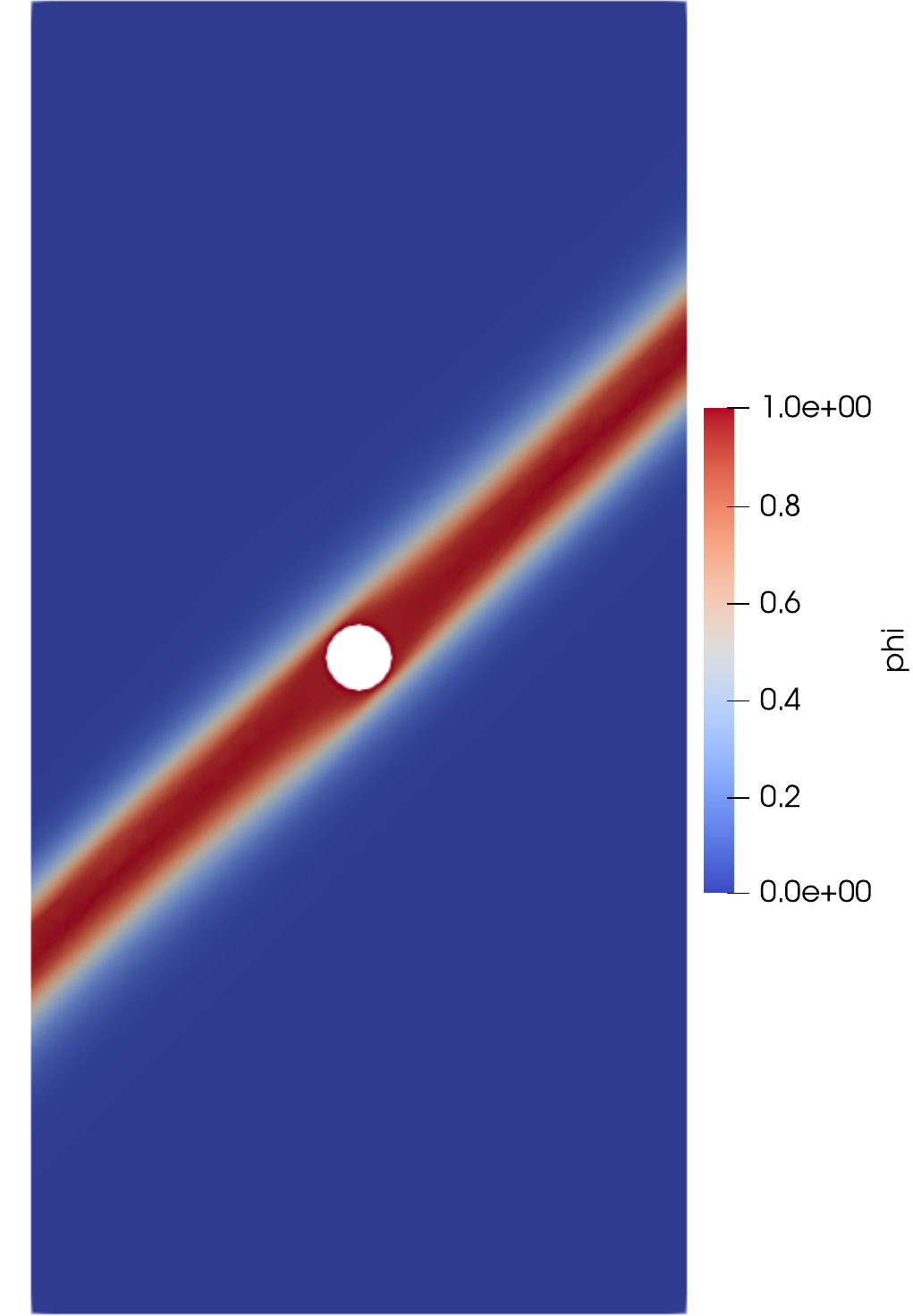}}
    \hfill
    \subfloat[Crack growth for the layers at $60^{\circ}$ in the homogenized domain.]
	{\includegraphics[width=0.25\textwidth]{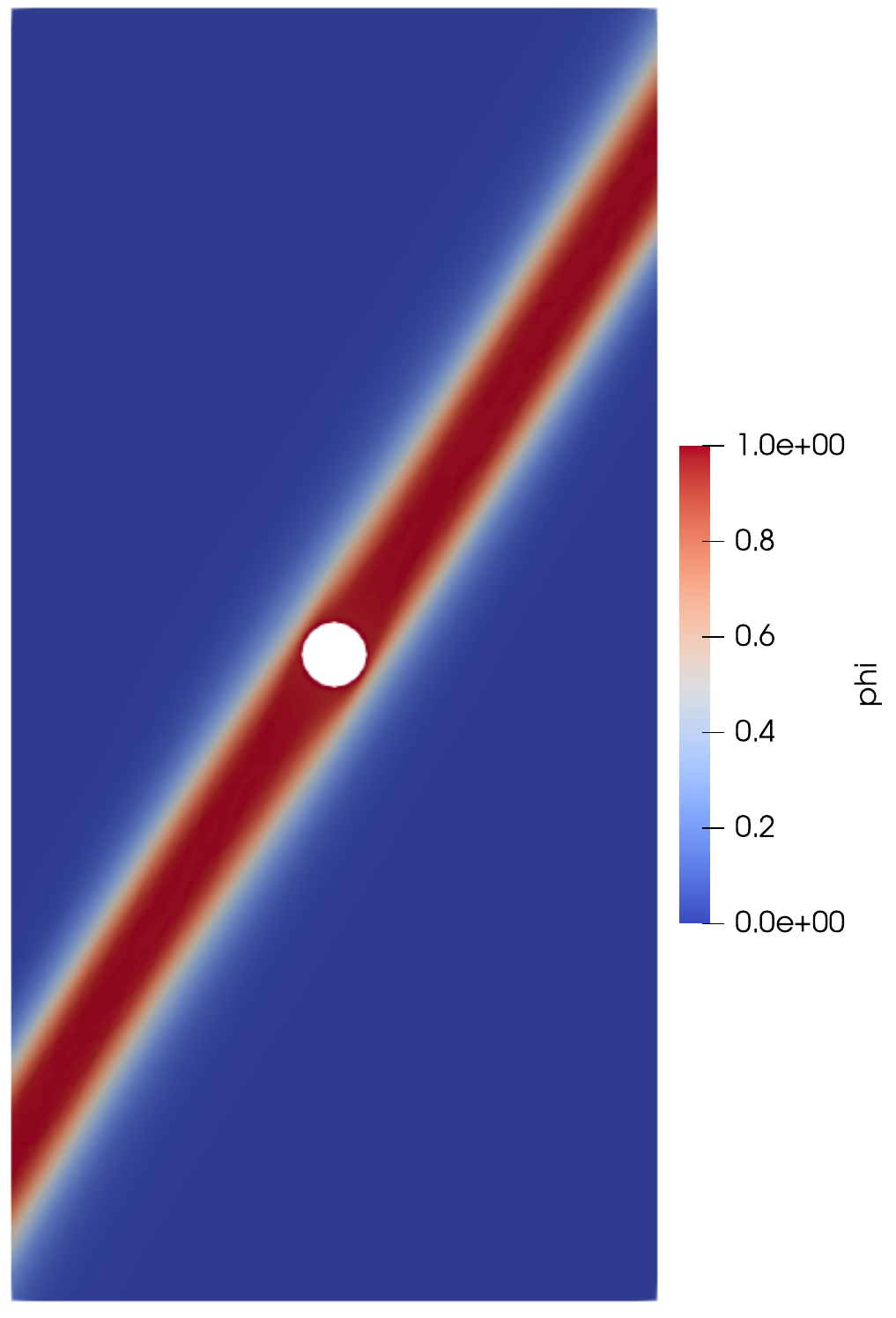}}
    \hfill
    \subfloat[Crack growth for the layers at $75^{\circ}$ in the homogenized domain.]
	{\includegraphics[width=0.25\textwidth]{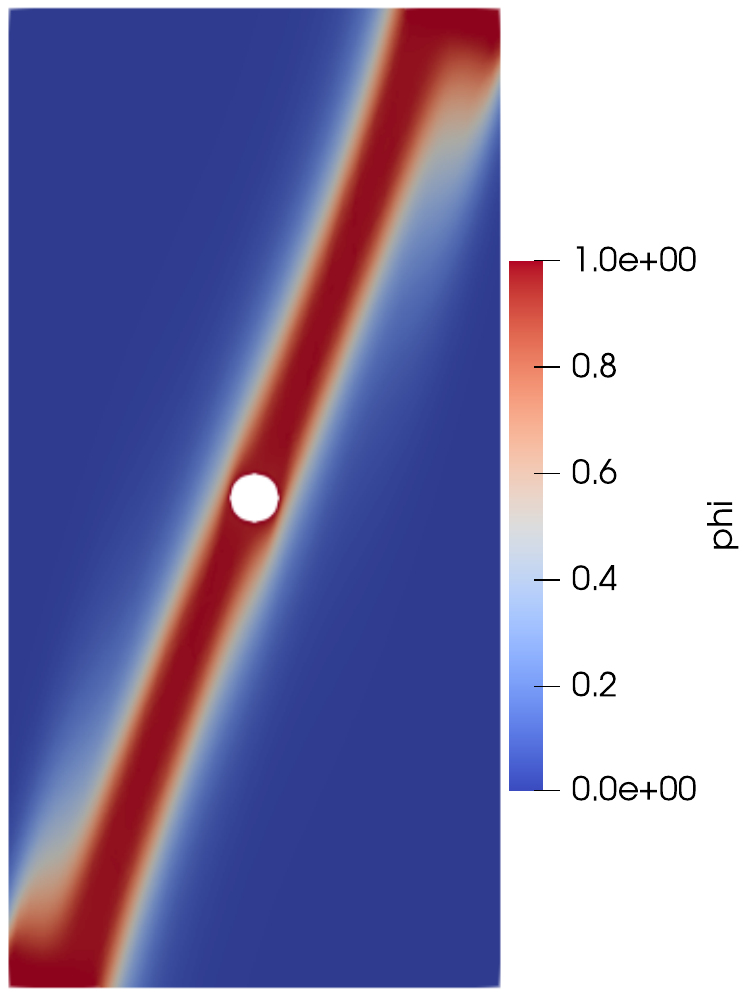}}
    \\
    \subfloat[Total force versus displacement for a specimen with layers oriented at \(45^\circ\) compared in the fully-resolved and homogenized settings.]
	{\includegraphics[width=0.3\textwidth]{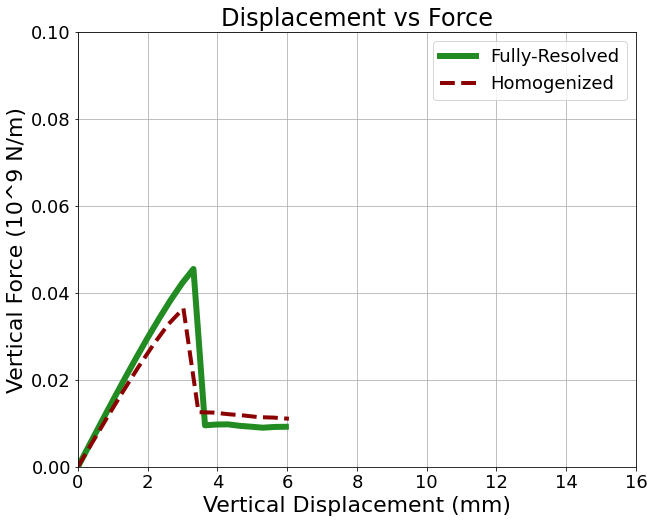}}
    \hfill
    \subfloat[Total force versus displacement for a specimen with layers oriented at \(60^\circ\) compared in the fully-resolved and homogenized settings.]
	{\includegraphics[width=0.3\textwidth]{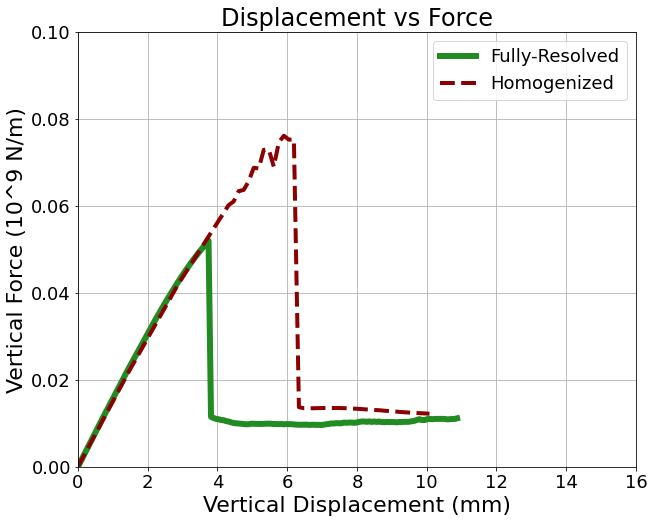}}
    \hfill
    \subfloat[Total force versus displacement for a specimen with layers oriented at \(75^\circ\) compared in the fully-resolved and homogenized settings.]
	{\includegraphics[width=0.3\textwidth]{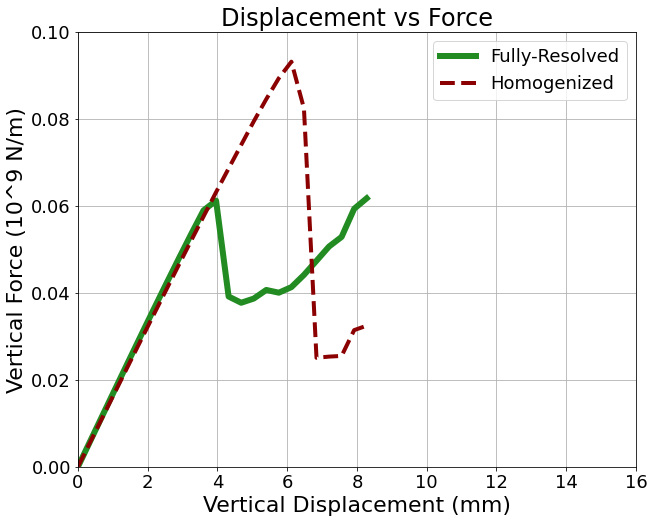}}
    \caption{
        Comparison of crack growth in the fully-resolved and homogenized settings at various layer orientations under uniaxial compression. The horizontal faces are subject to displacement-controlled compression, while the vertical faces are traction-free, following \cite{tien2006experimental}.
    }
        \label{crack compr}
\end{figure}

Figure \ref{crack compr} compares the fully-resolved and homogenized specimens for layer orientations of $45^{\circ}$, $60^{\circ}$, and $75^{\circ}$.
The results show the overall agreement in crack growth path between fully-resolved and homogenized domain, supporting the efficacy of our homogenization method in representing the fracture response of the rock layers. 
It validates the model's ability to replicate crack paths observed in the experimental studies of \cite{tien2006experimental} and to realistically simulate crack growth under compressive loading, as explained in \cite{hakimzadeh2022phase}. 
This is particularly notable given the challenges that previous models faced in accurately representing the response of cracks when the faces come into contact under compression, as described in the Appendix of \cite{hakimzadeh2022phase}.

\begin{figure}[htb!]
	\subfloat[Crack growth for the layers at $45^{\circ}$ in the fully-resolved domain.]
	{\includegraphics[width=0.25\textwidth]{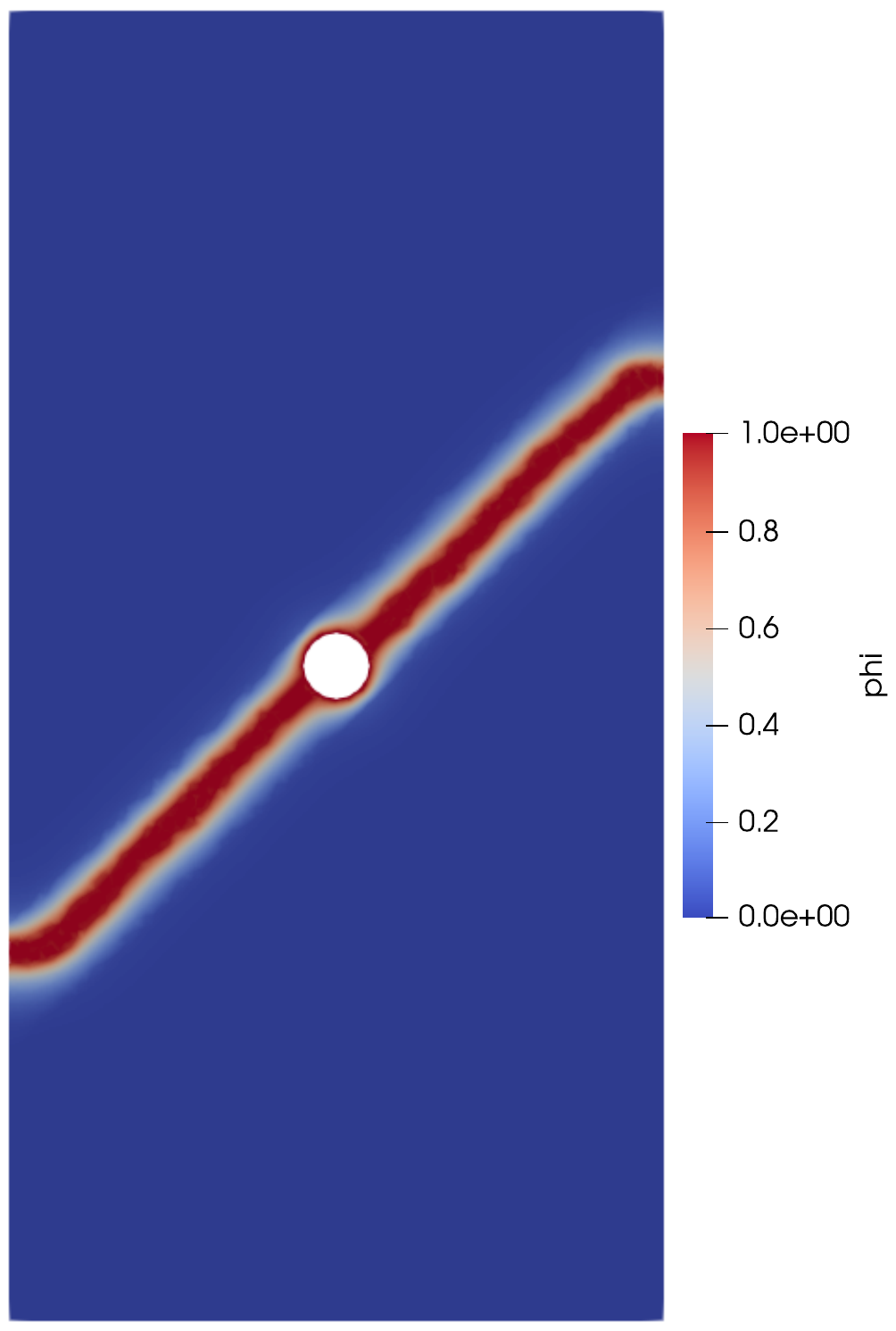}}
    \hfill
    \subfloat[Crack growth for the layers at $60^{\circ}$ in the fully-resolved domain.]
	{\includegraphics[width=0.25\textwidth]{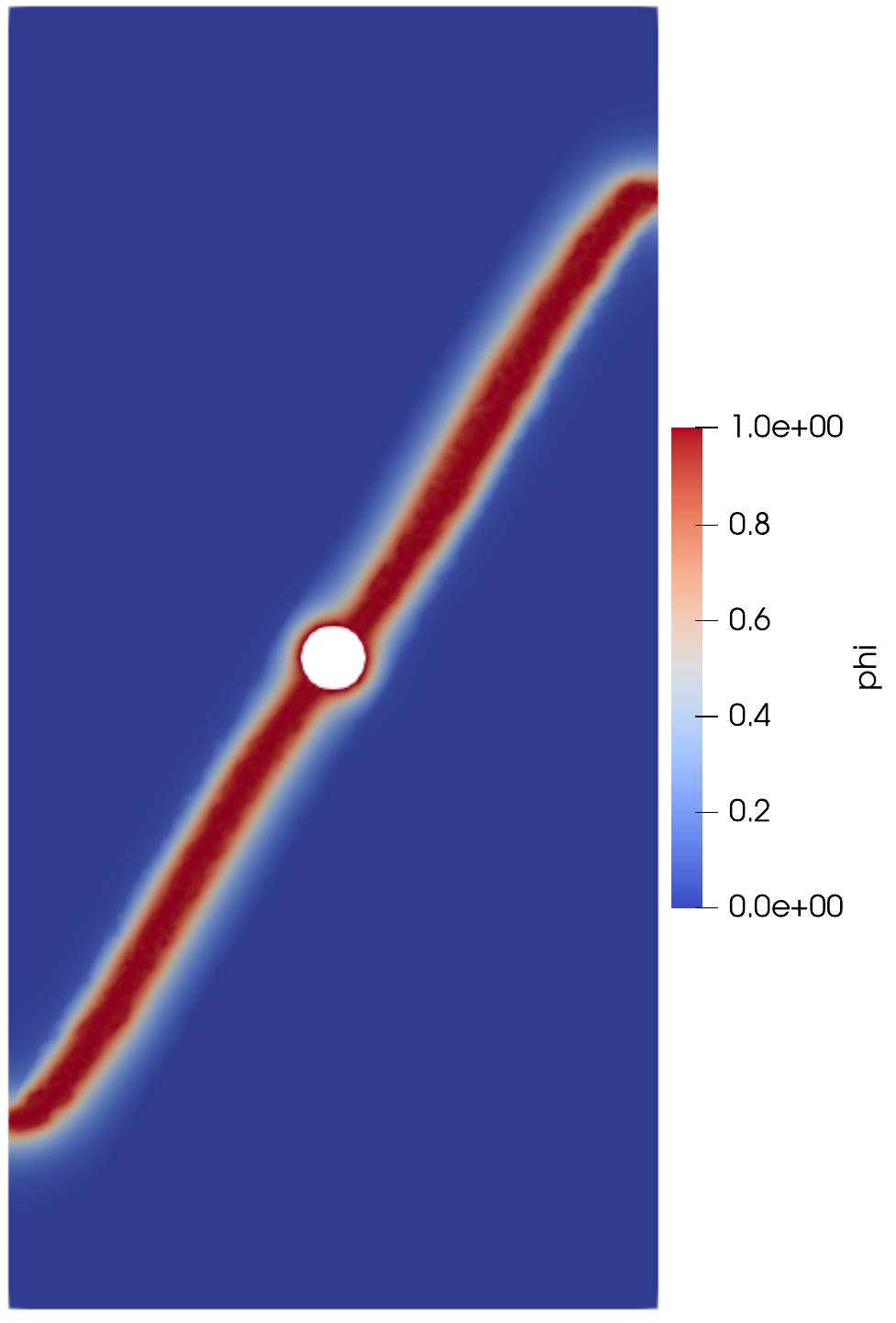}}
    \hfill
    \subfloat[Crack growth for the layers at $75^{\circ}$ in the fully-resolved domain.]
	{\includegraphics[width=0.25\textwidth]{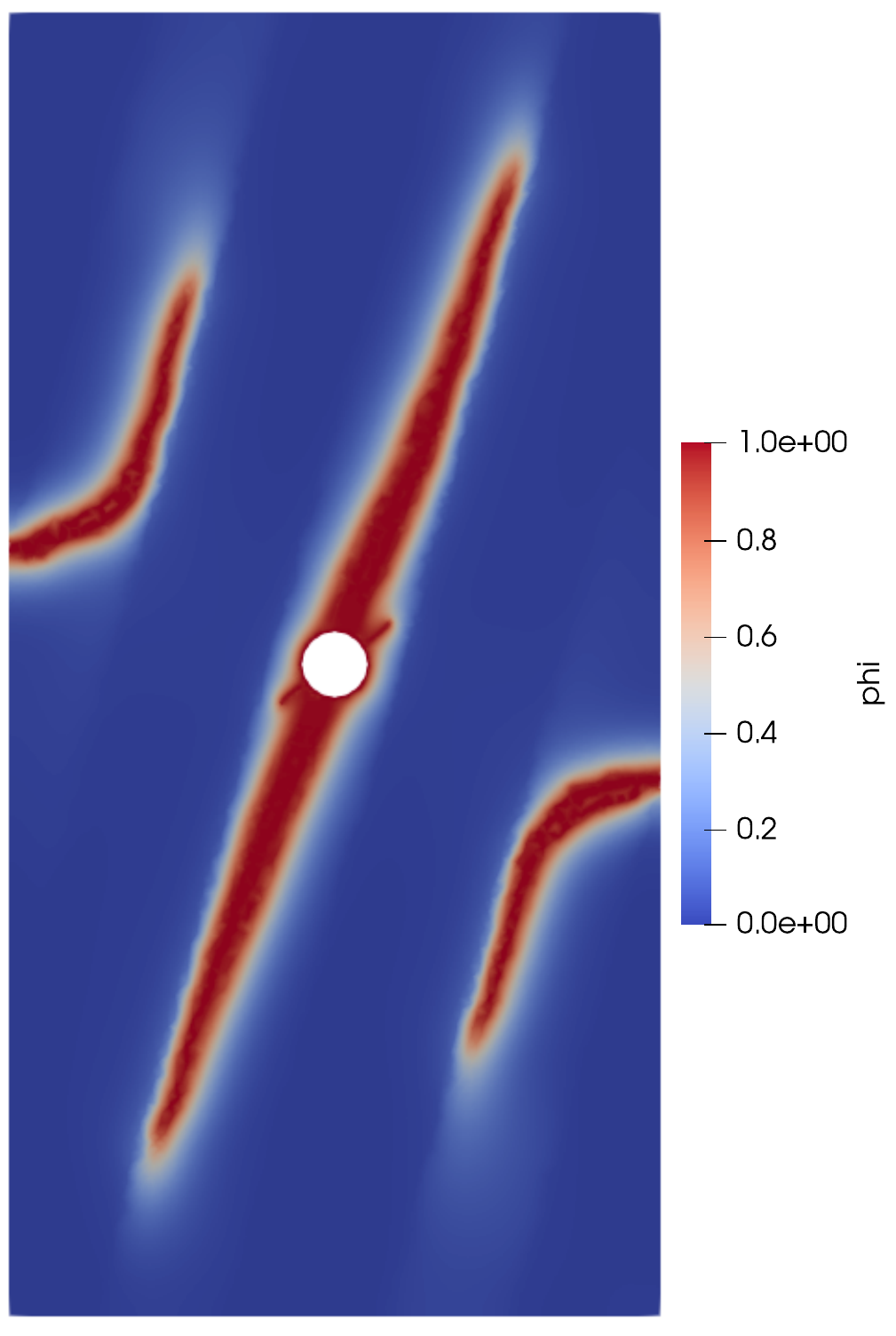}}
    \hfill
    \\
    \subfloat[Crack growth for the layers at $45^{\circ}$ in the homogenized domain.]
	{\includegraphics[width=0.25\textwidth]{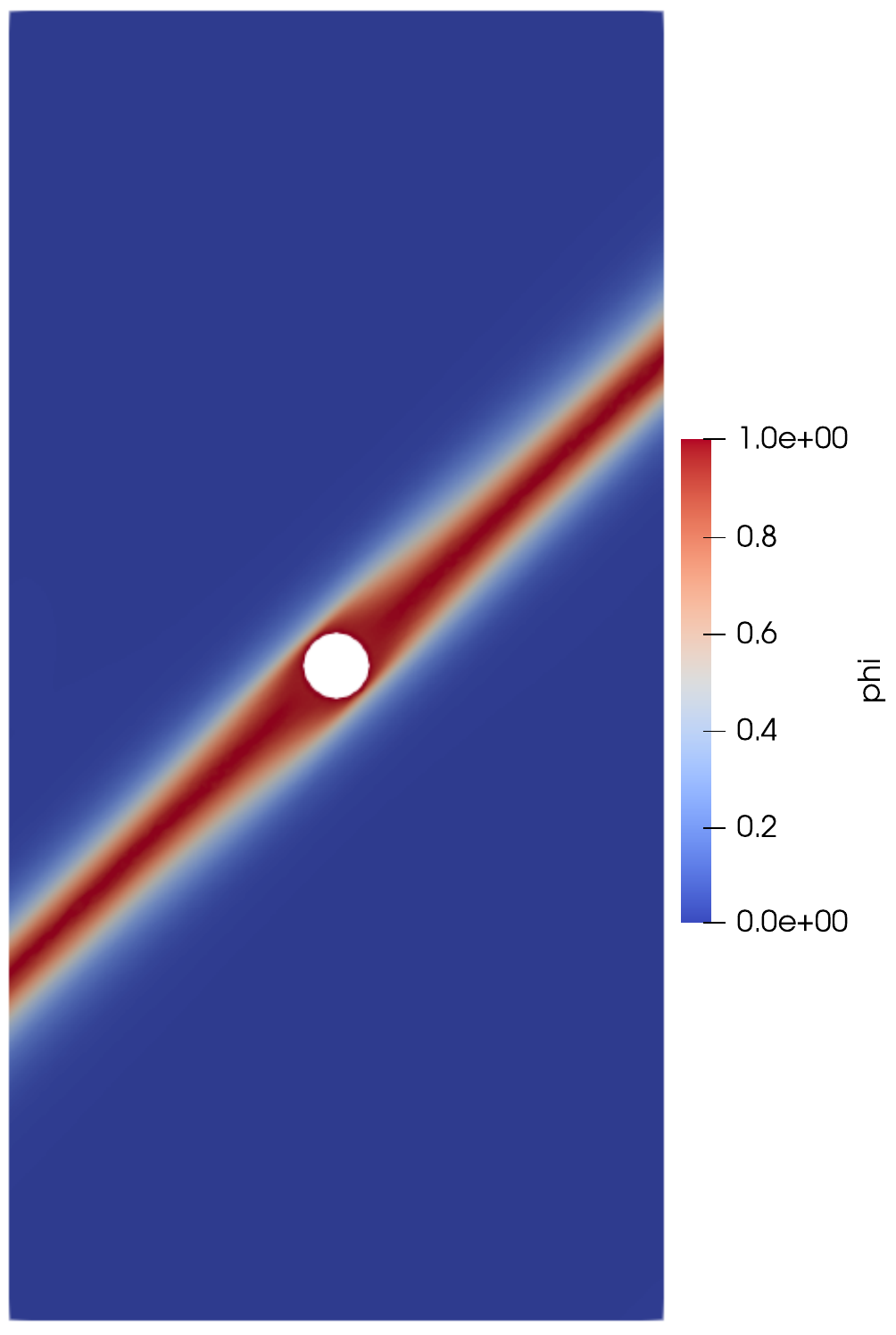}}
    \hfill
	\subfloat[Crack growth for the layers at $60^{\circ}$ in the homogenized domain.]
	{\includegraphics[width=0.25\textwidth]{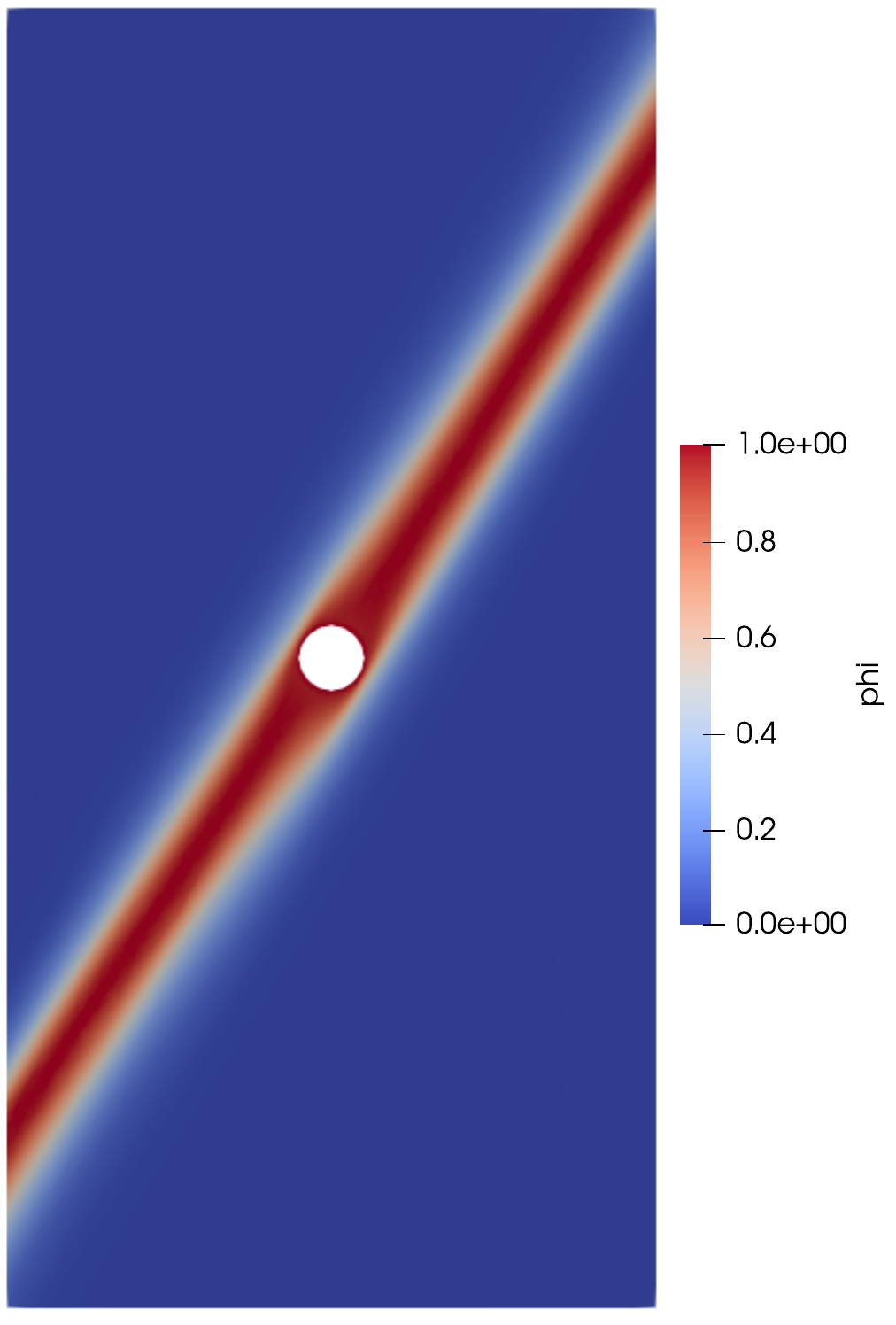}}
    \hfill
	\subfloat[Crack growth for the layers at $75^{\circ}$ in the homogenized domain.]
	{\includegraphics[width=0.25\textwidth]{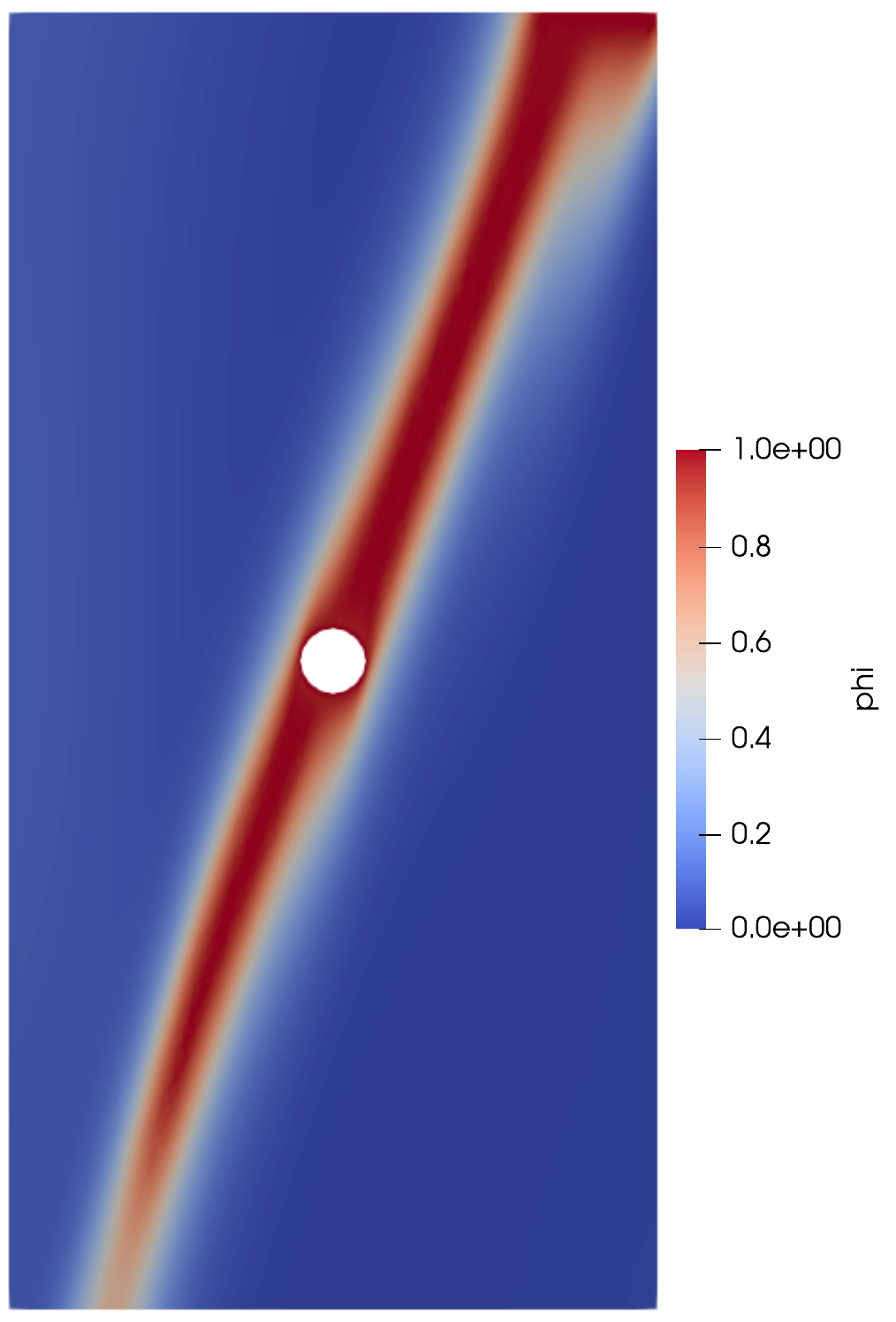}}
 \\
     \subfloat[Total force versus displacement for a specimen with layers oriented at \(45^\circ\) compared in the fully-resolved and homogenized settings.]
	{\includegraphics[width=0.3\textwidth]{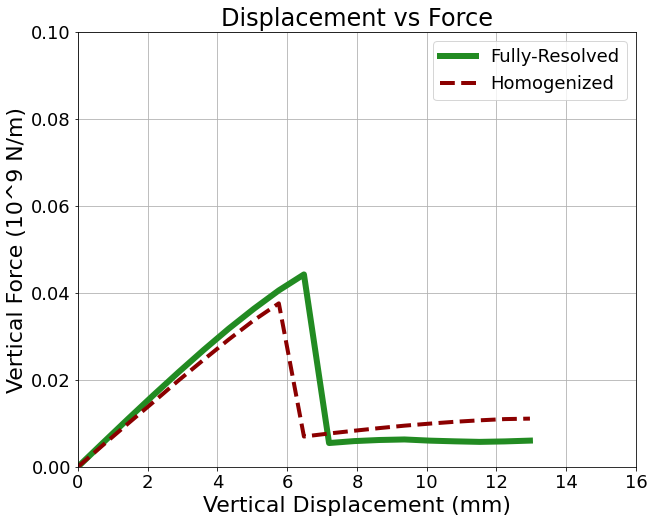}}
    \hfill
    \subfloat[Total force versus displacement for a specimen with layers oriented at \(60^\circ\) compared in the fully-resolved and homogenized settings.]
	{\includegraphics[width=0.3\textwidth]{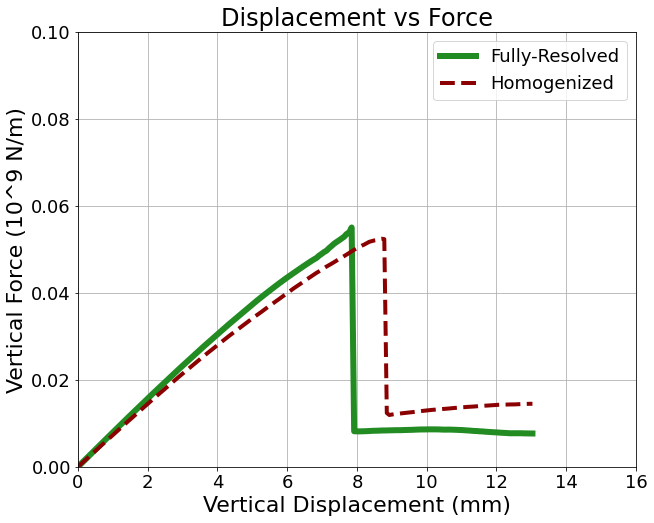}}
    \hfill
    \subfloat[Total force versus displacement for a specimen with layers oriented at \(75^\circ\) compared in the fully-resolved and homogenized settings.]
	{\includegraphics[width=0.3\textwidth]{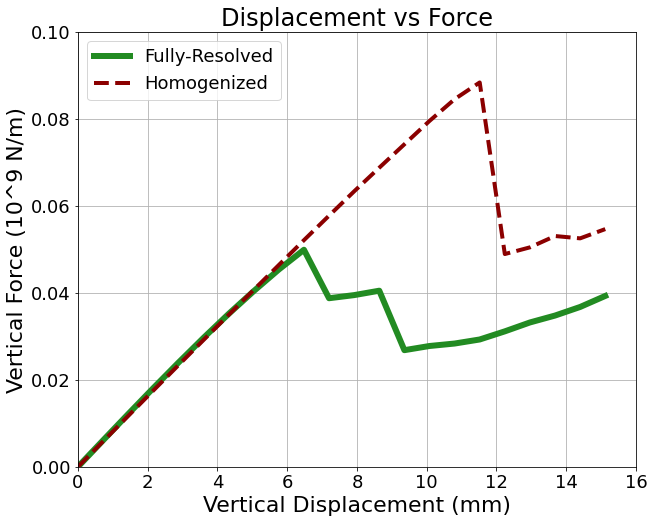}}
    \caption{Comparison of crack growth in the fully-resolved and homogenized settings at various layer orientations with imposed vertical extension to drive Mode I failure, keeping the vertical faces traction-free.}
         \label{crack open layer}
\end{figure}

Keeping the values of $\alpha$ and $\alpha_1$ fixed going forward, we next assess crack growth when the loading leads to Mode I failure. 
For these tests, the displacement at the bottom face of the domain is fixed. 
Subsequently, a displacement-controlled load is applied in the vertical direction on the top face. 
Figure \ref{crack open layer} compares the crack paths for both the fully-resolved and homogenized domains under Mode I failure at angles of $45^{\circ}$, $60^{\circ}$, and $75^{\circ}$. 
The results again show good agreement.

\begin{figure}[htb!] 
	\subfloat[Crack growth for the layers at $45^{\circ}$ in the fully-resolved domain.]
	{\includegraphics[width = 0.25\textwidth]{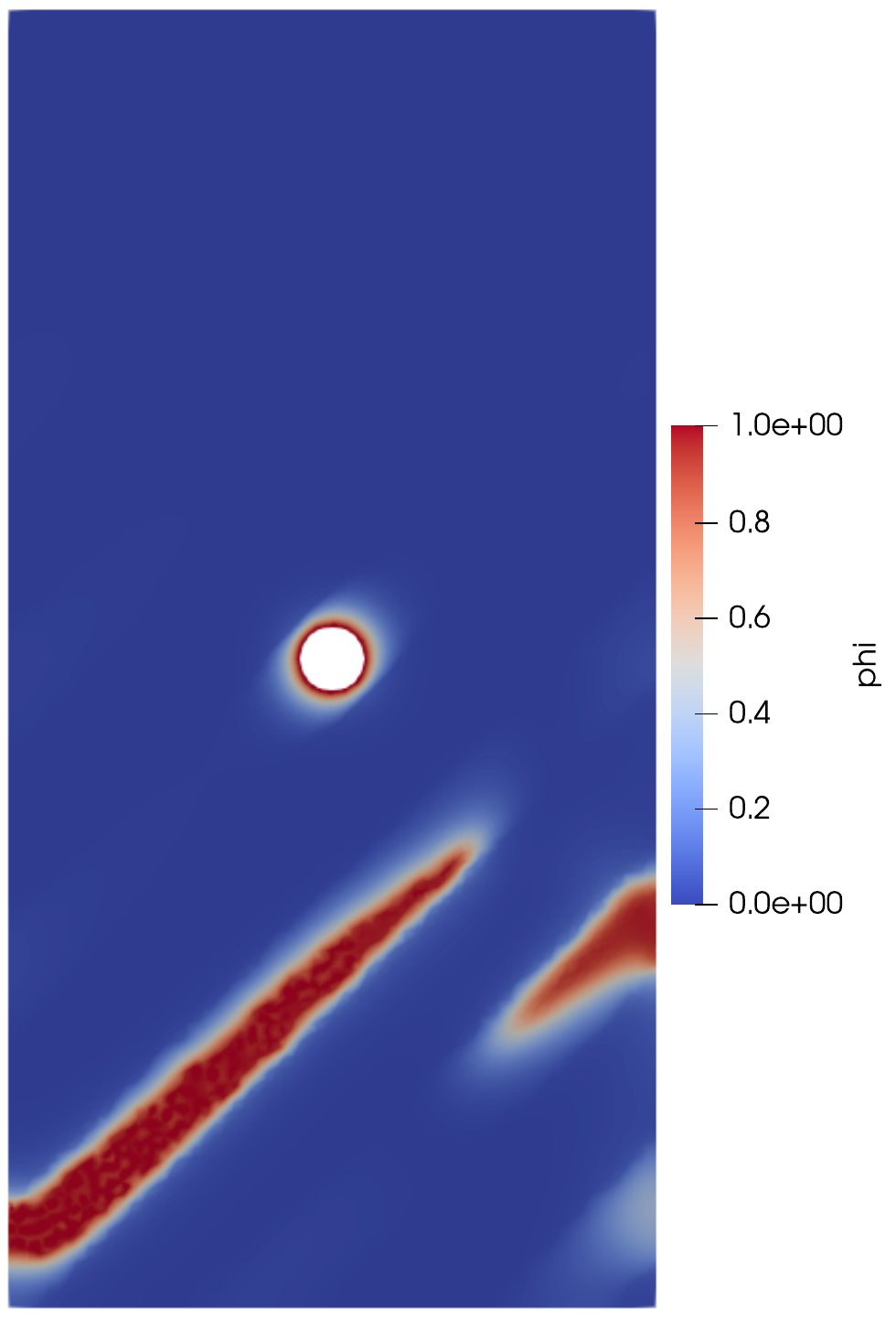}}
    \hfill
    \subfloat[Crack growth for the layers at $60^{\circ}$ in the fully-resolved domain.]
	{\includegraphics[width=0.25\textwidth]{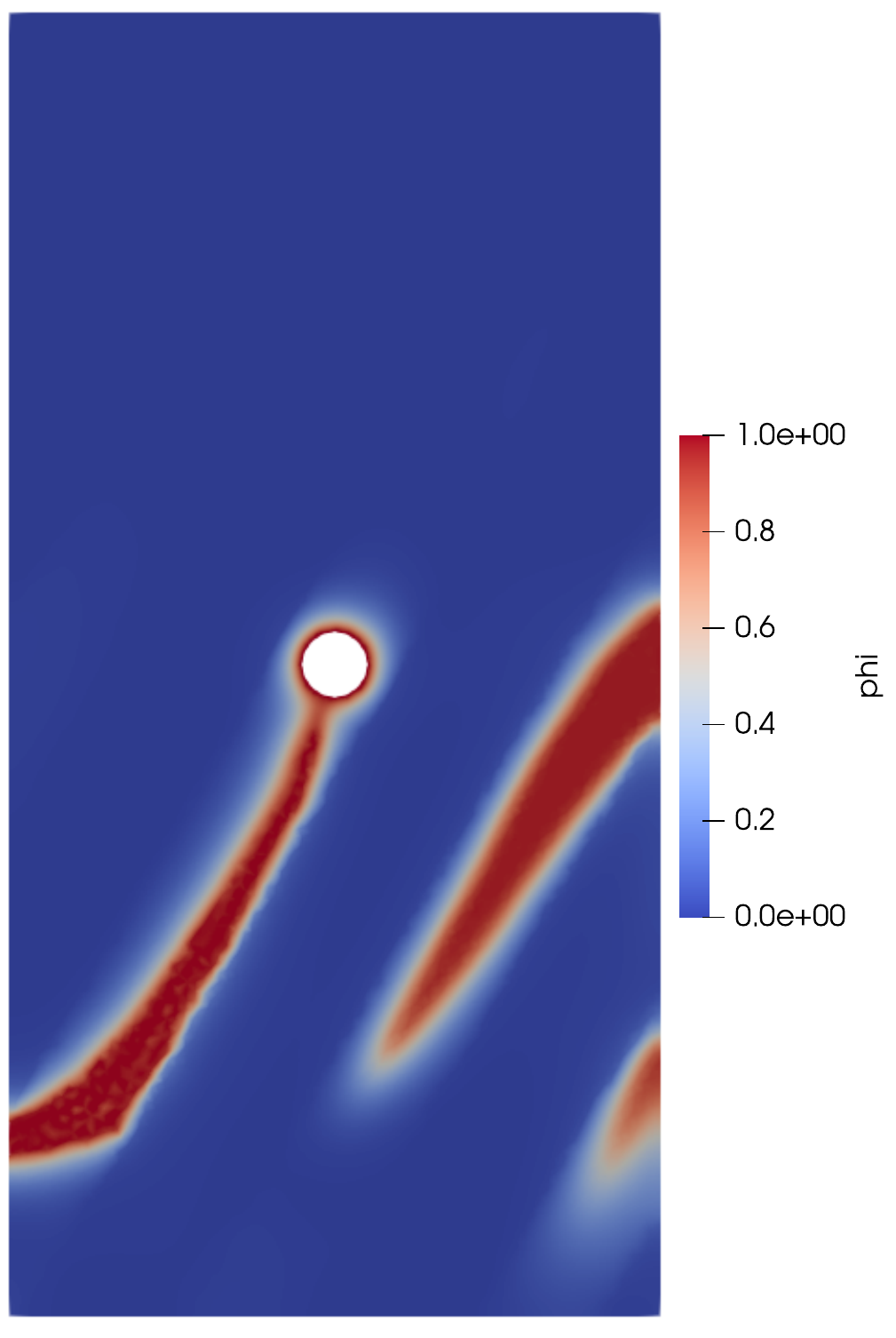}}
    \hfill
    \subfloat[Crack growth for the layers at $75^{\circ}$ in the fully-resolved domain.]
	{\includegraphics[width=0.25\textwidth]{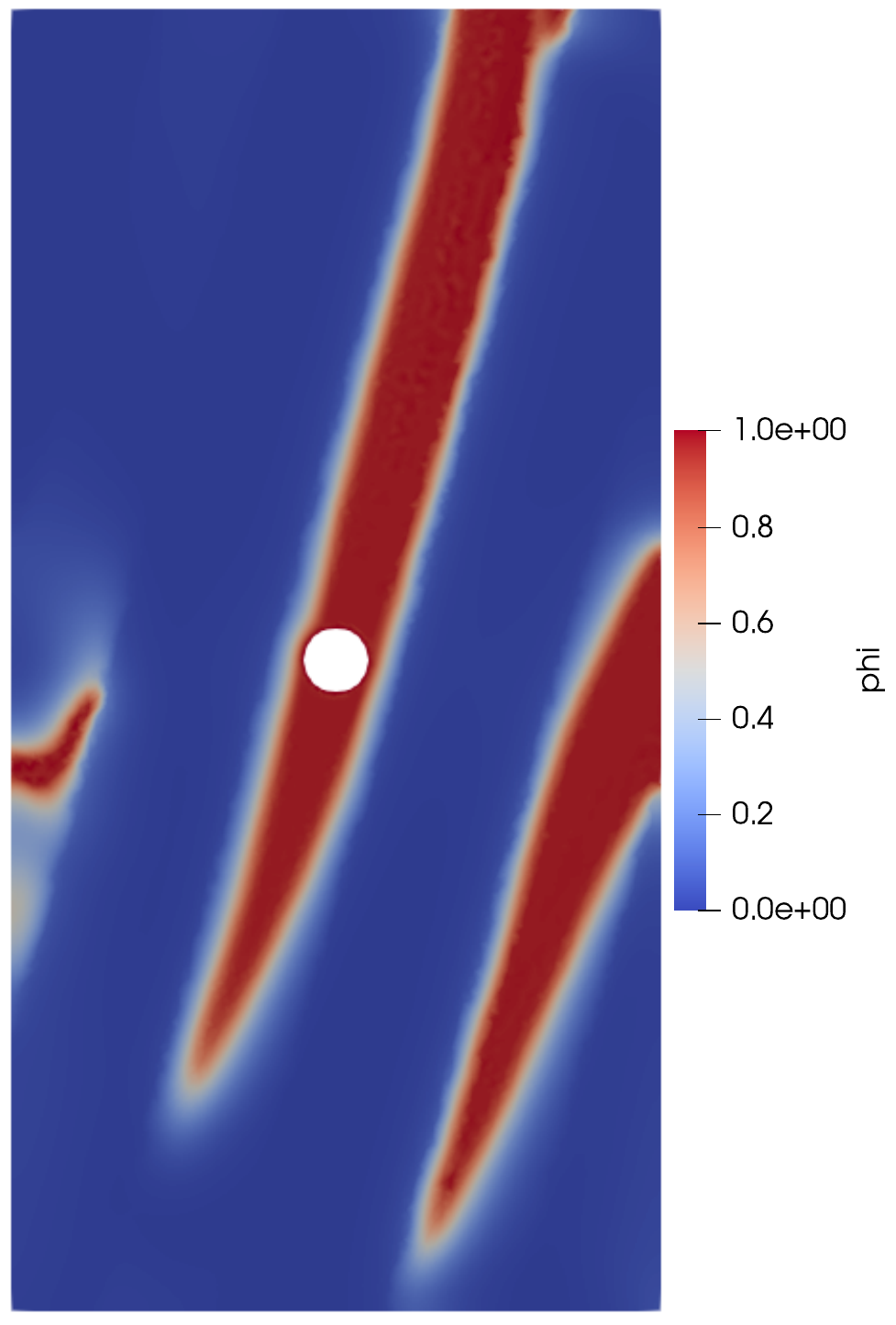}}
    \\
    \subfloat[Crack growth for the layers at $45^{\circ}$ in the homogenized domain.]
	{\includegraphics[width=0.25\textwidth]{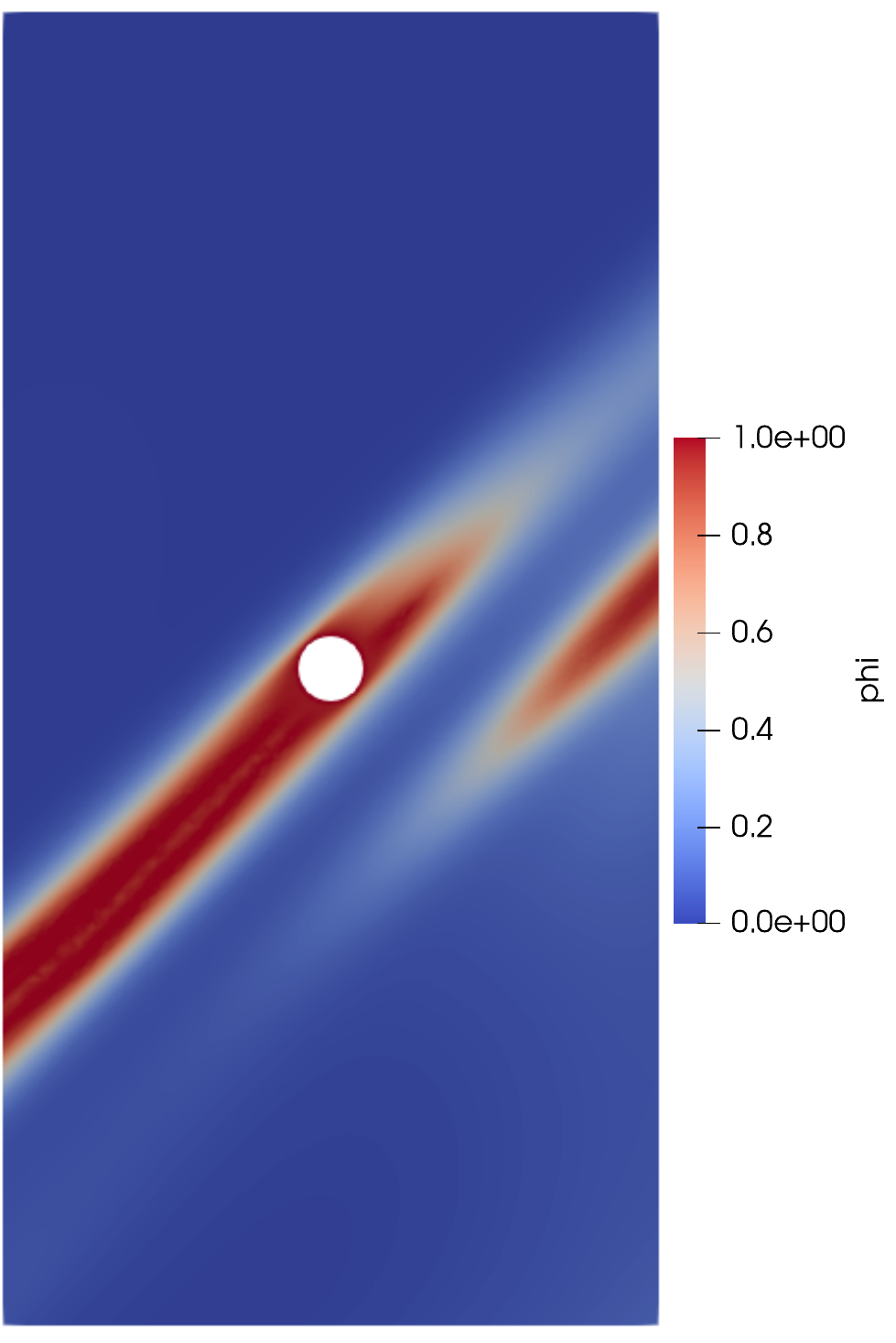}}
    \hfill
    \subfloat[Crack growth for the layers at $60^{\circ}$ in the homogenized domain.]
	{\includegraphics[width=0.25\textwidth]{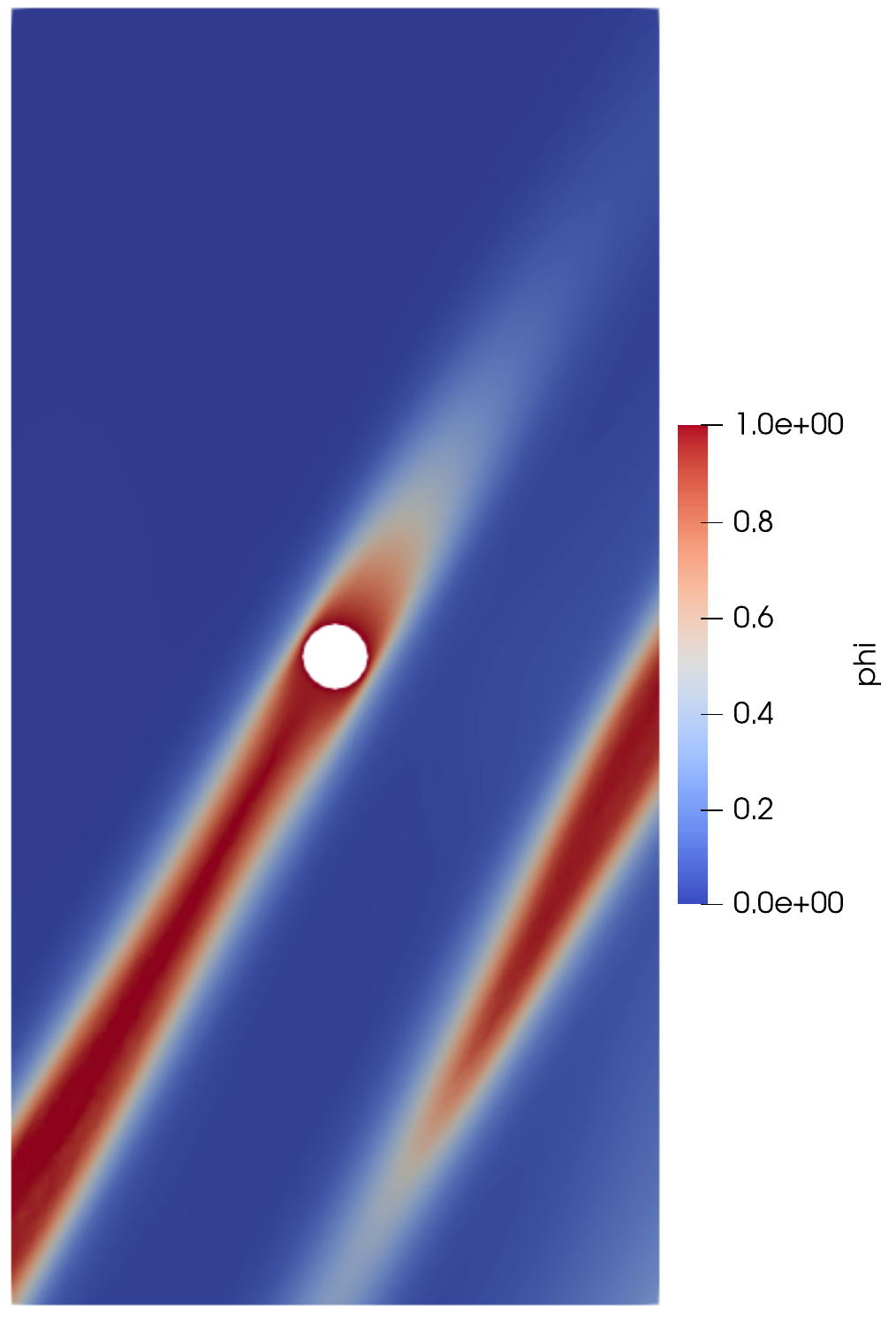}}
	\hfill
	\subfloat[Crack growth for the layers at $75^{\circ}$ in the homogenized domain.]
	{\includegraphics[width=0.25\textwidth]{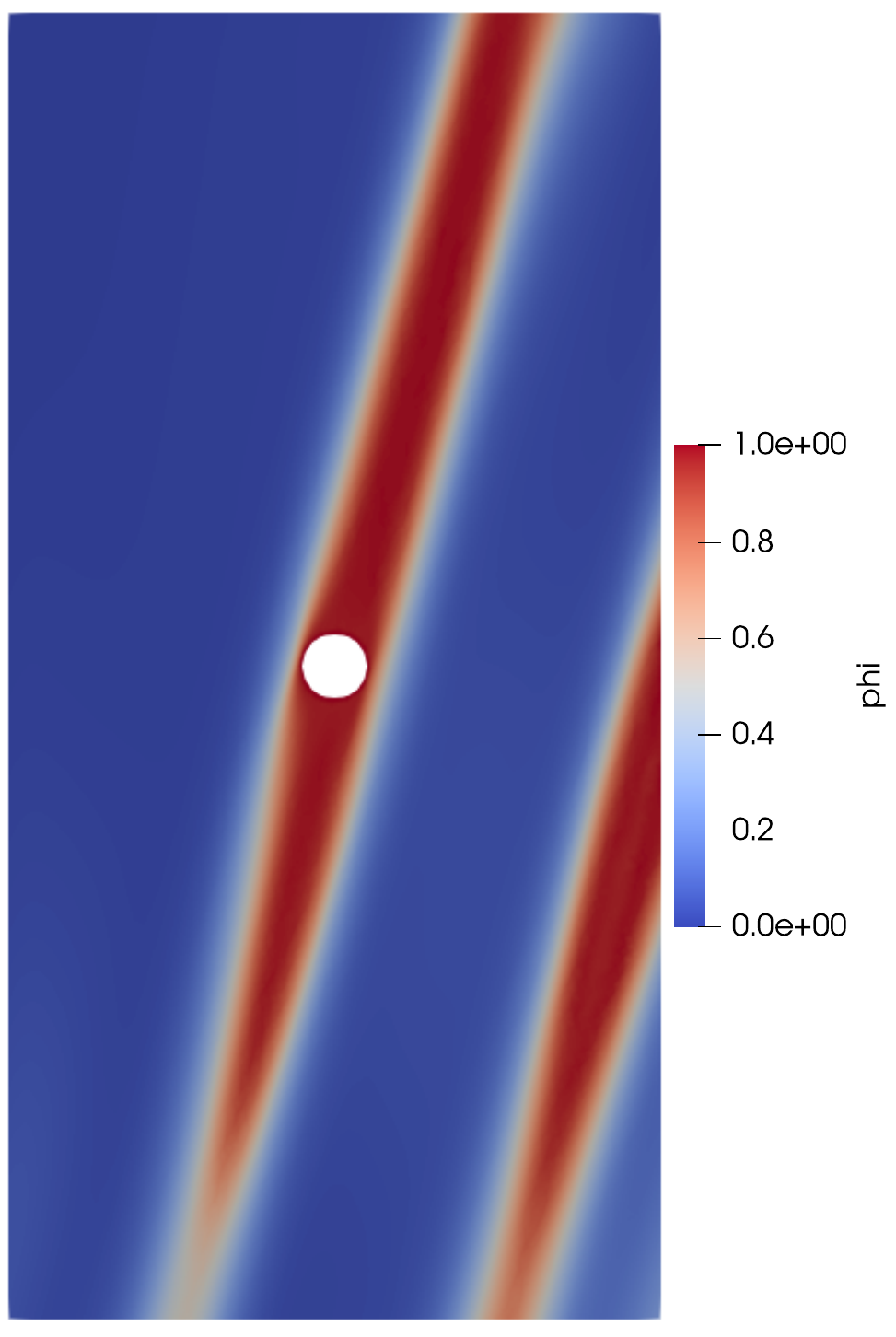}}
    \\
    \subfloat[Total force versus displacement for a specimen with layers oriented at \(45^\circ\) compared in the fully-resolved and homogenized settings.]
	{\includegraphics[width=0.3\textwidth]{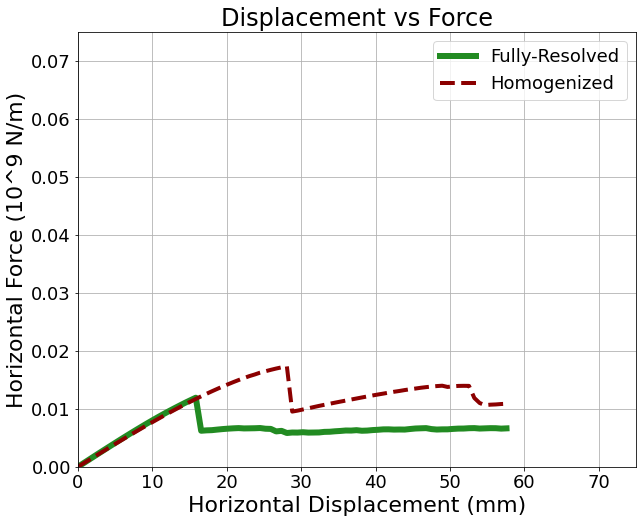}}
    \hfill
    \subfloat[Total force versus displacement for a specimen with layers oriented at \(60^\circ\) compared in the fully-resolved and homogenized settings.]
	{\includegraphics[width=0.3\textwidth]{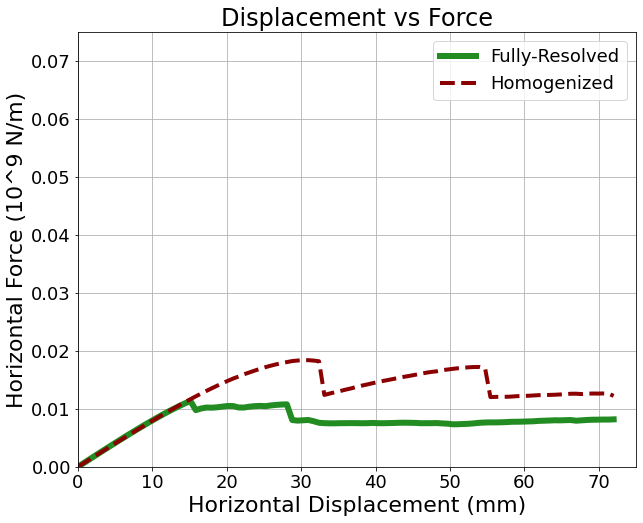}}
    \hfill
    \subfloat[Total force versus displacement for a specimen with layers oriented at \(75^\circ\) compared in the fully-resolved and homogenized settings.]
	{\includegraphics[width=0.3\textwidth]{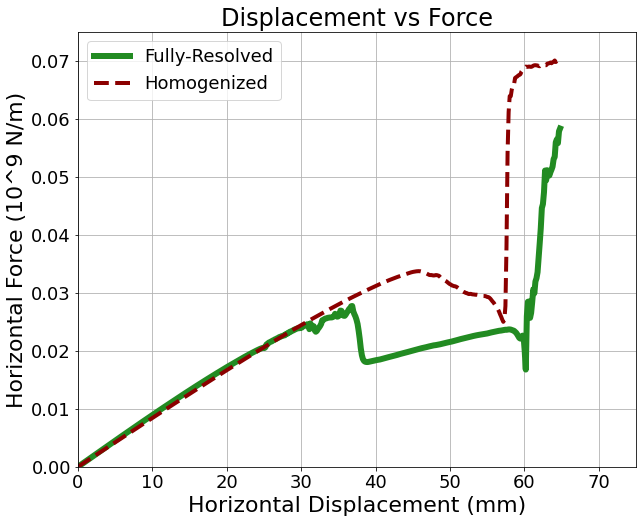}}
    \caption{Comparison of crack growth in the fully-resolved and homogenized settings at various layer orientations with shear displacement applied to the top face to drive Mode II failure, keeping the vertical faces traction-free.}
	\label{crack shear layer}
\end{figure}

We next compare crack growth in the fully-resolved and homogenized models when the fracture is driven by a dominant shear loading (Mode II fracture).
In these calculations, the bottom face is fixed, and the top face is subjected to displacement-controlled horizontal shearing to the right. The left and right faces remain traction-free. 
Figure \ref{crack shear layer} shows the results of these calculations.
The results show that the crack path in the fully-resolved and homogenized models matches well for layers at $45^\circ$,  $60^\circ$ and $75^\circ$ even in this complex loading scenario.

We highlight some key observations that holds broadly across all of the calculations described above.
First, the homogenized specimen typically nucleates a single dominant crack at the hole that drives the stress concentration, whereas the fully-resolved specimen sometimes nucleates secondary cracks in multiple layers.
It is expected that the homogenized description cannot capture this.
The effect of the nucleation of these additional cracks can be seen in the sudden load drops in the force-displacement curves in Figures \ref{crack compr}, \ref{crack open layer}, \ref{crack shear layer}.
There are further smaller oscillations in the force-displacement curves; these have been observed previously, e.g. \cite{wu2020phase,dammass2023phase}, and correspond to small discrete advances of the cracks in the case of displacement control.
Second, the force-displacement relations in the homogenized and fully-resolved setting match well in the linear regime before fracture initiation, as expected since classical homogenization is dominant in this regime.
Further, the force-displacement relations match reasonably well in terms of the trends and slopes, showing that the crack face contact model performs well.
However, there is a significant discrepancy in the peak load at which the primary fracture nucleates, and this is a well-known difficulty in phase-field fracture.
Third, we see that when the layers are oriented such that failure occurs in opening mode, the response is essentially brittle, as expected in this setting where we have not accounted for plastic effects.
However, when the layers are oriented such that the failure is driven by shear zones that are inclined with respect to the loading direction, the brittle response is seen to be attenuated.

\section{Formation of Wing Cracks Under Compressive Loading}
\label{sec:wing cracks}

Wing cracks in anisotropic materials are tensile cracks that propagate from pre-existing flaws under compressive loads.
They contribute significantly  to crack propagation and material failure in various applications, including in rock mechanics \cite{nemat1982compression, park2010crack, bobet1998fracture}.
These cracks typically initiate at the tips of the flaws and grow at an angle to the direction of applied compressive stress. 
The propagation of wing cracks is strongly influenced by material anisotropy where the fracture toughness and energy release rates vary with direction. 
Under compressive loads in an anisotropic setting, there are complex stress concentrations at flaw tips, and this drives the growth of complex fracture patterns, including secondary shear cracks. 
A key aspect of wing cracks is that the compressive stress causes crack face contact. It is essential to account for this appropriately, however standard phase-field fracture models do not distinguish between compressive traction across the crack face that causes contact and compressive traction that is parallel to the crack face \cite{steinke2022energetically,hakimzadeh2022phase}. However, the model proposed in this paper distinguishes between these tractions and is able to appropriately model crack face contact.

Several prior works have applied phase-field fracture mechanics to investigate the formation and propagation of wing cracks \cite{xu2024adaptive, li2021phase, spetz2021modified, bryant2018mixed, zhang2017modification}, and have proven effective in capturing these complex propagation modes, demonstrating good agreement with the expected physical behavior.
However, an important shortcoming in these prior studies is that they have modeled the initial crack as a boundary of the domain.
That is, the initial crack is treated as an external surface and is a geometric feature of the computational domain.
In this work, we are able to treat the initial crack through the damage field $\phi$: our computational domain is a simple rectangle, and the interior crack is prescribed through the initial condition for $\phi$.
This provides an important advantage: namely, by treating the crack as a damaged zone rather than as a special surface, it enables us to model wing cracks in a completely general way.
For instance, if the initial crack was not simply imposed but rather was the outcome of a prior loading, our approach is seamless and does not require any special treatment to then further observe the growth of wing cracks.
On the other hand, in the current method of treating this as an external surface, it would require a much more difficult process of identifying the crack geometry --- which can itself be complex --- replacing it with an external boundary and so on, which is extremely challenging.
Here, we highlight that our approach is enabled by the crack response model that enables us to obtain the correct crack response under crack face contact in the regularized setting without needing to use a cumbersome description of the crack faces as external surfaces.

\begin{figure}[h!] 
    \subfloat[]{\includegraphics[width=0.38\textwidth]{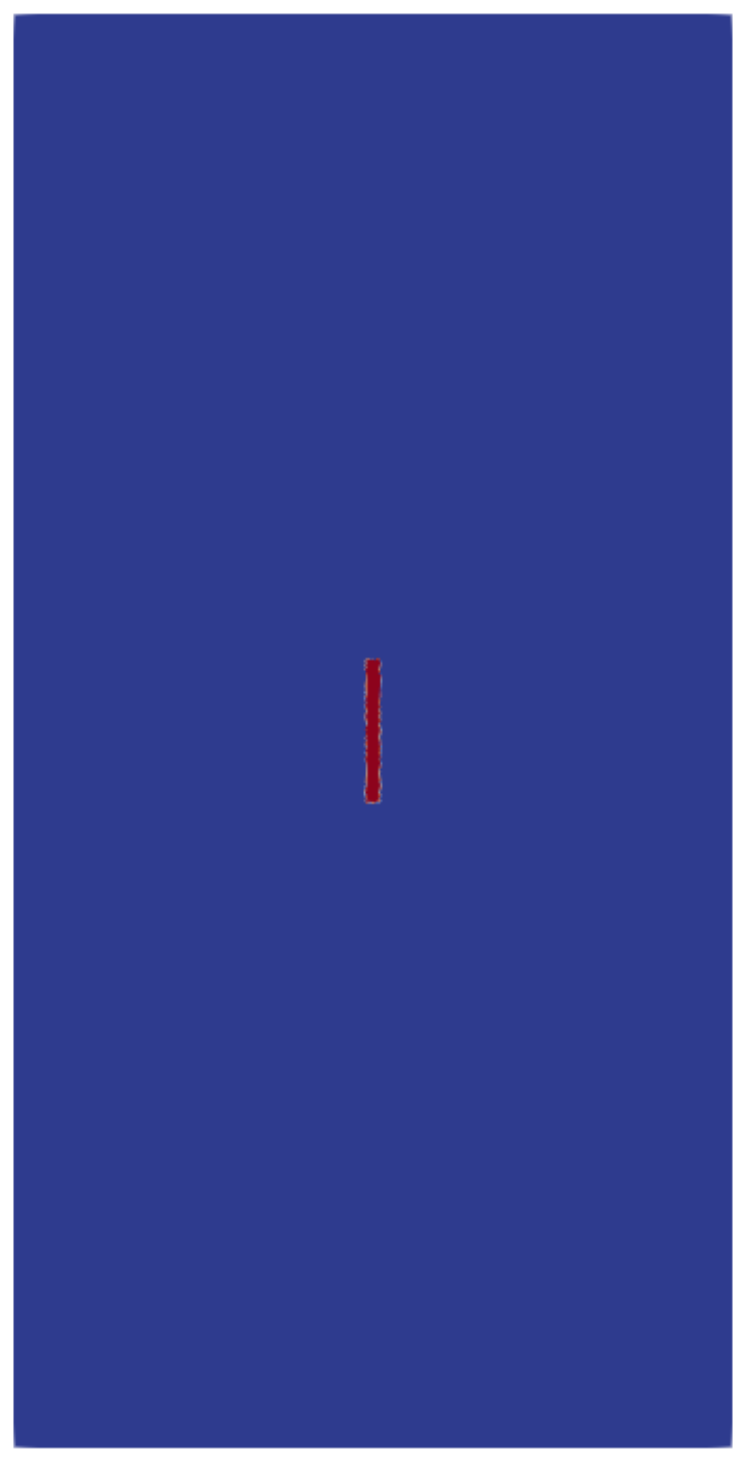}}
    \hfill
    \subfloat[]{\includegraphics[width=0.5\textwidth]{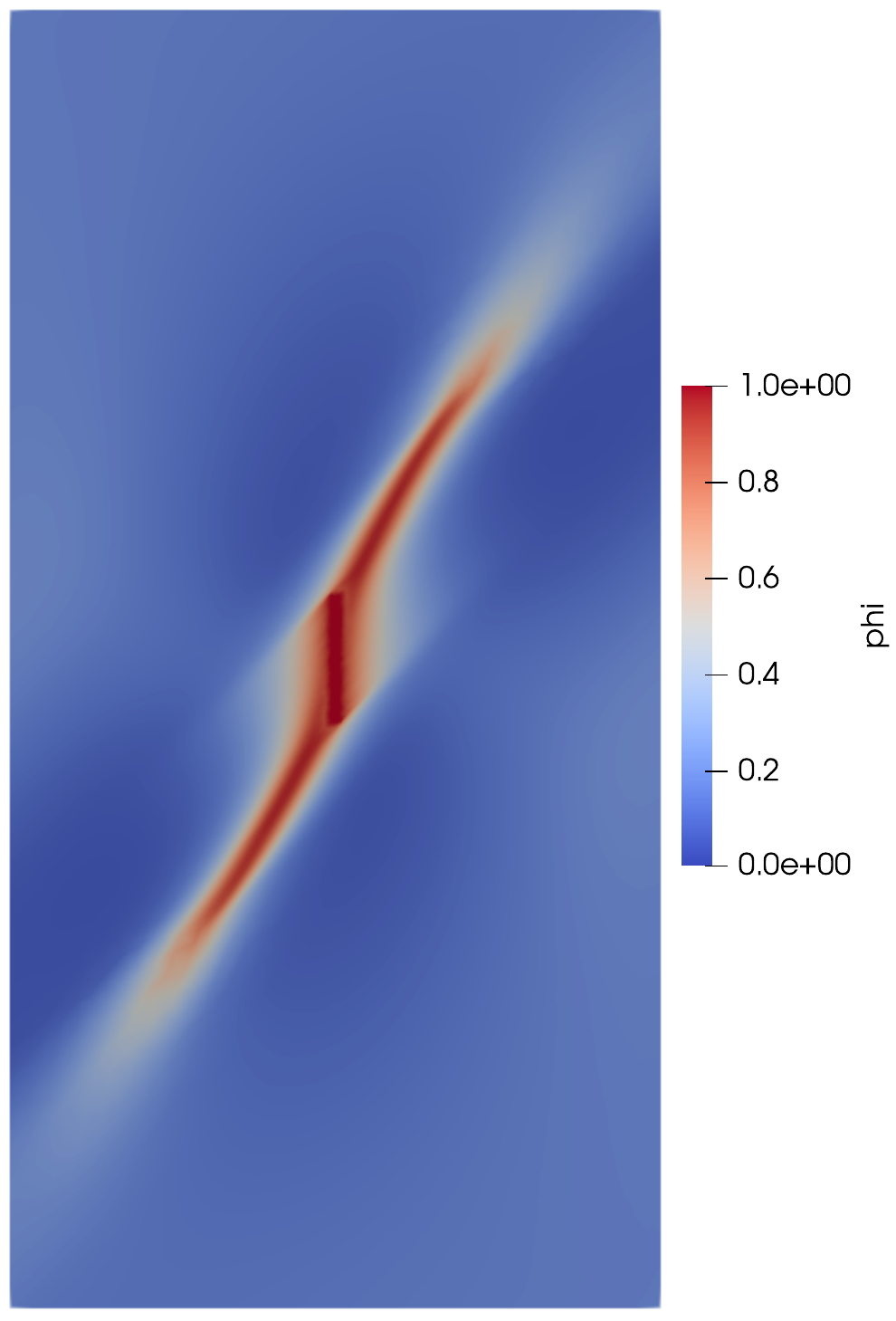}}
    \caption{
        (a) Initial crack before loading. 
        (b) Growth of wing cracks driven by compressive loading under confinement.
        We highlight that we use a simple rectangular domain for the the computations with no internal surfaces; the initial crack in (a) is modeled simply by the initial condition for $\phi$.
    }
    \label{fig:wings}
\end{figure}

Figure \ref{fig:wings} shows the results of wing crack growth in an anisotropic material.
We use the homogenized rock properties corresponding to layers oriented at $45^\circ$, as in Section \ref{sec:comparison}, with the difference that we set $\alpha = \alpha_1=200$. 
We apply displacement-prescribed compression and confinement on all faces using affine boundary conditions.
The results in Figure \ref{fig:wings} demonstrate that, under compressive loads, wing cracks initiate and propagate from the tips of pre-existing cracks, consistent with the physical expectations of brittle fracture behavior in anisotropic materials \cite{nemat2013micromechanics}.

We also highlight that we have idealized the crack faces as frictionless; however, several studies have demonstrated the significant role of friction  \cite{bobet1998fracture, horii1985compression, steif1984crack}.
The kinematic decomposition in our approach --- into crack opening, crack parallel, and crack shearing modes --- potentially enables the explicit modeling of friction in future work, unlike the conventional energy splitting methods. 

\section{Application to Confinement Experiments} \label{sec:Application to Layered Rock}

\begin{figure}[htb!]
    \centering
    \includegraphics[width=0.94\textwidth]{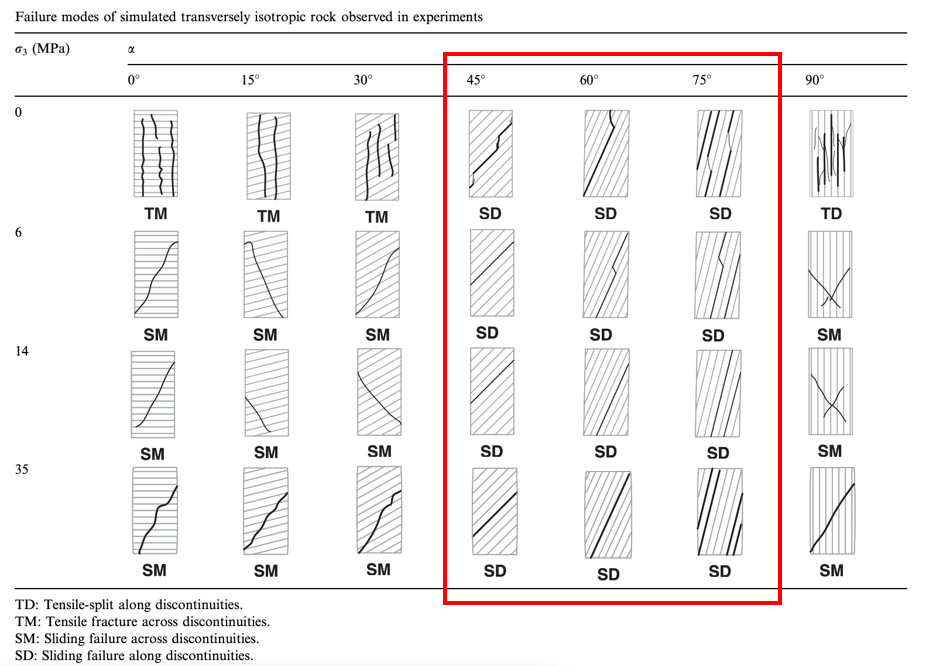}
    \caption{Failure modes of simulated transversely isotropic rock under different values of the confining compression, as observed in the experiments conducted by \cite{tien2006experimental}.}
    \label{fig:Tien_experimental}
\end{figure}

\begin{figure}[htb!]
    \begin{tabular}{c|cccccc}
         Anisotropy Angle & $45^\circ$ & Deformation ($\times 5$) & $60^\circ$ &  Deformation ($\times 5$) & $75^\circ$  &  Deformation ($\times 2$)  \\
         \hline
        \raisebox{2cm}{$\sigma_3 = \SI{0}{MPa}$} & 	\includegraphics[height=0.18\textwidth]{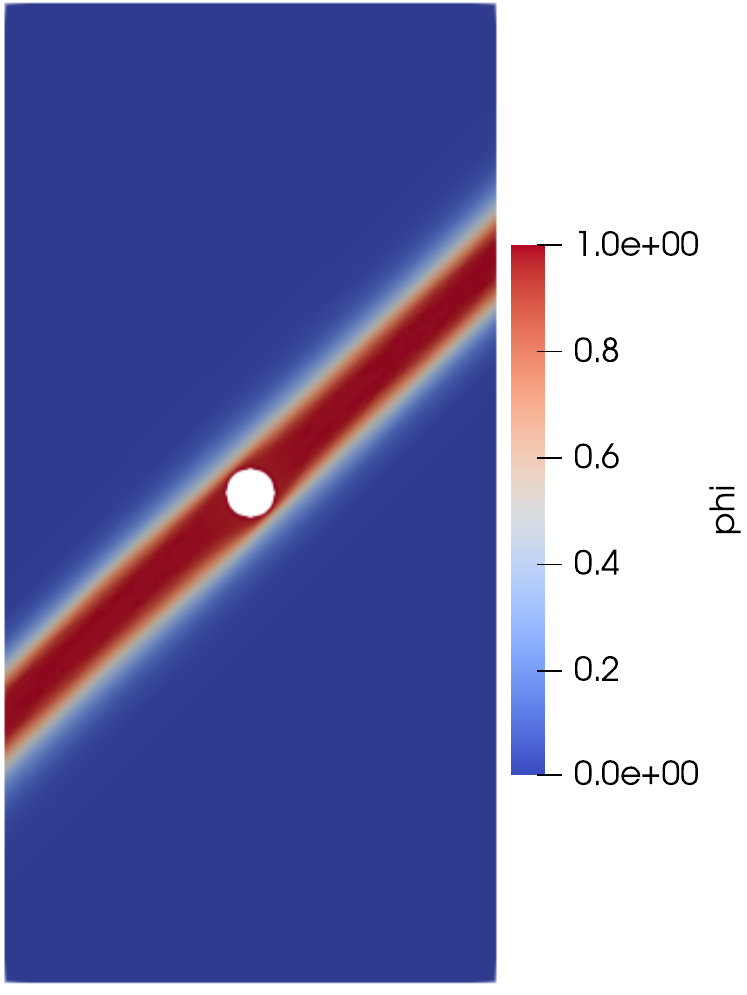} & \includegraphics[height=0.18\textwidth]{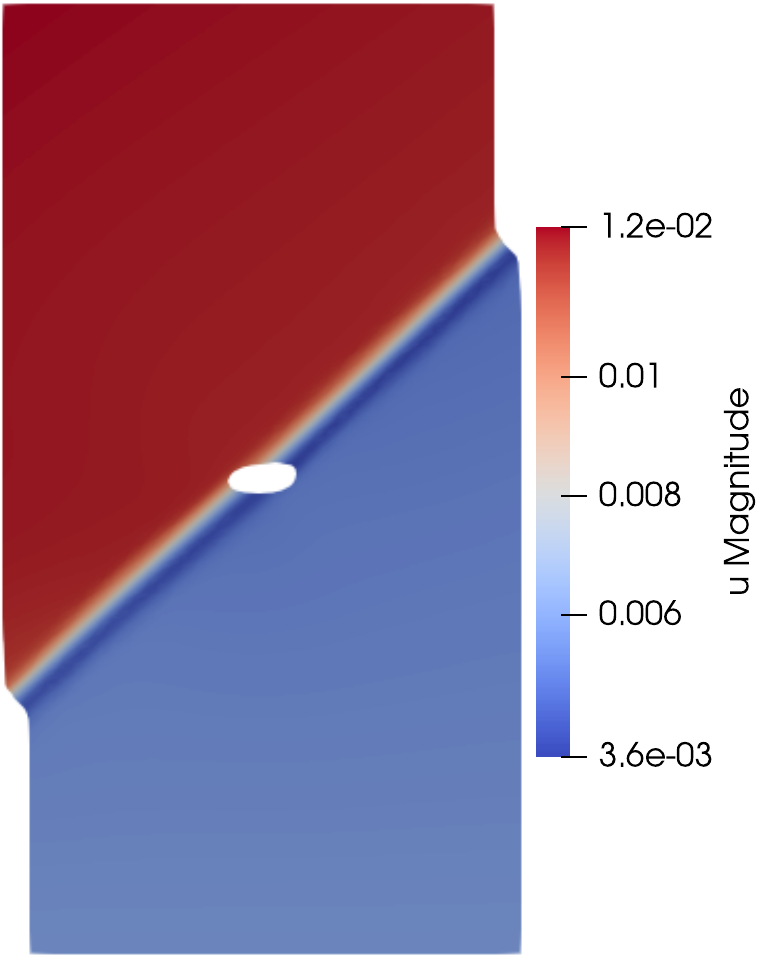} &	\includegraphics[height=0.18\textwidth]{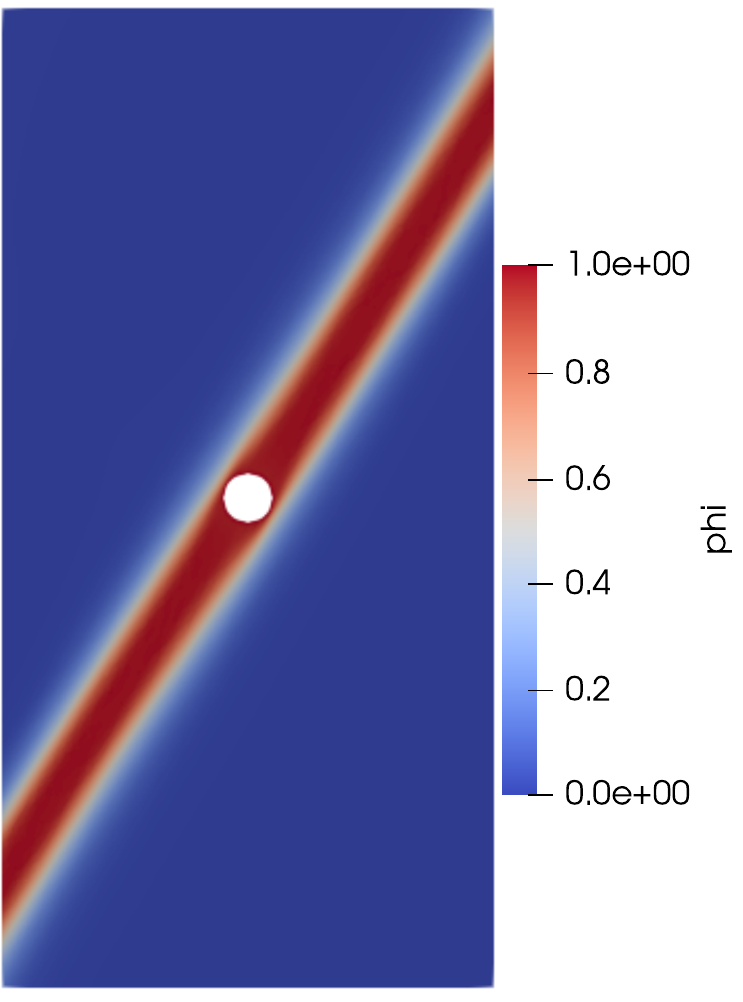}& \includegraphics[height=0.18\textwidth]{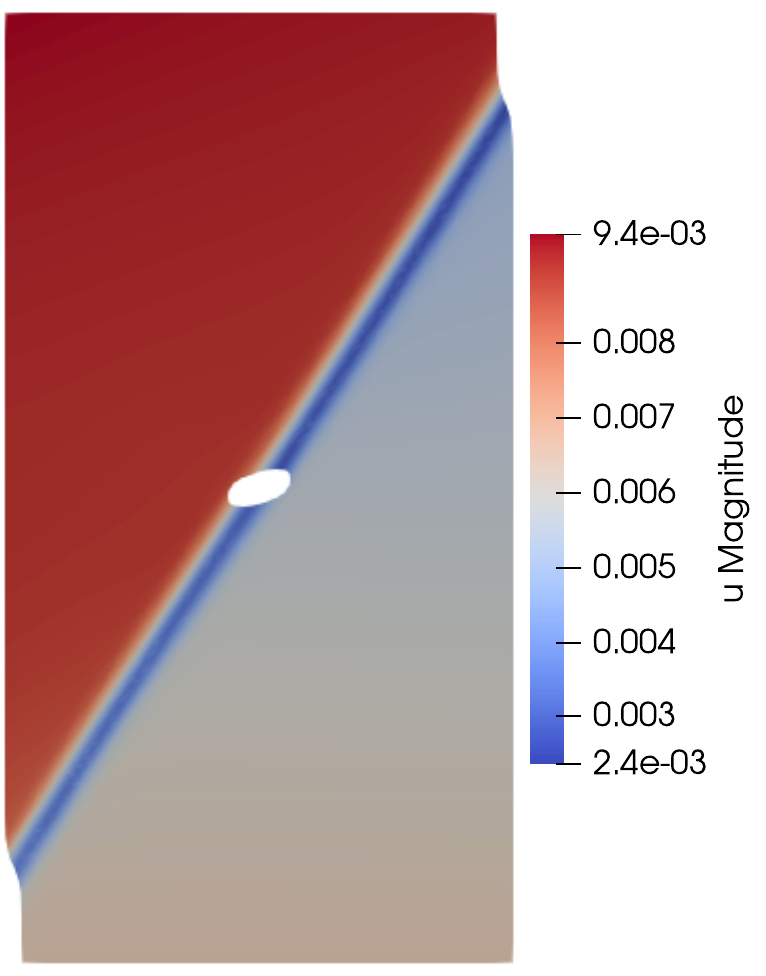} & \includegraphics[height=0.18\textwidth]{media/75Sig0.pdf} & \includegraphics[height=0.18\textwidth]{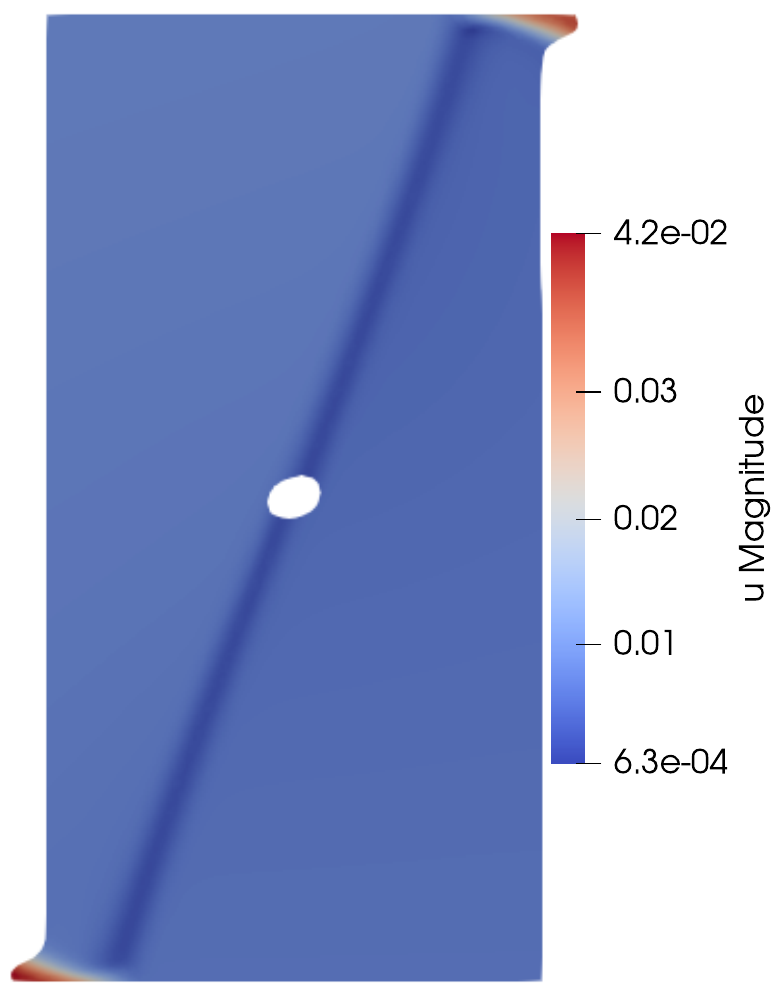} \\
         \hline
         \raisebox{2cm}{$\sigma_3 = \SI{6}{MPa}$}& 	\includegraphics[height=0.18\textwidth]{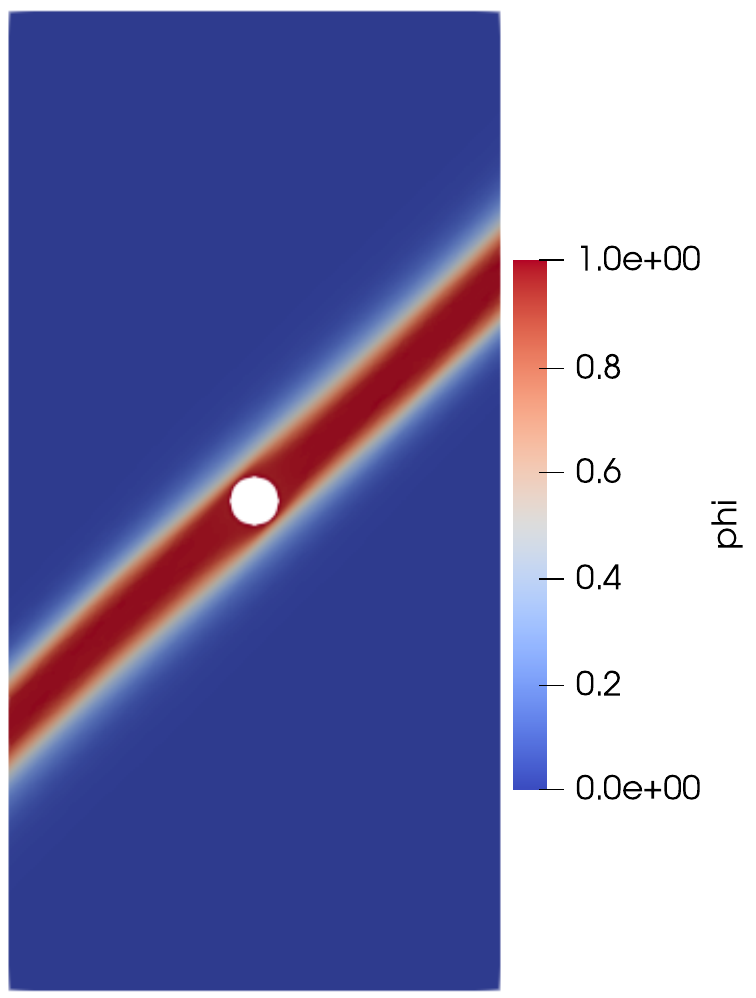} & \includegraphics[height=0.18\textwidth]{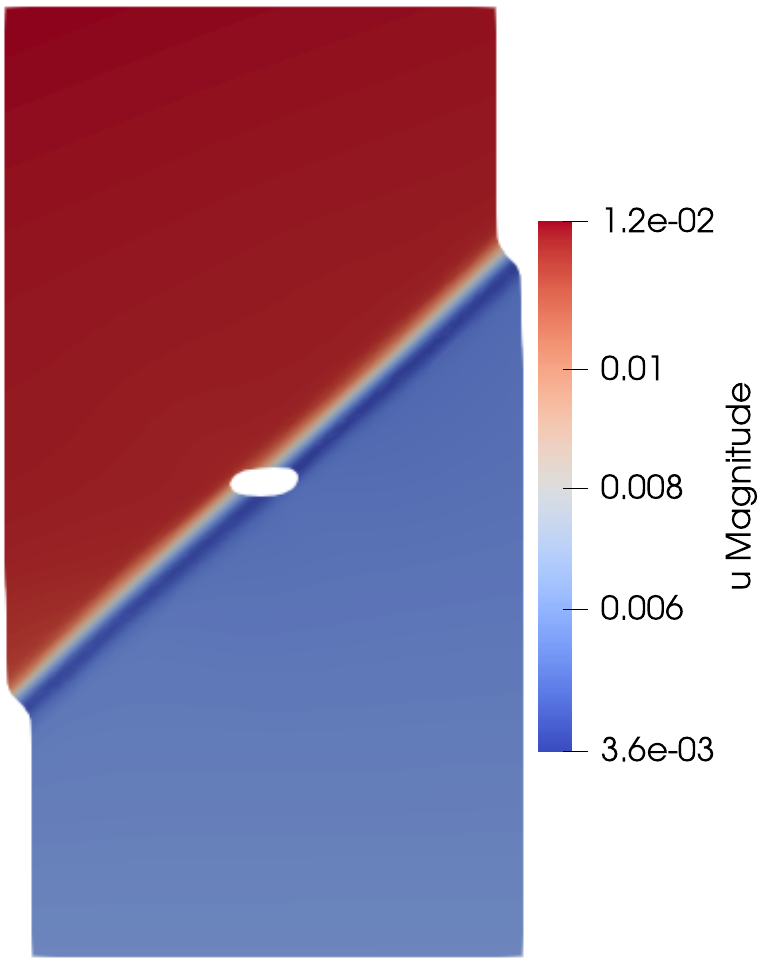}  &	\includegraphics[height=0.18\textwidth]{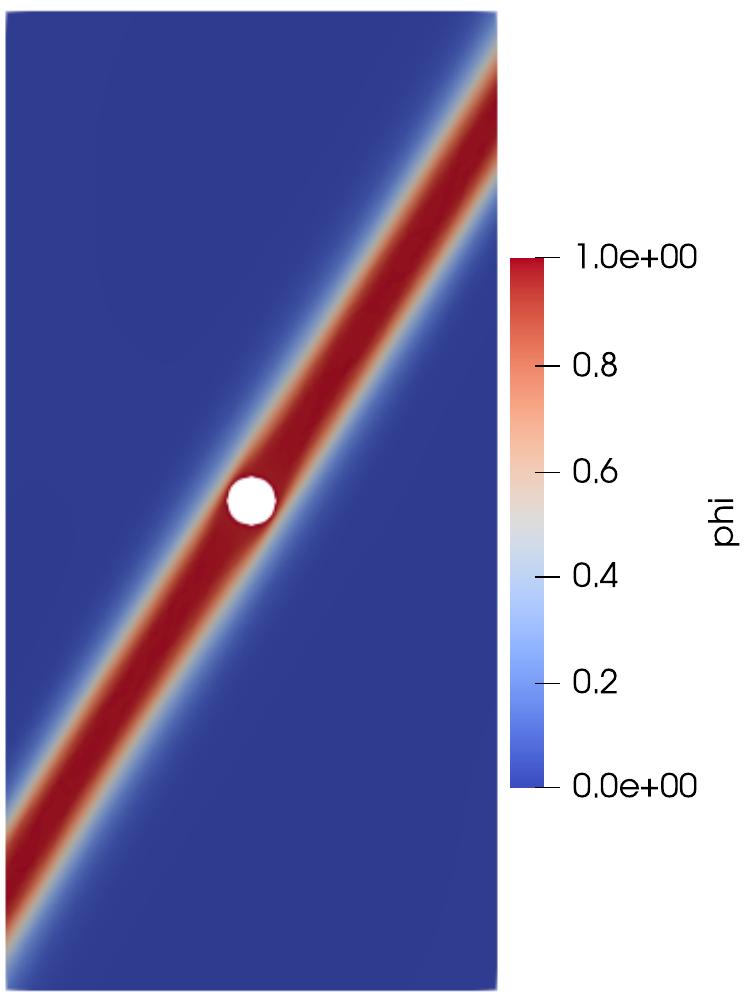}& \includegraphics[height=0.18\textwidth]{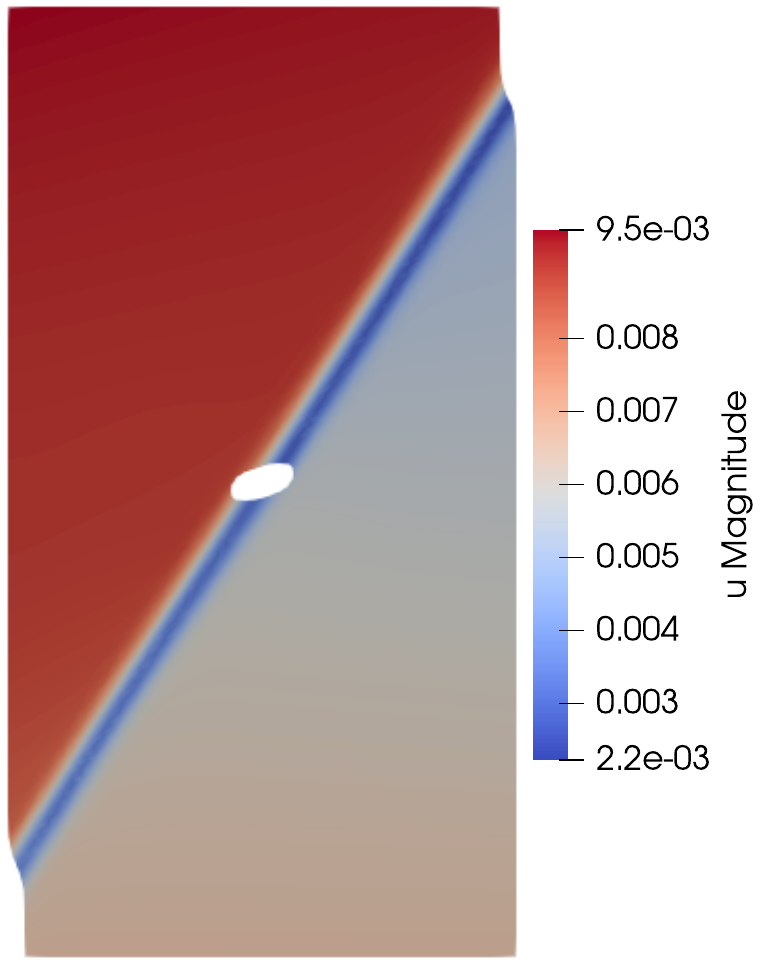}  & \includegraphics[height=0.18\textwidth]{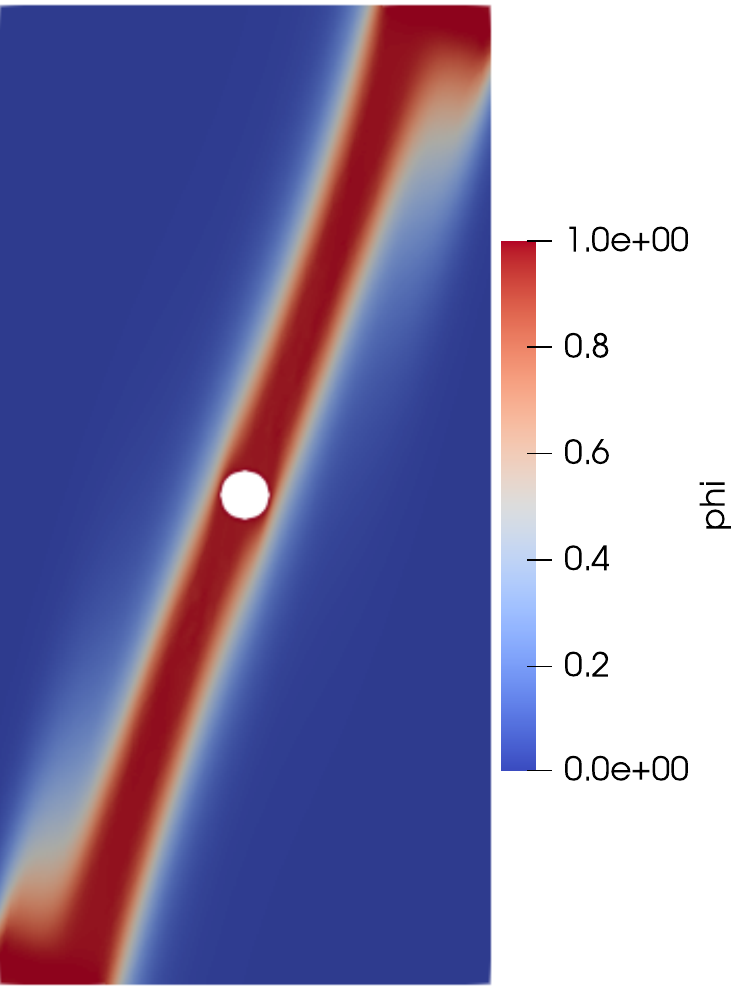} & \includegraphics[height=0.18\textwidth]{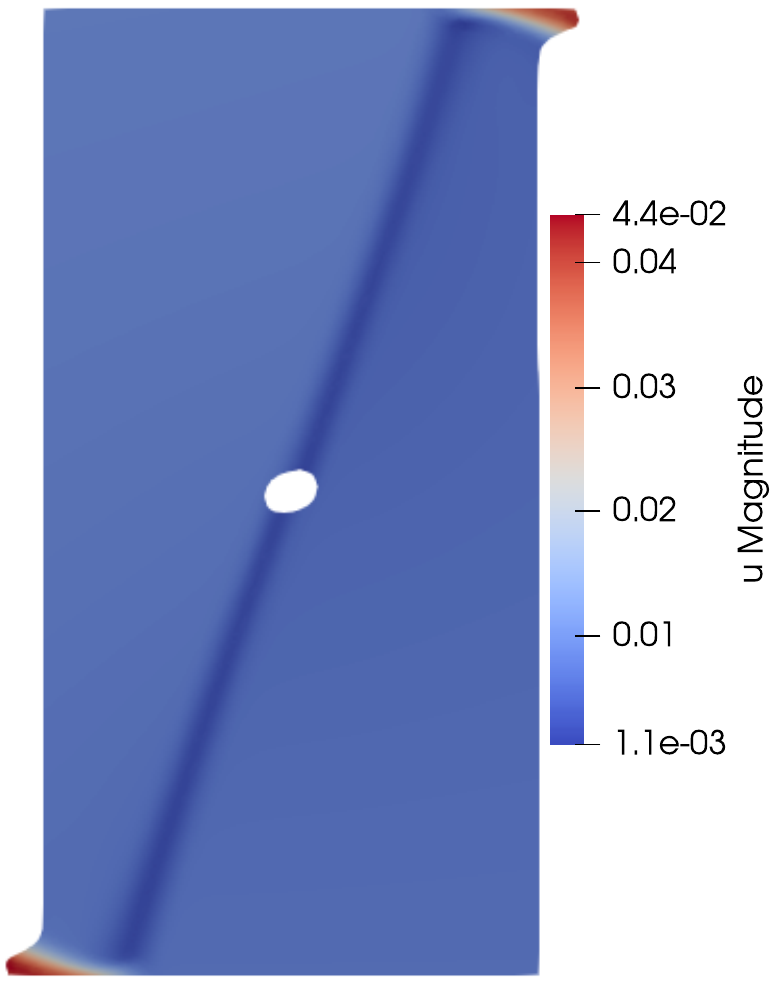}  \\
          \hline
          \raisebox{2cm}{$\sigma_3 = \SI{14}{MPa}$} & 	\includegraphics[height=0.18\textwidth]{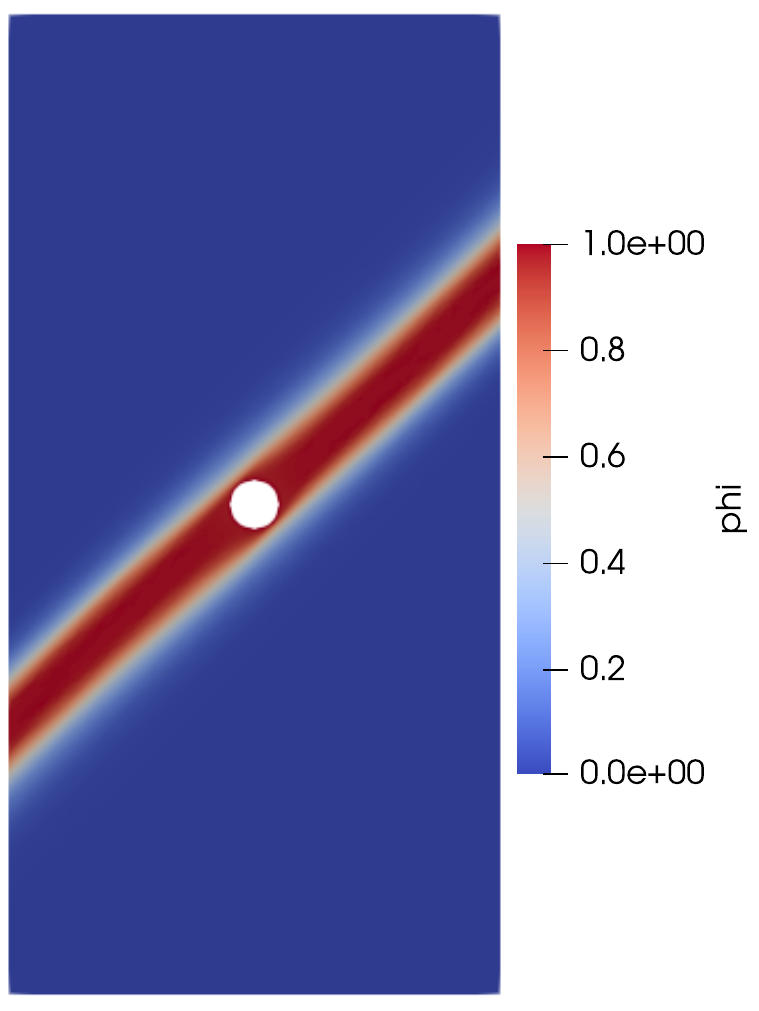} & \includegraphics[height=0.18\textwidth]{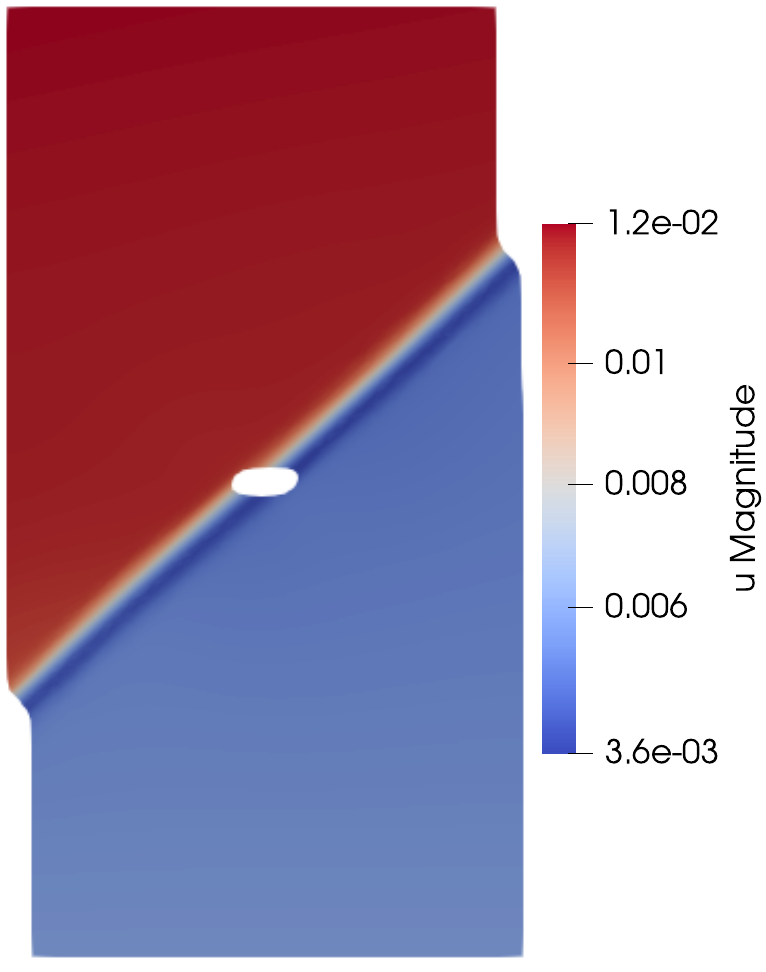} & 	\includegraphics[height=0.18\textwidth]{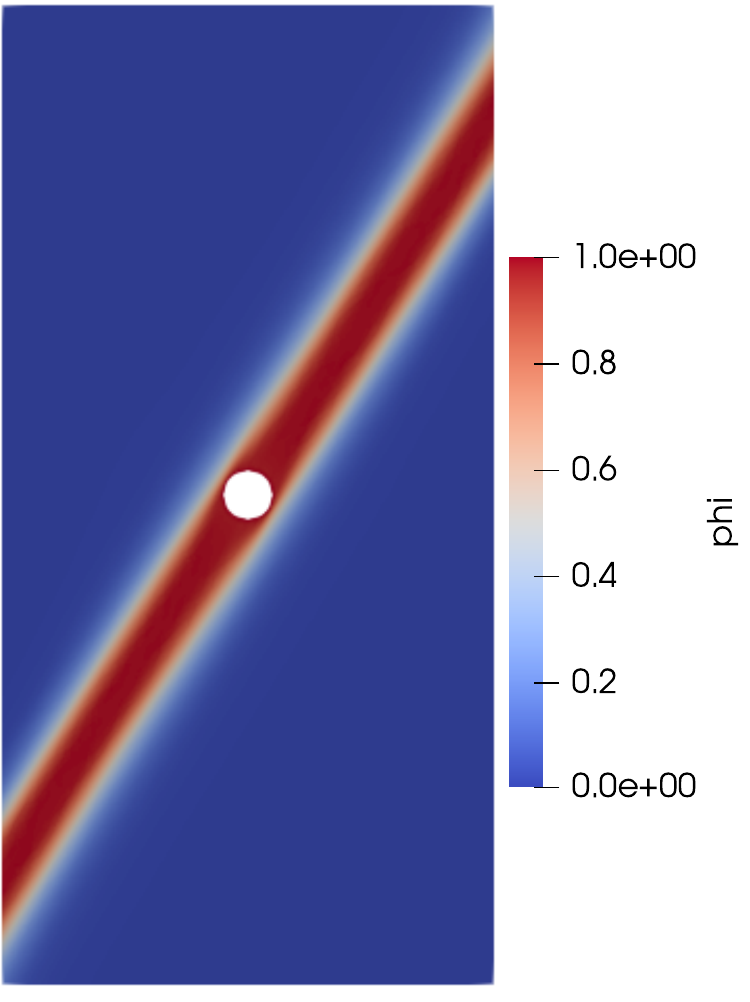}& \includegraphics[height=0.18\textwidth]{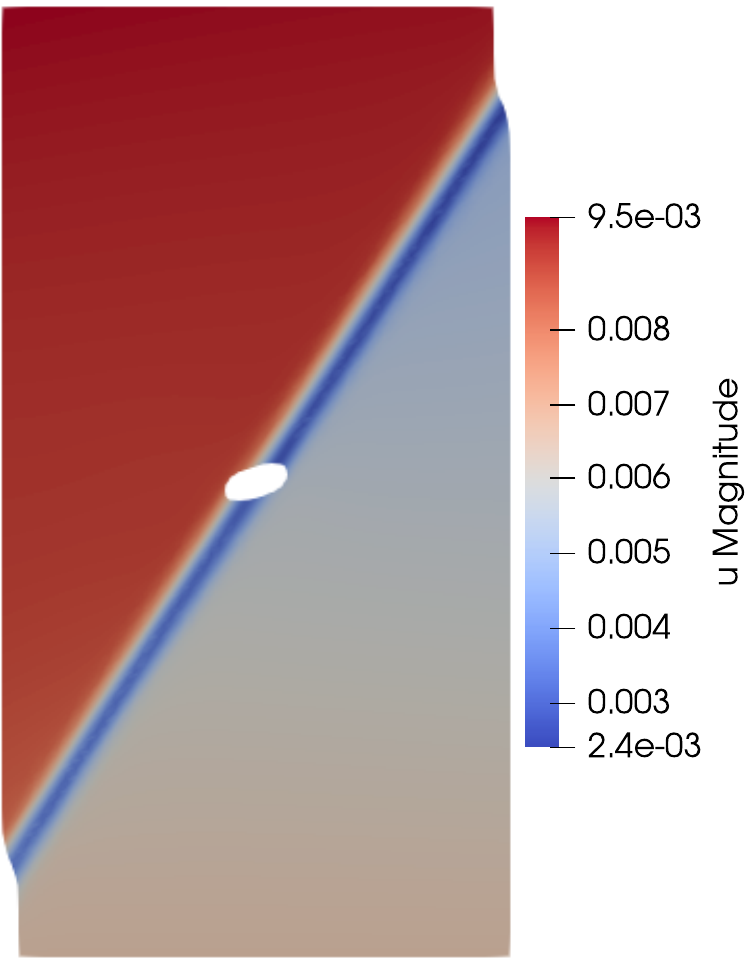} & \includegraphics[height=0.18\textwidth]{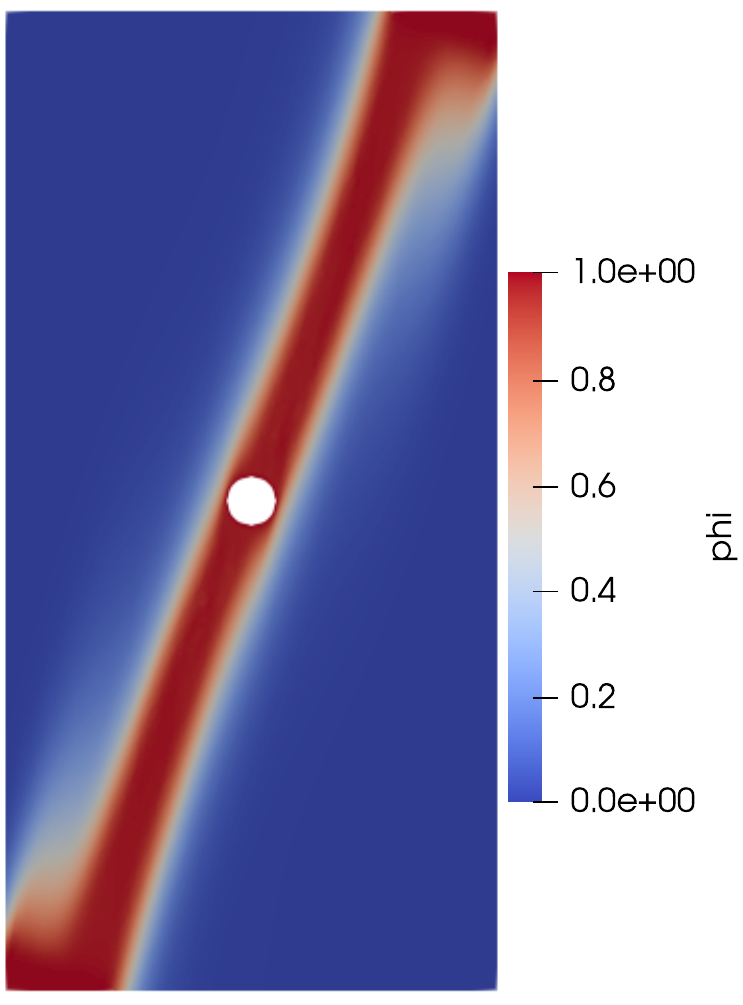} & \includegraphics[height=0.18\textwidth]{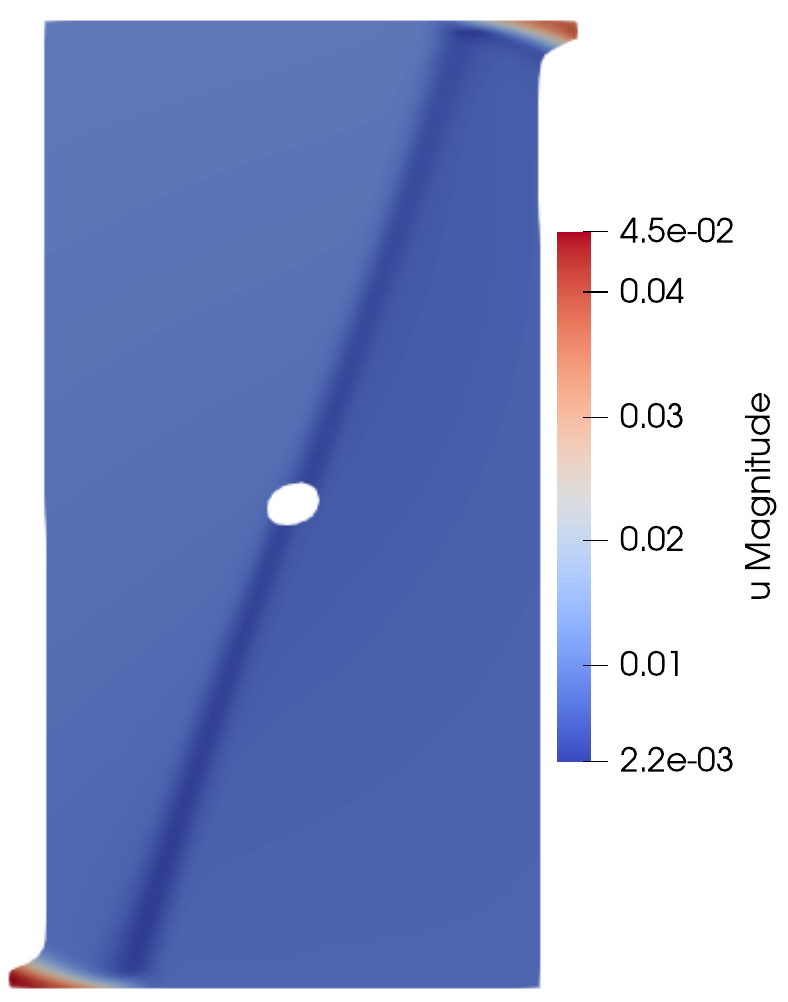} \\
          \hline
          \raisebox{2cm}{$\sigma_3 = \SI{35}{MPa}$} & 	\includegraphics[height=0.18\textwidth]{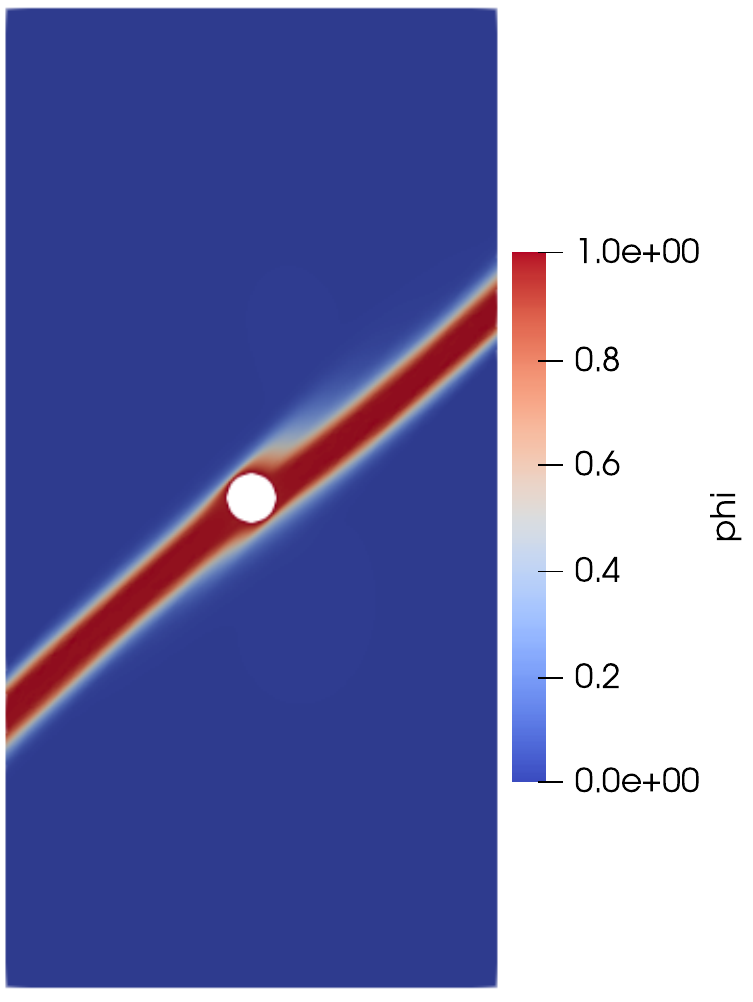} & \includegraphics[height=0.18\textwidth]{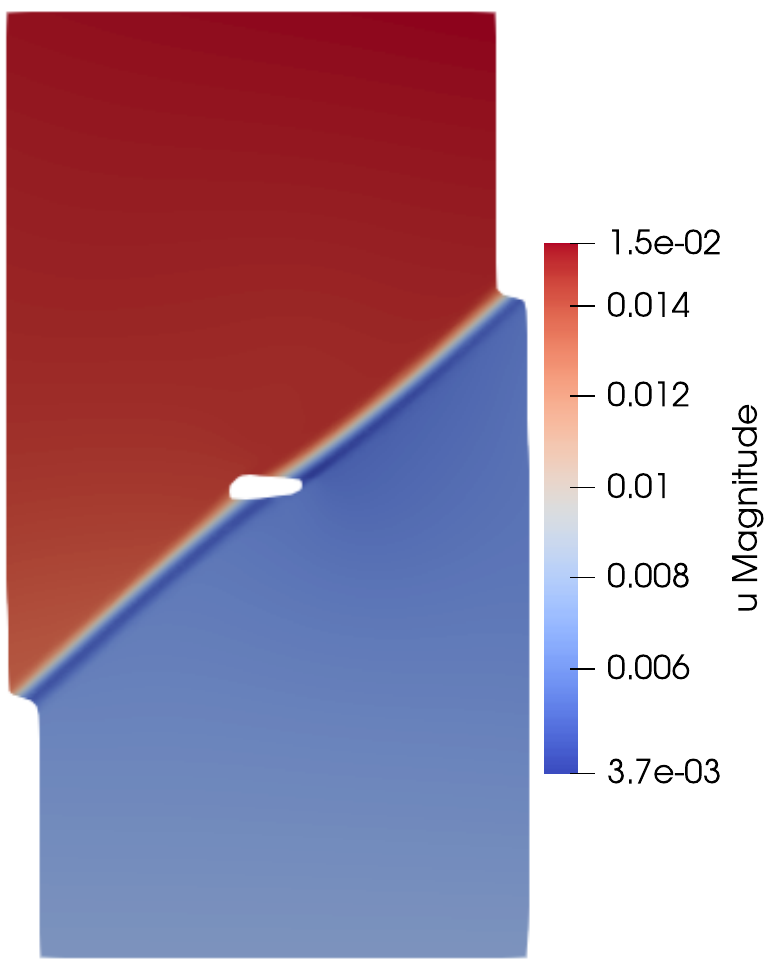} &	\includegraphics[height=0.18\textwidth]{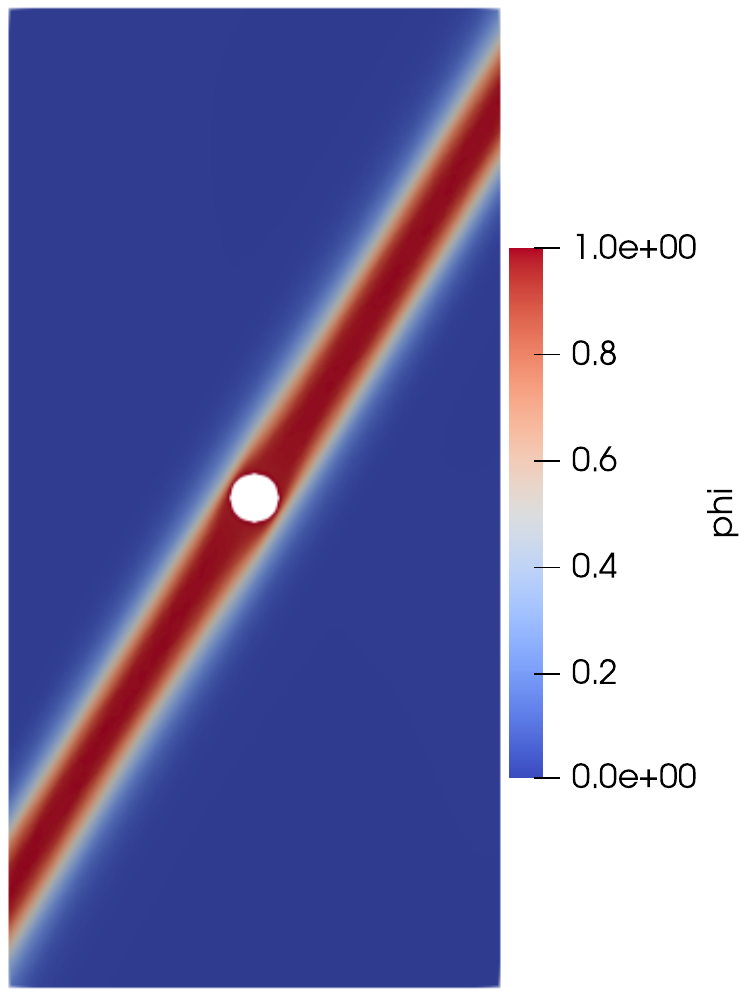}& \includegraphics[height=0.18\textwidth]{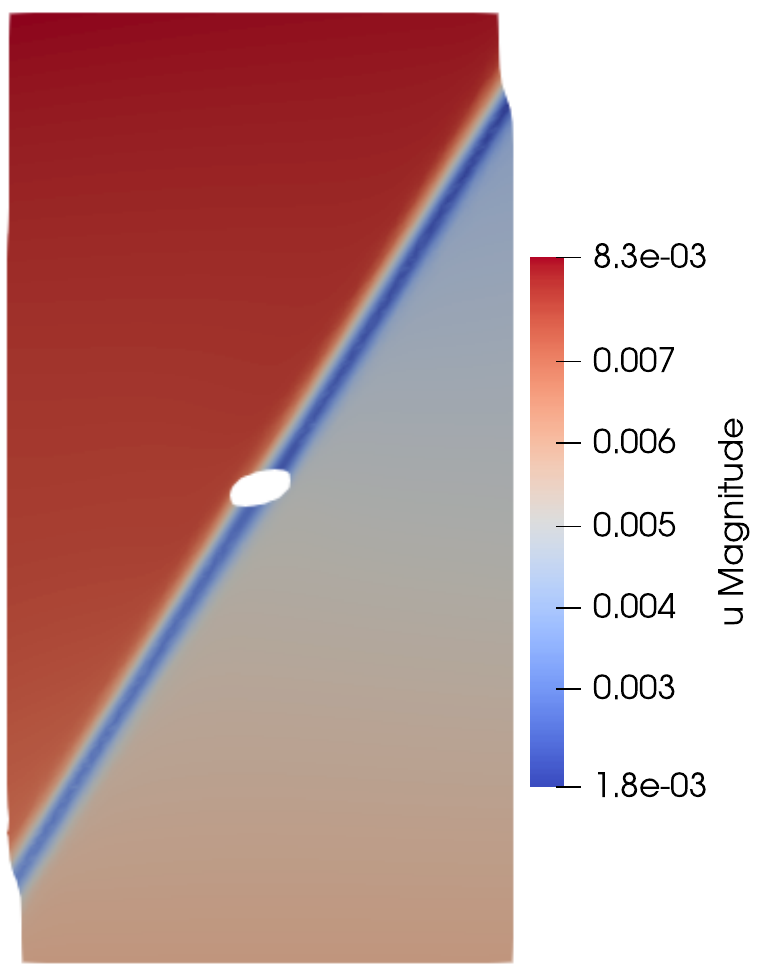} & \includegraphics[height=0.18\textwidth]{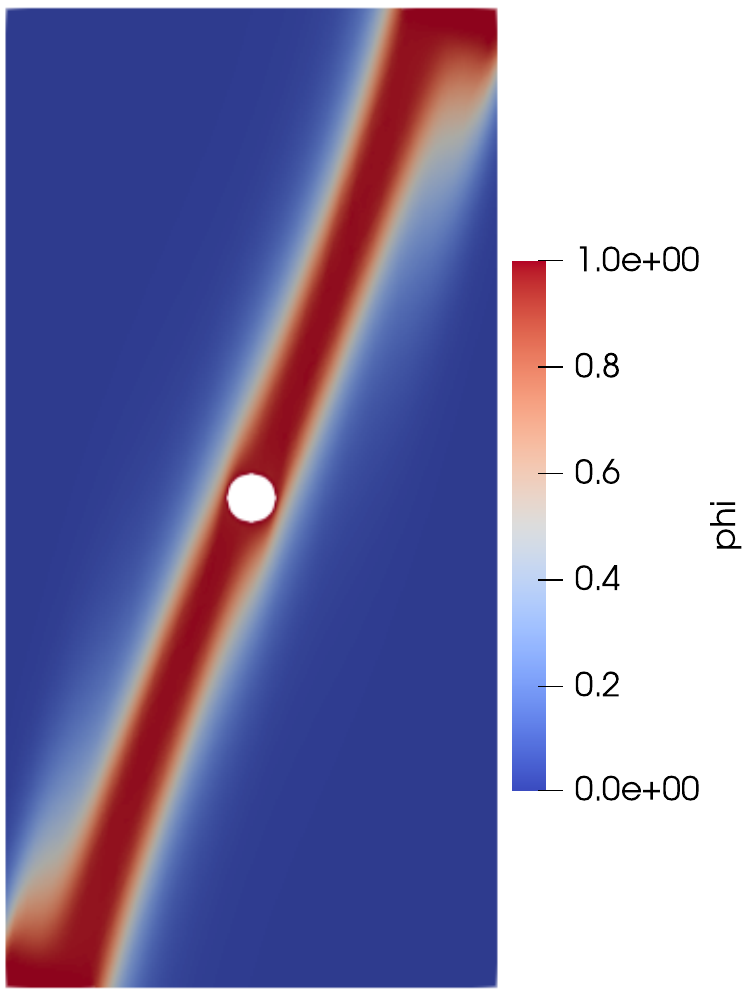} & \includegraphics[height=0.18\textwidth]{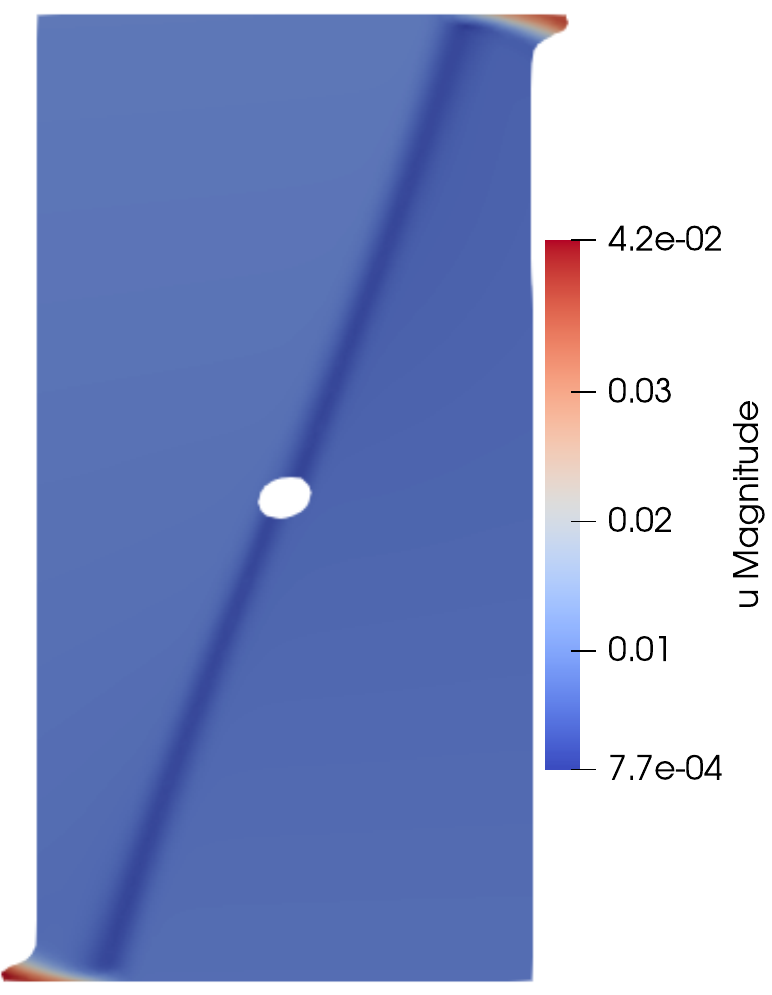} \\
    \end{tabular}
    \caption{
        Crack growth under compression with various confining pressures (\(\sigma_3\)) at layer orientations of \(45^\circ\), \(60^\circ\), and \(75^\circ\) generally aligns with the experimental observations depicted in Figure \ref{fig:Tien_experimental}. The phase-field parameter $\phi$ and the shape of the deformed specimen are both shown, with the color in the deformation plots corresponding to the norm of the displacement.
    }
 \label{fig:Reconstruct Tien}
\end{figure}

Figure \ref{fig:Tien_experimental} shows the fracture patterns experimentally observed for various rock layers under differing confining compression levels in \cite{tien2006experimental}.
We focus on reproducing the failure modes denoted SD with the homogenized model, without any further tuning of the fracture anisotropy parameters beyond that described in the previous section.
Specifically, we focus on the homogenized model for layer orientations at $45^\circ$, $60^\circ$, and $75^\circ$, under various confining pressures.
As shown in Figure \ref{fig:Reconstruct Tien}, the crack path in our simulations matches well with the experimental observations of \cite{tien2006experimental} for several, but not all, layer orientations and confining pressures.

We highlight that the stress at which the fracture propagates across the entire specimen is roughly $4$ times smaller than that reported in the experiments of \cite{tien2006experimental}. 
The issue of initiation of fracture in phase-field models is a well-known open problem with the model, e.g. \cite{kuhn2013,Maurini2018,agrawal-dayal-2015a,agrawal-dayal-2015b}.
In brief, it depends on the choice of the regularization parameter $\epsilon$, which is chosen based on considerations related to the smearing out of the crack following the general principles of phase-field modeling \cite{agrawal-dayal-2015a,agrawal-dayal-2015b}.
Since the ratio between the elastic energy and the work to fracture --- given roughly by the quantity  $\frac{G_c}{\epsilon}$ --- plays a leading role in determining the propagation of fractures, we can conclude that the value of $\epsilon$ that we used in our calculations should be approximately $4$ times smaller to align with the experimental values observed in \cite{tien2006experimental}.
Additionally, the presence of a hole in the middle of the domain in our simulations, which is meant to deterministically initiate the fracture at the center of the domain following \cite{zhao2018strength}, causes stress concentrations that lead to fracture at lower loads.
However, the primary focus of this paper is on the propagation modes and not initiation, and we will therefore not further explore this issue.

We also highlight that the experimental study of \cite{tien2006experimental} examines various types of failures, labeled as TD (Tensile-split along discontinuities), TM (Tensile fracture across discontinuities), SM (Sliding failure across discontinuities), and SD (Sliding failure along discontinuities) in Fig. \ref{fig:Tien_experimental}. 
In this work, we specifically focus on replicating failures corresponding to the SD failure mode, as our current model does not seem to adequately capture the other types of failures. 
Consequently, we focus on layer angles of $45^\circ$, $60^\circ$, and $90^\circ$.

As depicted in Figure \ref{fig:Reconstruct Tien}, the overall path of crack growth in the experimental data from \cite{tien2006experimental} is effectively captured by our homogenized model. However, specific details in the experimental results, such as the jump in the crack path at \(\sigma_3 = \SI{0}{\mega\pascal}\), are not fully replicated by the homogenized model, which is to be expected.
We highlight an important feature that emerges: the confining pressure has a minor effect on the overall crack path and deformation. As in the experiments of \cite{tien2006experimental}, the crack path is dominated by the anisotropy, and consequently by the orientation of the layering.

\section{Discussion}

In this study, we have developed a homogenized phase-field model for fracture that is anisotropic in both elasticity and fracture energy and is able to account for compressive loads across the crack face building on our previous work \cite{hakimzadeh2022phase}.
We applied this approach to model crack growth in transverse isotropic layered materials under various complex loadings, and are able to capture the significance of anisotropic crack growth when compared to fully-resolved calculations and experimental observations. 

We have further studied the important and challenging case of wing cracking. 
We have shown that the model is able to capture wing cracking under compression using a phase-field approach. 
Unlike prior phase-field models of wing cracks that use an external boundary to model the initial crack, we demonstrate that we can capture this by modeling the initial crack through the phase-field.
This provides much more generality since we do not have to introduce external boundaries within the domain.

We highlight that we have assumed that there is zero friction between crack faces for simplicity. 
For future studies aiming to infuse more realistic physics into the model, it is important to further incorporate friction in the model \cite{fei2020phase-cmame,fei2020phase-ijnme}. 
The QR decomposition, that is used in our model \cite{hakimzadeh2022phase}, offers a method to transparently separate the relative slip of the crack faces from other crack deformation modes. 
Therefore, we anticipate that the kinematics introduced in our model will enable this as a straightforward extension.
Other important directions for the future are to couple the phase-field fracture method to, first, plasticity following \cite{choo2020anisotropic,semnani2016thermoplasticity,semnani2020inelastic}; second, poromechanics following \cite{karimi2022energetic,karimi2023high,chua2024deformation}; and, third, accurate and realistic crack nucleation models following \cite{agrawal-dayal-2015a,agrawal-dayal-2015b}.

\paragraph*{Authorship Contribution.}

Maryam Hakimzadeh: Conceptualization; Investigation; Writing – Original Draft Preparation.
Carlos Mora-Corral: Conceptualization; Writing – Review \& Editing.
Noel Walkington: Conceptualization; Writing – Review \& Editing.
Giuseppe Buscarnera: Conceptualization; Writing – Review \& Editing.
Kaushik Dayal: Conceptualization; Investigation; Writing – Original Draft Preparation; Writing – Review \& Editing; Supervision.

\paragraph*{Software Availability.}

A version of the code developed for this work, along with a summary of the constants used in this paper, is available at \\ \url{https://github.com/maryhzd/Phase-Field-Anisotropic-Fracture-Mechanics} .

\paragraph*{Acknowledgments.}
    We acknowledge financial support from NSF (2108784, 2012259), ARO (MURI W911NF-19-1-0245, W911NF-24-2-0184), AEI (PID2021-124195NB-C32, CEX2019-000904-S), ERC (834728), and PRICIT; and NSF for XSEDE computing resources provided by Pittsburgh Supercomputing Center.


\newcommand{\etalchar}[1]{$^{#1}$}

\end{document}

%% file: preamble.tex
\usepackage{isomath}
\usepackage{amsmath,amsthm}
\usepackage{amsbsy}
\usepackage{amssymb}
\usepackage{amscd}
\usepackage{amsfonts}
\usepackage{stmaryrd}
\usepackage{siunitx}
\usepackage{euscript}
\usepackage[utf8]{inputenc}
\usepackage[T1]{fontenc}
\usepackage{newtxtext} 
\everymath{\displaystyle}
\usepackage{exscale}
\usepackage{microtype}
\usepackage{hyperref}
\usepackage{booktabs}
\usepackage{algorithm}
\usepackage{algpseudocode}

\usepackage{graphicx}
\usepackage{boxedminipage}
\usepackage{calc}
\usepackage[usenames,dvipsnames]{xcolor}
\graphicspath{ {media/} }
\usepackage[caption=false,justification=centerlast]{subfig}

\usepackage{setspace}
\usepackage{enumitem}
\setitemize{noitemsep,topsep=0pt,parsep=0pt,partopsep=0pt}
\setenumerate{noitemsep,topsep=0pt,parsep=0pt,partopsep=0pt}
\setdescription{noitemsep,topsep=0pt,parsep=0pt,partopsep=0pt}

\usepackage{soul} 
\usepackage[normalem]{ulem}

\usepackage{orcidlink}
\usepackage{siunitx}
\usepackage[small]{titlesec}

\titlespacing*{\section}{0pt}{12pt plus 4pt minus 2pt}{2pt plus 2pt minus 2pt}
\titlespacing*{\subsection}{0pt}{12pt plus 4pt minus 2pt}{2pt plus 2pt minus 2pt}
\titlespacing*\subsubsection{0pt}{12pt plus 4pt minus 2pt}{2pt plus 2pt minus 2pt}
\titlespacing*\paragraph{0pt}{12pt plus 4pt minus 2pt}{2pt plus 2pt minus 2pt}

\makeatletter

    \renewcommand*{\p@subsection}{}
    
    \renewcommand*{\p@subsubsection}{}
\makeatother

\input{definitions}

%% file: definitions.tex
\usepackage{isomath}
\usepackage{amsmath}
\usepackage{amssymb}
\usepackage{amscd}
\usepackage{amsfonts}
\usepackage{upgreek}
\usepackage{todonotes}

\theoremstyle{definition}

\AtEndEnvironment{definition}{\null\hfill\qedsymbol}

\AtEndEnvironment{remark}{\null\hfill\qedsymbol}

\AtEndEnvironment{example}{\null\hfill\qedsymbol}

\AtEndEnvironment{assumption}{\null\hfill\qedsymbol}

\DeclareMathOperator*{\argmin}{argmin}

\newcommand{\dm}{\ \mathrm{d}}

\newcommand{\variation}[2]{\updelta_{#2} #1}

\newcommand{\bfa}{{\mathbold a}}

\newcommand{\bfd}{{\mathbold d}}
\newcommand{\bfe}{{\mathbold e}}

\newcommand{\bfp}{{\mathbold p}}

\newcommand{\bfx}{{\mathbold x}}
\newcommand{\bfy}{{\mathbold y}}

\newcommand{\bfA}{{\mathbold A}}

\newcommand{\bfI}{{\mathbold I}}

%% file: manuscript.bbl
\begin{thebibliography}{HADMC22}
	
	\bibitem[AD15a]{agrawal-dayal-2015a}
	Vaibhav Agrawal and Kaushik Dayal.
	\newblock A dynamic phase-field model for structural transformations and
	twinning: Regularized interfaces with transparent prescription of complex
	kinetics and nucleation. part {I}: Formulation and one-dimensional
	characterization.
	\newblock {\em Journal of the Mechanics and Physics of Solids}, 2015.
	
	\bibitem[AD15b]{agrawal-dayal-2015b}
	Vaibhav Agrawal and Kaushik Dayal.
	\newblock A dynamic phase-field model for structural transformations and
	twinning: Regularized interfaces with transparent prescription of complex
	kinetics and nucleation. part {II}: Two-dimensional characterization and
	boundary kinetics.
	\newblock {\em Journal of the Mechanics and Physics of Solids}, 2015.
	
	\bibitem[AD17]{agrawal2017dependence}
	Vaibhav Agrawal and Kaushik Dayal.
	\newblock Dependence of equilibrium {G}riffith surface energy on crack speed in
	phase-field models for fracture coupled to elastodynamics.
	\newblock {\em International Journal of Fracture}, 207:243--249, 2017.
	
	\bibitem[Agr16]{agrawal2016multiscale}
	Vaibhav Agrawal.
	\newblock {\em Multiscale phase-field model for phase transformation and
		fracture}.
	\newblock PhD thesis, Carnegie Mellon University, 2016.
	
	\bibitem[Ama12]{amadei2012rock}
	Bernard Amadei.
	\newblock {\em Rock anisotropy and the theory of stress measurements},
	volume~2.
	\newblock Springer Science \& Business Media, 2012.
	
	\bibitem[AMM09]{amor2009regularized}
	Hanen Amor, Jean-Jacques Marigo, and Corrado Maurini.
	\newblock Regularized formulation of the variational brittle fracture with
	unilateral contact: Numerical experiments.
	\newblock {\em Journal of the Mechanics and Physics of Solids},
	57(8):1209--1229, 2009.
	
	\bibitem[AT90]{AmTo90}
	Luigi Ambrosio and Vincenzo~Maria Tortorelli.
	\newblock Approximation of functionals depending on jumps by elliptic
	functionals via {$\Gamma$}-convergence.
	\newblock {\em Comm. Pure Appl. Math.}, 43(8):999--1036, 1990.
	
	\bibitem[AT92]{AmTo92}
	Luigi Ambrosio and Vincenzo~Maria Tortorelli.
	\newblock On the approximation of free discontinuity problems.
	\newblock {\em Boll. Un. Mat. Ital. B (7)}, 6(1):105--123, 1992.
	
	\bibitem[Bac62]{backus1962long}
	George~E Backus.
	\newblock Long-wave elastic anisotropy produced by horizontal layering.
	\newblock {\em Journal of Geophysical Research}, 67(11):4427--4440, 1962.
	
	\bibitem[BBNB15]{bennett2015instrumented}
	Kane~C Bennett, Lucas~A Berla, William~D Nix, and Ronaldo~I Borja.
	\newblock Instrumented nanoindentation and 3d mechanistic modeling of a shale
	at multiple scales.
	\newblock {\em Acta Geotechnica}, 10:1--14, 2015.
	
	\bibitem[BE98]{bobet1998fracture}
	Antonio Bobet and HH~Einstein.
	\newblock Fracture coalescence in rock-type materials under uniaxial and
	biaxial compression.
	\newblock {\em International Journal of Rock Mechanics and Mining Sciences},
	35(7):863--888, 1998.
	
	\bibitem[BS18]{bryant2018mixed}
	Eric~C Bryant and WaiChing Sun.
	\newblock A mixed-mode phase field fracture model in anisotropic rocks with
	consistent kinematics.
	\newblock {\em Computer Methods in Applied Mechanics and Engineering},
	342:561--584, 2018.
	
	\bibitem[BSC23]{bijaya2023consistent}
	Ananya Bijaya, Shiv Sagar, and Shubhankar~Roy Chowdhury.
	\newblock A consistent multi-phase-field formulation for anisotropic brittle
	fracture.
	\newblock {\em Engineering Fracture Mechanics}, page 109825, 2023.
	
	\bibitem[BYT{\etalchar{+}}21]{barchiesi2021computation}
	Emilio Barchiesi, Hua Yang, Chuong~Anthony Tran, Luca Placidi, and Wolfgang~H
	M{\"u}ller.
	\newblock Computation of brittle fracture propagation in strain gradient
	materials by the fenics library.
	\newblock {\em Mathematics and Mechanics of Solids}, 26(3):325--340, 2021.
	
	\bibitem[CAB{\etalchar{+}}22]{chua2022phase}
	Janel Chua, Vaibhav Agrawal, Timothy Breitzman, George Gazonas, and Kaushik
	Dayal.
	\newblock Phase-field modeling and peridynamics for defect dynamics, and an
	augmented phase-field model with viscous stresses.
	\newblock {\em Journal of the Mechanics and Physics of Solids}, 159:104716,
	2022.
	
	\bibitem[CKK{\etalchar{+}}24]{chua2024deformation}
	Janel Chua, Mina Karimi, Patrick Kozlowski, Mehrdad Massoudi, Santosh
	Narasimhachary, Kai Kadau, George Gazonas, and Kaushik Dayal.
	\newblock Deformation decomposition versus energy decomposition for chemo-and
	poro-mechanics.
	\newblock {\em Journal of Applied Mechanics}, 91(1):014501, 2024.
	
	\bibitem[Cla21]{clayton2021nonlinear}
	JD~Clayton.
	\newblock Nonlinear thermodynamic phase field theory with application to
	fracture and dynamic inelastic phenomena in ceramic polycrystals.
	\newblock {\em Journal of the Mechanics and Physics of Solids}, 157:104633,
	2021.
	
	\bibitem[CLK23]{clayton2023phase}
	JD~Clayton, RB~Leavy, and J~Knap.
	\newblock Phase field theory for pressure-dependent strength in brittle solids
	with dissipative kinetics.
	\newblock {\em Mechanics Research Communications}, 129:104097, 2023.
	
	\bibitem[CQH22]{cui2022role}
	Zhendong Cui, Shengwen Qi, and Weige Han.
	\newblock The role of weak bedding planes in the cross-layer crack growth paths
	of layered rocks.
	\newblock {\em Geomechanics and Geophysics for Geo-Energy and Geo-Resources},
	8:1--21, 2022.
	
	\bibitem[CSW20]{choo2020anisotropic}
	Jinhyun Choo, Shabnam~J Semnani, and Joshua~A White.
	\newblock An anisotropic viscoplasticity model for shale based on layered
	microstructure homogenization.
	\newblock {\em arXiv preprint arXiv:2008.10910}, 2020.
	
	\bibitem[CZCF22]{chang2022crack}
	Xu~Chang, Xu~Zhang, Long Cheng, and Lei Fu.
	\newblock Crack path at bedding planes of cracked layered rocks.
	\newblock {\em Journal of Structural Geology}, 154:104504, 2022.
	
	\bibitem[DKAK23]{dammass2023phase}
	Franz Damma{\ss}, Karl~A Kalina, Marreddy Ambati, and Markus K{\"a}stner.
	\newblock Phase-field modelling and analysis of rate-dependent fracture
	phenomena at finite deformation.
	\newblock {\em Computational Mechanics}, 72(5):859--883, 2023.
	
	\bibitem[DLM22]{de2022nucleation}
	Laura De~Lorenzis and Corrado Maurini.
	\newblock Nucleation under multi-axial loading in variational phase-field
	models of brittle fracture.
	\newblock {\em International Journal of Fracture}, 237(1-2):61--81, 2022.
	
	\bibitem[Eri91]{ericksen1989liquid}
	Jerald~L Ericksen.
	\newblock Liquid crystals with variable degree of orientation.
	\newblock {\em Archive for Rational Mechanics and Analysis}, 113:97--120, 1991.
	
	\bibitem[FC20a]{fei2020phase-ijnme}
	Fan Fei and Jinhyun Choo.
	\newblock A phase-field method for modeling cracks with frictional contact.
	\newblock {\em International Journal for Numerical Methods in Engineering},
	121(4):740--762, 2020.
	
	\bibitem[FC20b]{fei2020phase-cmame}
	Fan Fei and Jinhyun Choo.
	\newblock A phase-field model of frictional shear fracture in geologic
	materials.
	\newblock {\em Computer Methods in Applied Mechanics and Engineering},
	369:113265, 2020.
	
	\bibitem[FM98]{francfort1998revisiting}
	Gilles~A Francfort and J-J Marigo.
	\newblock Revisiting brittle fracture as an energy minimization problem.
	\newblock {\em Journal of the Mechanics and Physics of Solids},
	46(8):1319--1342, 1998.
	
	\bibitem[GND{\etalchar{+}}24]{gupta2024damage}
	Abhinav Gupta, Duc~Tien Nguyen, Ravindra Duddu, et~al.
	\newblock Damage mechanics challenge: Predictions from an adaptive finite
	element implementation of the stress-based phase-field fracture model.
	\newblock {\em Engineering Fracture Mechanics}, page 110252, 2024.
	
	\bibitem[Goo93]{thomas1993book}
	RE~Goodman.
	\newblock {\em Engineering Geology — Rock in Engineering Construction},
	volume~15.
	\newblock Wiley, Chichester, UK, 1993.
	
	\bibitem[HADMC22]{hakimzadeh2022phase}
	Maryam Hakimzadeh, Vaibhav Agrawal, Kaushik Dayal, and Carlos Mora-Corral.
	\newblock Phase-field finite deformation fracture with an effective energy for
	regularized crack face contact.
	\newblock {\em Journal of the Mechanics and Physics of Solids}, 167:104994,
	2022.
	
	\bibitem[Hil48]{hill1948theory}
	Rodney Hill.
	\newblock A theory of the yielding and plastic flow of anisotropic metals.
	\newblock {\em Proceedings of the Royal Society of London. Series A.
		Mathematical and Physical Sciences}, 193(1033):281--297, 1948.
	
	\bibitem[HK05]{hakim2005crack}
	Vincent Hakim and Alain Karma.
	\newblock Crack path prediction in anisotropic brittle materials.
	\newblock {\em Physical review letters}, 95(23):235501, 2005.
	
	\bibitem[HK09]{hakim2009laws}
	Vincent Hakim and Alain Karma.
	\newblock Laws of crack motion and phase-field models of fracture.
	\newblock {\em Journal of the Mechanics and Physics of Solids}, 57(2):342--368,
	2009.
	
	\bibitem[HNN85]{horii1985compression}
	H~Horii and S~Nemat-Nasser.
	\newblock Compression-induced microcrack growth in brittle solids: Axial
	splitting and shear failure.
	\newblock {\em Journal of Geophysical Research: Solid Earth},
	90(B4):3105--3125, 1985.
	
	\bibitem[Jae60]{jaeger1960shear}
	JC~Jaeger.
	\newblock Shear failure of anistropic rocks.
	\newblock {\em Geological magazine}, 97(1):65--72, 1960.
	
	\bibitem[KM13]{kuhn2013}
	Charlotte Kuhn and Ralf M{\"u}ller.
	\newblock Crack nucleation in phase field fracture models.
	\newblock {\em ICF13}, 579, 2013.
	
	\bibitem[KMDP23]{karimi2023high}
	Mina Karimi, Mehrdad Massoudi, Kaushik Dayal, and Matteo Pozzi.
	\newblock High-dimensional nonlinear bayesian inference of poroelastic fields
	from pressure data.
	\newblock {\em Mathematics and Mechanics of Solids}, 28(9):2108--2131, 2023.
	
	\bibitem[KMW{\etalchar{+}}22]{karimi2022energetic}
	Mina Karimi, Mehrdad Massoudi, Noel Walkington, Matteo Pozzi, and Kaushik
	Dayal.
	\newblock Energetic formulation of large-deformation poroelasticity.
	\newblock {\em International Journal for Numerical and Analytical Methods in
		Geomechanics}, 46(5):910--932, 2022.
	
	\bibitem[KV93]{khan1993anisotropy}
	AA~Khan and JFV Vincent.
	\newblock Anisotropy in the fracture properties of apple flesh as investigated
	by crack-opening tests.
	\newblock {\em Journal of materials science}, 28:45--51, 1993.
	
	\bibitem[LMW12]{logg2012automated}
	Anders Logg, Kent-Andre Mardal, and Garth Wells.
	\newblock {\em Automated solution of differential equations by the finite
		element method: The FEniCS book}, volume~84.
	\newblock Springer Science \& Business Media, 2012.
	
	\bibitem[LPM{\etalchar{+}}15]{li2015phase}
	Bin Li, Christian Peco, Daniel Mill{\'a}n, Irene Arias, and Marino Arroyo.
	\newblock Phase-field modeling and simulation of fracture in brittle materials
	with strongly anisotropic surface energy.
	\newblock {\em International Journal for Numerical Methods in Engineering},
	102(3-4):711--727, 2015.
	
	\bibitem[LWCL21]{li2021phase}
	Haifeng Li, Wei Wang, Yajun Cao, and Shifan Liu.
	\newblock Phase-field modeling fracture in anisotropic materials.
	\newblock {\em Advances in Civil Engineering}, 2021(1):4313755, 2021.
	
	\bibitem[LZW{\etalchar{+}}20]{liu2020dynamic}
	K~Liu, J~Zhao, G~Wu, A~Maksimenko, A~Haque, and QB~Zhang.
	\newblock Dynamic strength and failure modes of sandstone under biaxial
	compression.
	\newblock {\em International Journal of Rock Mechanics and Mining Sciences},
	128:104260, 2020.
	
	\bibitem[MM77]{MoMo77}
	Luciano Modica and Stefano Mortola.
	\newblock Un esempio di {$\Gamma ^{-}$}-convergenza.
	\newblock {\em Boll. Un. Mat. Ital. B (5)}, 14(1):285--299, 1977.
	
	\bibitem[Mod87]{Modica87}
	Luciano Modica.
	\newblock The gradient theory of phase transitions and the minimal interface
	criterion.
	\newblock {\em Arch. Rational Mech. Anal.}, 98(2):123--142, 1987.
	
	\bibitem[MSH22]{mader2022gradient}
	Thomas Mader, Magdalena Schreter, and G{\"u}nter Hofstetter.
	\newblock A gradient enhanced transversely isotropic damage plasticity model
	for rock-formulation and comparison of different approaches.
	\newblock {\em International Journal for Numerical and Analytical Methods in
		Geomechanics}, 46(5):933--960, 2022.
	
	\bibitem[MWH10]{miehe2010thermodynamically}
	Christian Miehe, Fabian Welschinger, and Martina Hofacker.
	\newblock Thermodynamically consistent phase-field models of fracture:
	Variational principles and multi-field {FE} implementations.
	\newblock {\em International journal for numerical methods in engineering},
	83(10):1273--1311, 2010.
	
	\bibitem[NAMP{\etalchar{+}}19]{natarajan2019phase}
	Sundararajan Natarajan, Ratna~K Annabattula, Emilio Mart{\'\i}nez-Pa{\~n}eda,
	et~al.
	\newblock Phase field modelling of crack propagation in functionally graded
	materials.
	\newblock {\em Composites Part B: Engineering}, 169:239--248, 2019.
	
	\bibitem[NM08]{nasseri2008fracture}
	MHB Nasseri and B~Mohanty.
	\newblock Fracture toughness anisotropy in granitic rocks.
	\newblock {\em International Journal of Rock Mechanics and Mining Sciences},
	45(2):167--193, 2008.
	
	\bibitem[NNH82]{nemat1982compression}
	S~Nemat-Nasser and H~Horii.
	\newblock Compression-induced nonplanar crack extension with application to
	splitting, exfoliation, and rockburst.
	\newblock {\em Journal of Geophysical Research: Solid Earth},
	87(B8):6805--6821, 1982.
	
	\bibitem[NNH13]{nemat2013micromechanics}
	Siavouche Nemat-Nasser and Muneo Hori.
	\newblock {\em Micromechanics: overall properties of heterogeneous materials}.
	\newblock Elsevier, 2013.
	
	\bibitem[NS24]{najmeddine2024efficient}
	Aimane Najmeddine and Maryam Shakiba.
	\newblock Efficient bfgs quasi-newton method for large deformation phase-field
	modeling of fracture in hyperelastic materials.
	\newblock {\em Engineering Fracture Mechanics}, page 110463, 2024.
	
	\bibitem[NSHF97]{niandou1997laboratory}
	H~Niandou, JF~Shao, JP~Henry, and D~Fourmaintraux.
	\newblock Laboratory investigation of the mechanical behaviour of tournemire
	shale.
	\newblock {\em International Journal of Rock Mechanics and Mining Sciences},
	34(1):3--16, 1997.
	
	\bibitem[NWD21]{naghibzadeh2021surface}
	S~Kiana Naghibzadeh, Noel Walkington, and Kaushik Dayal.
	\newblock Surface growth in deformable solids using an {E}ulerian formulation.
	\newblock {\em Journal of the Mechanics and Physics of Solids}, 154:104499,
	2021.
	
	\bibitem[NWD22]{naghibzadeh2022accretion}
	S~Kiana Naghibzadeh, Noel Walkington, and Kaushik Dayal.
	\newblock Accretion and ablation in deformable solids with an {E}ulerian
	description: examples using the method of characteristics.
	\newblock {\em Mathematics and Mechanics of Solids}, 27(6):989--1010, 2022.
	
	\bibitem[Par68]{pariseau1968plasticity}
	William~G Pariseau.
	\newblock Plasticity theory for anisotropic rocks and soil.
	\newblock In {\em ARMA US Rock Mechanics/Geomechanics Symposium}, pages
	ARMA--68. ARMA, 1968.
	
	\bibitem[PB10]{park2010crack}
	CH~Park and A~Bobet.
	\newblock Crack initiation, propagation and coalescence from frictional flaws
	in uniaxial compression.
	\newblock {\em Engineering Fracture Mechanics}, 77(14):2727--2748, 2010.
	
	\bibitem[PF90]{penrose1990thermodynamically}
	Oliver Penrose and Paul~C Fife.
	\newblock Thermodynamically consistent models of phase-field type for the
	kinetic of phase transitions.
	\newblock {\em Physica D: Nonlinear Phenomena}, 43(1):44--62, 1990.
	
	\bibitem[QA03]{qiao2003cleavage}
	Y~Qiao and AS~Argon.
	\newblock Cleavage cracking resistance of high angle grain boundaries in
	{F}e--3\% {S}i alloy.
	\newblock {\em Mechanics of materials}, 35(3-6):313--331, 2003.
	
	\bibitem[Rah22]{rahaman2022open}
	Mohammad~Masiur Rahaman.
	\newblock An open-source implementation of a phase-field model for brittle
	fracture using gridap in julia.
	\newblock {\em Mathematics and Mechanics of Solids}, 27(11):2404--2427, 2022.
	
	\bibitem[SB17]{semnani2017quantifying}
	Shabnam~J Semnani and Ronaldo~I Borja.
	\newblock Quantifying the heterogeneity of shale through statistical
	combination of imaging across scales.
	\newblock {\em Acta Geotechnica}, 12:1193--1205, 2017.
	
	\bibitem[SDTD21]{spetz2021modified}
	Alex Spetz, Ralf Denzer, Erika Tudisco, and Ola Dahlblom.
	\newblock A modified phase-field fracture model for simulation of mixed mode
	brittle fractures and compressive cracks in porous rock.
	\newblock {\em Rock Mechanics and Rock Engineering}, 54:5375--5388, 2021.
	
	\bibitem[SK19]{steinke2019phase}
	Christian Steinke and Michael Kaliske.
	\newblock A phase-field crack model based on directional stress decomposition.
	\newblock {\em Computational Mechanics}, 63:1019--1046, 2019.
	
	\bibitem[SSK22]{steinke2022energetically}
	Christian Steinke, Johannes Storm, and Michael Kaliske.
	\newblock Energetically motivated crack orientation vector for phase-field
	fracture with a directional split.
	\newblock {\em International Journal of Fracture}, 237(1):15--46, 2022.
	
	\bibitem[Ste84]{steif1984crack}
	Paul~S Steif.
	\newblock Crack extension under compressive loading.
	\newblock {\em Engineering Fracture Mechanics}, 20(3):463--473, 1984.
	
	\bibitem[SW20]{semnani2020inelastic}
	Shabnam~J Semnani and Joshua~A White.
	\newblock An inelastic homogenization framework for layered materials with
	planes of weakness.
	\newblock {\em Computer Methods in Applied Mechanics and Engineering},
	370:113221, 2020.
	
	\bibitem[SWB16]{semnani2016thermoplasticity}
	Shabnam~J Semnani, Joshua~A White, and Ronaldo~I Borja.
	\newblock Thermoplasticity and strain localization in transversely isotropic
	materials based on anisotropic critical state plasticity.
	\newblock {\em International Journal for Numerical and Analytical Methods in
		Geomechanics}, 40(18):2423--2449, 2016.
	
	\bibitem[TK01]{tien2001failure}
	Yong~Ming Tien and Ming~Chuan Kuo.
	\newblock A failure criterion for transversely isotropic rocks.
	\newblock {\em International Journal of Rock Mechanics and Mining Sciences},
	38(3):399--412, 2001.
	
	\bibitem[TKAK17]{teichtmeister2017phase}
	S~Teichtmeister, D~Kienle, F~Aldakheel, and M-A Keip.
	\newblock Phase field modeling of fracture in anisotropic brittle solids.
	\newblock {\em International Journal of Non-Linear Mechanics}, 97:1--21, 2017.
	
	\bibitem[TKJ06]{tien2006experimental}
	Yong~Ming Tien, Ming~Chuan Kuo, and Charng~Hsein Juang.
	\newblock An experimental investigation of the failure mechanism of simulated
	transversely isotropic rocks.
	\newblock {\em International journal of rock mechanics and mining sciences},
	43(8):1163--1181, 2006.
	
	\bibitem[TLB{\etalchar{+}}18]{Maurini2018}
	E.~Tann{\'{e}}, T.~Li, B.~Bourdin, J.-J. Marigo, and C.~Maurini.
	\newblock Crack nucleation in variational phase-field models of brittle
	fracture.
	\newblock 110:80--99, January 2018.
	
	\bibitem[TRB{\etalchar{+}}13]{takei2013forbidden}
	Atsushi Takei, Beno{\^\i}t Roman, Jos{\'e} Bico, Eugenio Hamm, and Francisco
	Melo.
	\newblock Forbidden directions for the fracture of thin anisotropic sheets: an
	analogy with the {W}ulff plot.
	\newblock {\em Physical review letters}, 110(14):144301, 2013.
	
	\bibitem[TZG{\etalchar{+}}19]{tang2019phase}
	Shan Tang, Gang Zhang, Tian~Fu Guo, Xu~Guo, and Wing~Kam Liu.
	\newblock Phase field modeling of fracture in nonlinearly elastic solids via
	energy decomposition.
	\newblock {\em Computer Methods in Applied Mechanics and Engineering},
	347:477--494, 2019.
	
	\bibitem[TZZL22]{tian2022mixed}
	Fucheng Tian, Jun Zeng, Mengnan Zhang, and Liangbin Li.
	\newblock Mixed displacement--pressure-phase field framework for finite strain
	fracture of nearly incompressible hyperelastic materials.
	\newblock {\em Computer Methods in Applied Mechanics and Engineering},
	394:114933, 2022.
	
	\bibitem[vDEEI20]{van2020strain}
	Nico~P van Dijk, Juan~Jos{\'e} Espadas-Escalante, and Per Isaksson.
	\newblock Strain energy density decompositions in phase-field fracture theories
	for orthotropy and anisotropy.
	\newblock {\em International Journal of Solids and Structures}, 196:140--153,
	2020.
	
	\bibitem[VZC{\etalchar{+}}24]{vicentini2023energy}
	F~Vicentini, C~Zolesi, P~Carrara, C~Maurini, and L~de~Lorenzis.
	\newblock On the energy decomposition in variational phase-field models for
	brittle fracture under multi-axial stress states.
	\newblock {\em International Journal of Fracture}, 2024.
	
	\bibitem[Wit90]{wittke1990rock}
	W~Wittke.
	\newblock {\em Rock Mechanics, Theory and Applications with case histories}.
	\newblock Springer, 1990.
	
	\bibitem[WNN{\etalchar{+}}20]{wu2020phase}
	Jian-Ying Wu, Vinh~Phu Nguyen, Chi~Thanh Nguyen, Danas Sutula, Sina Sinaie, and
	St{\'e}phane~PA Bordas.
	\newblock Phase-field modeling of fracture.
	\newblock {\em Advances in applied mechanics}, 53:1--183, 2020.
	
	\bibitem[Won82]{wong1982shear}
	Teng-fong Wong.
	\newblock Shear fracture energy of westerly granite from post-failure behavior.
	\newblock {\em Journal of Geophysical Research: Solid Earth}, 87(B2):990--1000,
	1982.
	
	\bibitem[WSH21]{wang2021phase}
	Feiyang Wang, Jianfu Shao, and Hongwei Huang.
	\newblock A phase-field modeling method for the mixed-mode fracture of brittle
	materials based on spectral decomposition.
	\newblock {\em Engineering Fracture Mechanics}, 242:107473, 2021.
	
	\bibitem[XXH{\etalchar{+}}24]{xu2024adaptive}
	Bin Xu, Tao Xu, Michael~J Heap, Alexandra~RL Kushnir, Bo-yi Su, and Xiao-cong
	Lan.
	\newblock An adaptive phase field approach to 3d internal crack growth in
	rocks.
	\newblock {\em Computers and Geotechnics}, 173:106551, 2024.
	
	\bibitem[XYSW23]{xing2023adaptive}
	Chen Xing, Tiantang Yu, Yulin Sun, and Yongxiang Wang.
	\newblock An adaptive phase-field model with variable-node elements for
	fracture of hyperelastic materials at large deformations.
	\newblock {\em Engineering Fracture Mechanics}, 281:109115, 2023.
	
	\bibitem[ZB22]{zhao2022double}
	Yang Zhao and Ronaldo~I Borja.
	\newblock A double-yield-surface plasticity theory for transversely isotropic
	rocks.
	\newblock {\em Acta Geotechnica}, 17(11):5201--5221, 2022.
	
	\bibitem[ZJT22]{zhang2022assessment}
	Shuaifang Zhang, Wen Jiang, and Michael~R Tonks.
	\newblock Assessment of four strain energy decomposition methods for phase
	field fracture models using quasi-static and dynamic benchmark cases.
	\newblock {\em Materials Theory}, 6(1):6, 2022.
	
	\bibitem[ZKJT21]{zhang2021phase}
	Shuaifang Zhang, Dong-Uk Kim, Wen Jiang, and Michael~R Tonk.
	\newblock A phase field model of crack propagation in anisotropic brittle
	materials with preferred fracture planes.
	\newblock {\em Computational Materials Science}, 193:110400, 2021.
	
	\bibitem[ZO22]{zhang2022energy}
	Zong-Xian Zhang and Finn Ouchterlony.
	\newblock Energy requirement for rock breakage in laboratory experiments and
	engineering operations: a review.
	\newblock {\em Rock mechanics and rock engineering}, 55(2):629--667, 2022.
	
	\bibitem[ZSVS17]{zhang2017modification}
	Xue Zhang, Scott~W Sloan, Chet Vignes, and Daichao Sheng.
	\newblock A modification of the phase-field model for mixed mode crack
	propagation in rock-like materials.
	\newblock {\em Computer Methods in Applied Mechanics and Engineering},
	322:123--136, 2017.
	
	\bibitem[ZSYB18]{zhao2018strength}
	Yang Zhao, Shabnam~J Semnani, Qing Yin, and Ronaldo~I Borja.
	\newblock On the strength of transversely isotropic rocks.
	\newblock {\em International Journal for Numerical and Analytical Methods in
		Geomechanics}, 42(16):1917--1934, 2018.
	
\end{thebibliography}
